\pdfoutput=1

\documentclass[]{aa}  
\usepackage{etex}
\usepackage[colorlinks]{hyperref}
\usepackage{graphicx}
\usepackage{babel}
\usepackage{caption}
\usepackage{subcaption}
\usepackage[varg]{txfonts}
\usepackage{rotating}
\usepackage{dcolumn}
\usepackage{fmtcount}
\usepackage{longtable, booktabs}
\usepackage{array,multirow}
\usepackage{color}
\usepackage{aas_macros}
\usepackage{multirow}
\usepackage{gensymb}
\usepackage{amssymb}
\usepackage{amsmath}
\usepackage{comment}
\bibpunct{(}{)}{;}{a}{}{,}
\usepackage{xcolor}
\definecolor{black}{rgb}{0.5,0.15,0.15}
\definecolor{dark-blue}{rgb}{0.15,0.15,0.5}
\definecolor{medium-blue}{rgb}{0,0,0.5}
\definecolor{medium-red}{rgb}{1,0,0}
\hypersetup{
    colorlinks={true}, linkcolor={medium-red},
    citecolor={dark-blue}, urlcolor={medium-blue}
}
\usepackage{upgreek}
\usepackage{tipa}

\usepackage{float}

\def\hi{$\rm{H\textsc{i}}$}
\def\h2{$\rm{H_2}$}
\def\ha{$\rm{H\alpha}$}
\def\sfr{$\rm{\dot{\Sigma}_\star}$}
\def\sfe{$\rm{SFE_{\rm{H_2}}}$}
\def\rmolp{$R_{\rm{mol}}/P_{\rm{tot}}$}
\def\rmol{$R_{\rm{mol}}$}
\def\sga{$\rm{\Sigma _{\rm{HI}}}$}
\def\sgm{$\rm{\Sigma _{\rm{H_2}}}$}
\def\msun{$\rm{M_\odot}$}
\def\msunpc{$\rm{M_\odot pc^{-2}}$}
\def\aco{$\rm{\alpha _{\rm{CO}}}$}
\def\acomw{$\rm{\alpha _{\rm{CO}} ^{\rm{MW}}}$}
\def\acou{$\rm{M_\odot(K\ \rm km\ s^{-1}\ pc^2)^{-1}}$}
\def\xco{$\rm{X_{\rm{CO}}}$}
\def\xcomw{$\rm{X_{\rm{CO}} ^{\rm{MW}}}$}
\def\xcou{$\rm{cm^{-2}(K\ \rm km\ s^{-1})^{-1}}$}
\def\mdot{$\dot{M}$}

\def\hsd{$\rm Hi\Sigma_{HI} \ region$}
\def\hsd{$\rm{high\ H\textsc{i}\ surface\ density\ region}$}
\def\sfr{$\rm{\dot{\Sigma}_\star}$}

\newcommand{\fig}[1]{Fig.~\ref{#1}}
\newcommand{\app}[1]{Appendix~\ref{#1}}
\newcommand{\tab}[1]{Table~\ref{#1}}
\newcommand{\eq}[1]{Eq.~\ref{#1}}
\newcommand{\sect}[1]{Sect.~\ref{#1}}

\begin{document}

\title{Gas compression and stellar feedback in the tidally interacting and ram-pressure stripped Virgo spiral galaxy NGC~4654}

   \author{T.~Liz\'ee\inst{1}, B.~Vollmer\inst{1},  J.~Braine\inst{2}, \& F.~Nehlig\inst{1}}

   \institute{Universit\'e de Strasbourg, CNRS, Observatoire astronomique de Strasbourg, UMR 7550, F-67000 Strasbourg, France \and 
          Laboratoire d'Astrophysique de Bordeaux, Univ. Bordeaux, CNRS, B18N, all\'ee Geoffroy Saint-Hilaire, 33615 Pessac, France}

 \abstract{
Due to an environment that promotes gravitational interactions and ram pressure stripping, galaxies within clusters are particularly likely to present unusual interstellar medium (ISM) properties. NGC~4654 is a Virgo cluster galaxy seen almost face-on, which undergoes nearly edge-on gas ram pressure stripping and a fly-by gravitational interaction with another massive galaxy, NGC~4639. %
NGC~4654 shows a strongly compressed gas region near the outer edge of the optical disk, with \hi\ surface densities (\hsd ) significantly exceeding the canonical value of 10-15 \msunpc. New IRAM 30m HERA CO(2-1) data of NGC~4654 are used to study the physical conditions of the ISM and its ability to form stars in the region where gas compression occurs. %
The CO-to-H$_2$ conversion factor was estimated by (i) simultaneously solving for the conversion factor and the dust-to-gas ratio by assuming that the latter is approximately constant on giant molecular cloud scales and (ii) by assuming that the dust-to-gas ratio is proportional to the metallicity. The CO-to-H$_2$ conversion factor was found to be one to two times the Galactic value. Based on the comparison with a region of similar properties in NGC~4501, we favor the higher value. %
We observe a significant decrease in the ratio between the molecular fraction and the total ISM pressure in the \hsd . The gas in this region is self-gravitating, with a Toomre parameter below the critical value of $Q=1$. However, the star-formation efficiency ($\rm SFE_{H_2} = \Sigma_{SFR}/\Sigma_{H_2}$) is 1.5 to 2 times higher, depending on the assumed conversion factor, in the \hsd\ than in the rest of the disk. Analytical models were used to reproduce radial profiles of the SFR and the atomic and molecular surface densities to better understand which physical properties are mandatory to maintain such \hsd s. We conclude that a Toomre parameter of $\rm Q \sim 0.8$ combined with an increase in the velocity dispersion of $\Delta v_{\rm disp} \sim 5\ \rm{km~s^{-1}}$ are necessary conditions to simultaneously reproduce the gas surface densities and the SFR. %
A dynamical model that takes into account both gravitational interactions and ram pressure stripping was used to reproduce the gas distribution of NGC~4654. While the ISM properties are well reproduced in the whole disk, we find that the model SFR is significantly underestimated in the \hsd\ due to the absence of gas cooling and stellar feedback. 
The comparison between the velocity dispersion given by the moment 2 map and the intrinsic 3D velocity dispersion from the model were used to discriminate between regions of broader linewidths caused by a real increase in the velocity dispersion and those caused by an unresolved velocity gradient only.
We found that the 5 km~s$^{-1}$ increase in the intrinsic velocity dispersion predicted by the model is compatible with the observed velocity dispersion measured in the \hsd . %
During a period of gas compression through external interactions, the gas surface density is enhanced, leading to an increased SFR and stellar feedback. Our observations and subsequent modeling suggest that, under the influence of stellar feedback, the gas density increases only moderately (by less than a factor of two). The stellar feedback acts as a regulator of star-formation, significantly increasing the turbulent velocity within the region. 
}


\authorrunning{Liz\'ee et al.}
\titlerunning{Gas compression and stellar feedback in NGC~4654}
\maketitle

%
\newpage

\section{Introduction}

To understand how galaxies form stars from gas, it is essential to study disturbed disk galaxies. This is because it is possible to observe the influence of the perturbations on the interstellar medium (ISM) and its ability to form stars in these systems. 

Galaxy clusters represent ideal laboratories for studying perturbations due to
environmental interactions. Proximity between individual galaxies promotes
gravitational interactions (slow galaxy-galaxy interactions or harassment)
that affect both the stellar and dense gas distribution on larges scales.
Moreover, the hydrodynamical interaction created by the motion of a galaxy
through the hot and tenuous gas that constitutes the intracluster medium (ICM)
also strongly affects the distribution of the interstellar medium (ISM). This effect - known as ram
pressure stripping - depends on the ICM density and the velocity of the galaxy
with respect to the cluster mean. Both quantities increase with a decreasing
distance to the cluster center.

The main ingredient required for star-formation is dense molecular gas. Within
spiral galaxies, a strong correlation has been identified between the molecular
gas (\h2) and the star-formation rate (SFR) { (e.g, \citealt{1998ApJ...498..541K}, \citealt{2008AJ....136.2846B}, \citealt{2012ARA&A..50..531K}, \citealt{2017ApJ...846..159B}).}
Therefore, one of the major quantities that has to be investigated is the
star-formation efficiency with respect to the molecular gas, \sfe. {Past studies
have shown that variations in the \sfe\ are wider between galaxies than within the same
galactic disk (e.g, \citealt{2008AJ....136.2782L},
\citealt{2008AJ....136.2846B}, \citealt{2010MNRAS.407.2091G}, \citealt{2011AJ....142...37S}, \citealt{2017ApJS..233...22S}, \citealt{2018ApJ...853..179T}).} The molecular ISM
usually presents a constant depletion time of $t_{\rm{dep}} ^{\rm{H_2}}$ $\rm \sim 2.35\ Gyr$
with a 1$\sigma$ scatter of 0.24dex (\citealt{2011ApJ...730L..13B}). Only few
extreme cases of interacting galaxies where the star-formation efficiency is significantly different have been found. {This is the case for NGC~4438, which undergoes a tidal interaction and ram pressure stripping with $t_{\rm{dep}} ^{\rm{H_2}}$ $\rm \sim 6\ Gyr$ (\citealt{2009A&A...496..669V}, \citealt{2012A&A...543A..33V}), as well as the Taffy system, a head-on collision between two massive spiral galaxies with $t_{\rm{dep}} ^{\rm{H_2}}$ $\rm \sim 6\ Gyr$ (\citealt{2012A&A...547A..39V}).}
\citet{2002ApJ...569..157W} highlighted a second correlation that describes the
division of the molecular and the atomic phases within the ISM. {The molecular fraction \rmol\ $\equiv$ \sgm/\sga\ is approximately proportional to the total ISM pressure, $P_{\rm tot}$ (see \eq{eq:ptot}).} The natural question that arises from this observation is whether or not we can find galaxies among the cases of disturbed disk galaxies, especially those affected by ram pressure stripping, where these relations do not hold.
\begin{figure}[ht!]
  \centering
  \includegraphics[width=0.8\hsize]{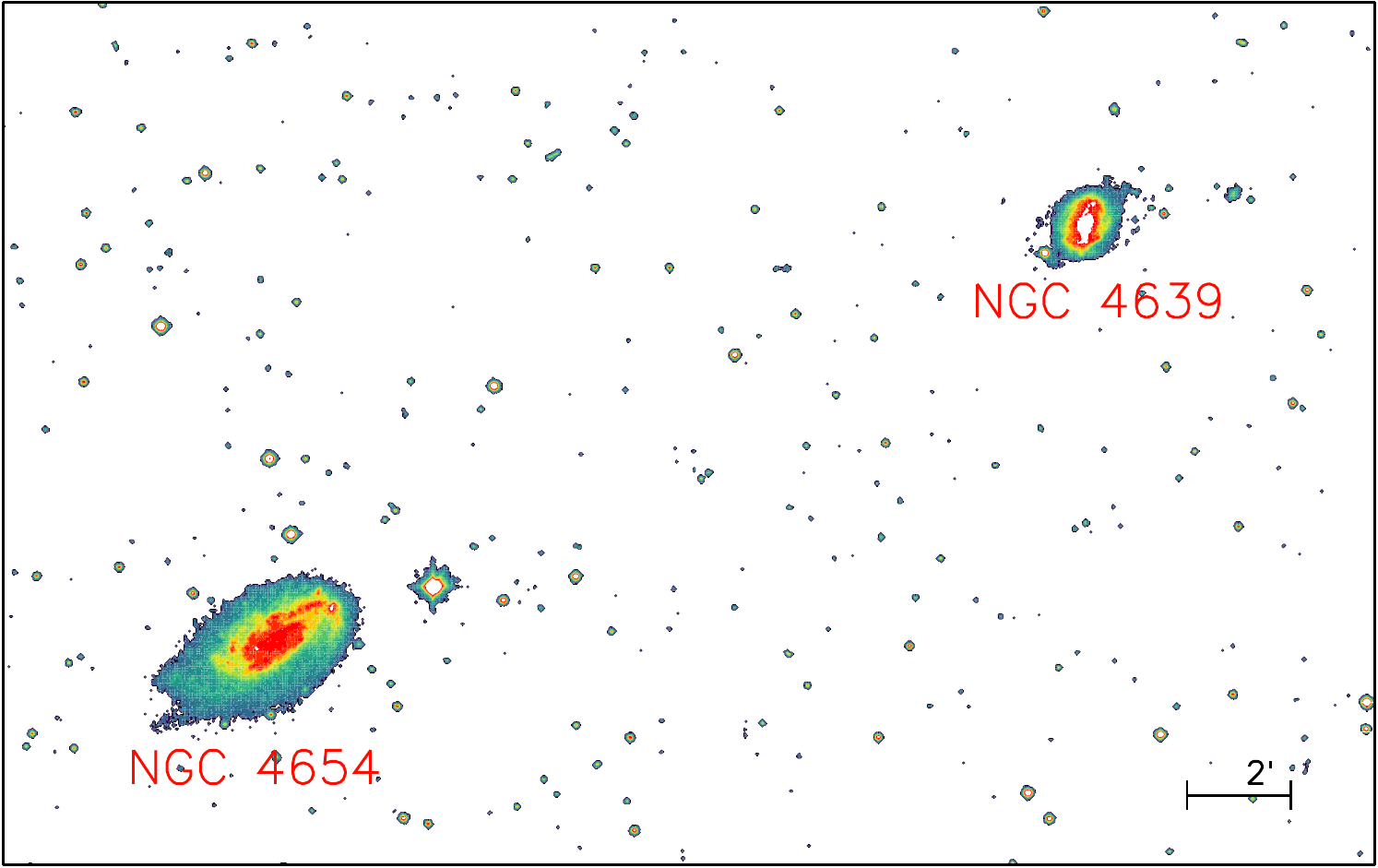}\label{fig:f1}
  \hfill
  \includegraphics[width=\hsize]{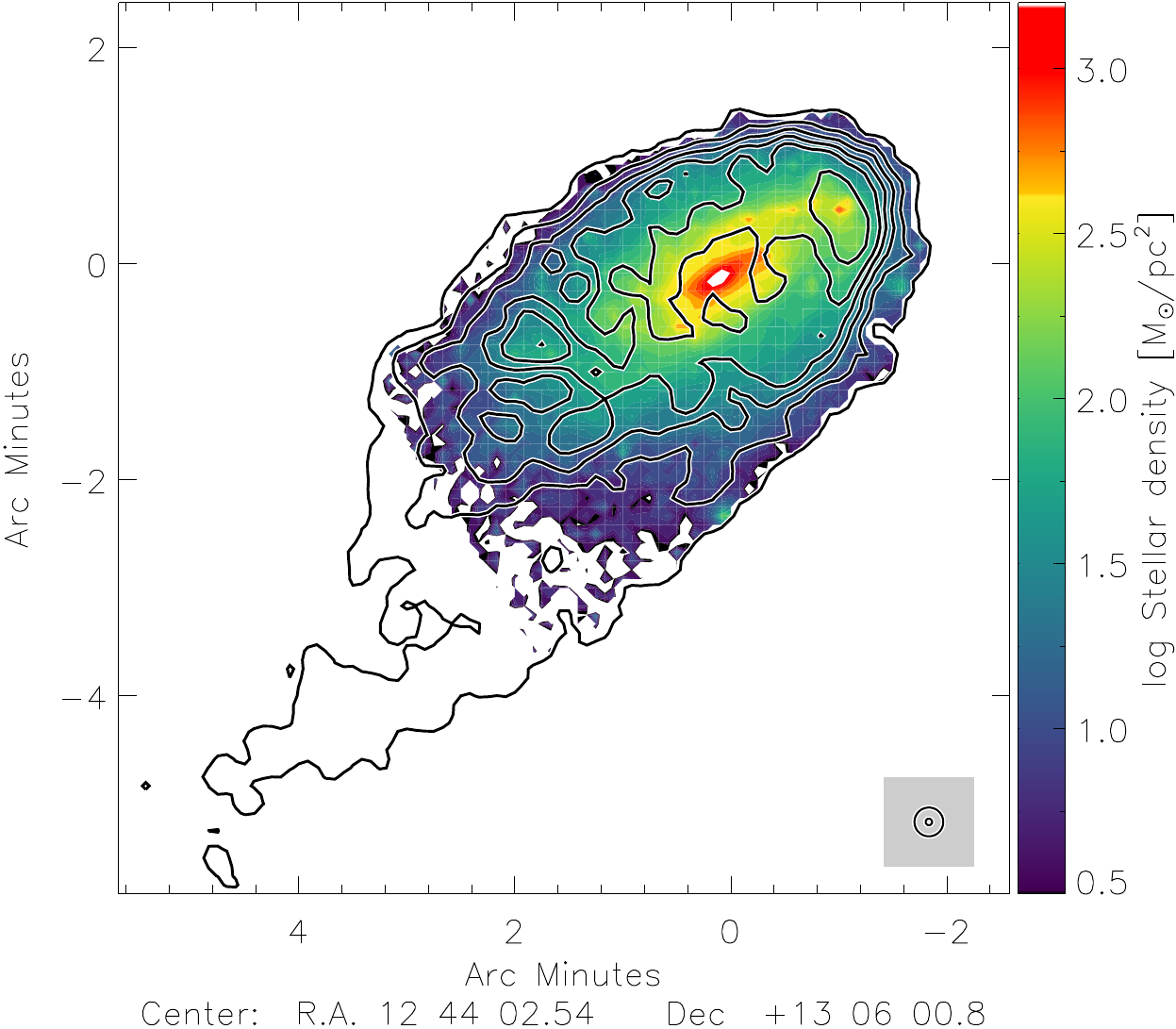}\label{fig:f2}
   \caption{{Environment and interstellar medium distribution of NGC 4654. \textit{Top panel:} NGC~4654 and its companion NGC~4639 (SDSS9 extracted via Aladin v10 (\citealt{2020ASPC..522...77N})). \textit{Bottom panel:} H\textsc{i} surface density (contours) on the stellar surface density based on 3.6 $\rm{\mu} m$ SPITZER data (color).  Contour levels are 1, 5, 10, 15, 20, and 30 \msunpc. The resolution is 16''.}}
   \label{fig:hi}
\end{figure}

\begin{figure}[ht!]
   \centering
   \includegraphics[width=\hsize]{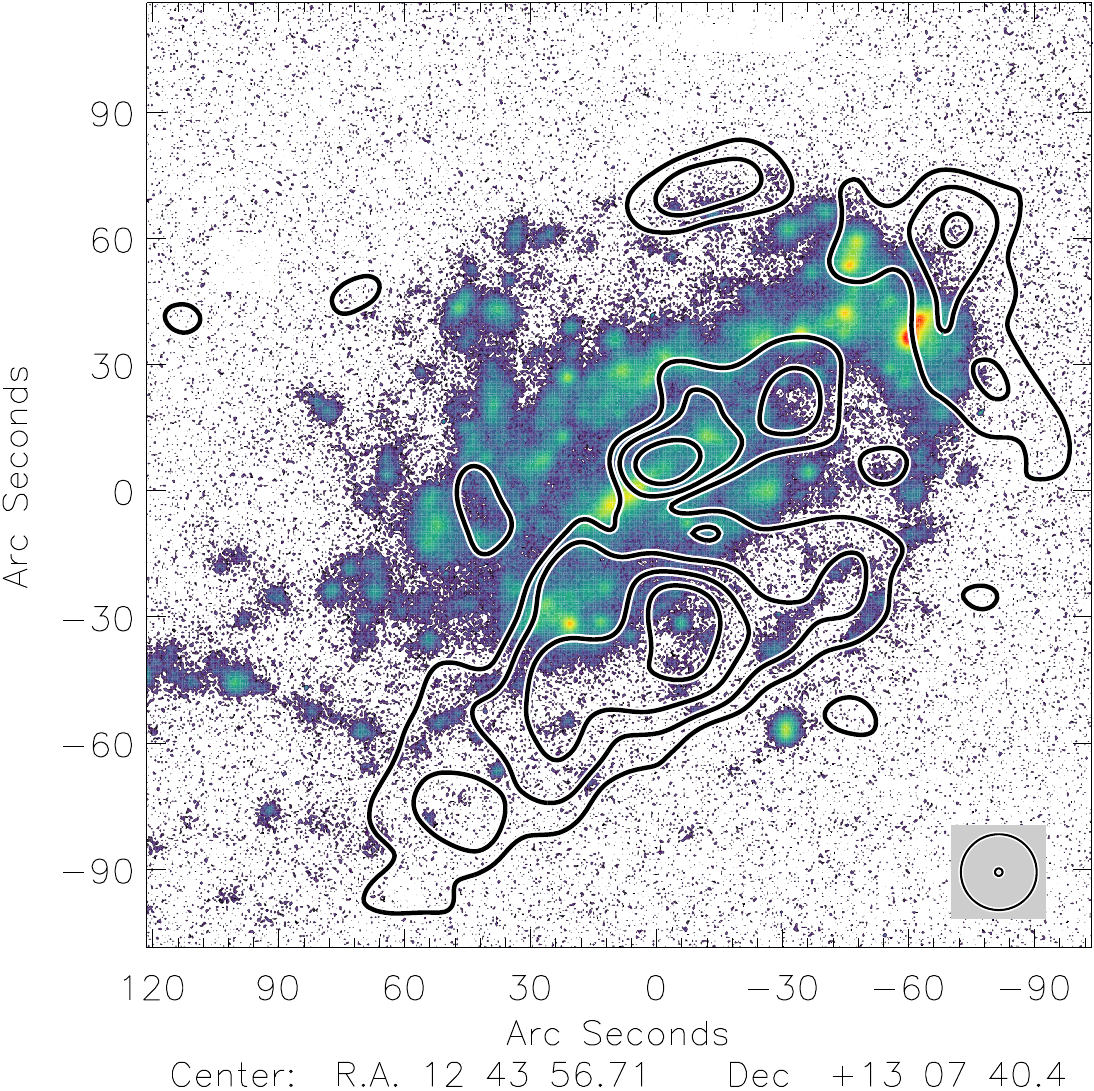}
   \caption{{NGC~4654: 6cm polarized radio continuum emission (contours) on the GOLDMINE H$\alpha$ emission map in color levels (\citealt{2003A&A...400..451G}). The spatial resolutions are 18'' and 1.8''.}}
   \label{fig:ha}
\end{figure}

Using the VIVA survey (VLA Imaging of Virgo galaxies in Atomic gas),
\citet{2007ApJ...659L.115C} revealed the presence of seven galaxies within the Virgo
cluster with truncated \hi\ disks and an extended \hi\ gas tail on the opposite
side. Among this sample, one galaxy in particular attracts our attention, NGC
4654. The \hi\ disk of NGC~4654 is sharply truncated in the northwestern side of
the disk and presents an unusually dense region of atomic hydrogen surface
density (about 25 \msunpc~including Helium) at the edge of the optical disk
(\fig{fig:hi}.). Such high surface densities are exceptional and have
only been observed in very rare cases (e.g, in the interacting Eyelid galaxy,
\citealt{2016ApJ...823...26E}). In addition to being affected by ram pressure
stripping, NGC~4654 underwent a gravitational interaction with another Virgo
galaxy, NGC~4639, about 500 millions years ago (\citealt{2003A&A...398..525V}).
NGC~4654 presents an asymmetric stellar distribution with a dense stellar arm
toward the northwest (\fig{fig:hi}). A strongly enhanced \ha\
emission is also observed in the region of the highest atomic gas surface
density (\fig{fig:ha}). The gravitational interaction and ram pressure
stripping gave rise to asymmetric ridges of polarized radio continuum emission
whereas the sudden ridge is most probably due to shear motions induced by the
gravitational interaction and the western ridge is caused by ram pressure
compression (\citealt{2006A&A...458..727S}). 

\begin{table}[ht!]
   \caption{NGC~4654 general properties}
   \label{tab:ngc}
   \centering
   \begin{tabular}{p{3.7cm}p{3.7cm}}
      \hline
      Morphological type          & \hfill SAB(rs)cd                        \\
      Optical diameter            & \hfill 5.17' $\times$ 1.41'             \\
      Distance                    & \hfill 17 Mpc                           \\
      $\alpha$(J2000)             & \hfill 12$^h$ 43$^m$ 56.6$^s$           \\
      $\delta$(J2000)             & \hfill 13$\degree$ 07' 36''             \\
      Inclination angle           & \hfill 51$\degree$                      \\
      Systemic velocity           & \hfill 1060 $\rm km\ s^{-1}$            \\
      Rotational velocity         & \hfill 170 $\rm km\ s^{-1}$             \\
      Total SFR\tablefootmark{a}  & \hfill 1.84 M$_\odot$/yr                \\
      \hi ~mass\tablefootmark{b}  & \hfill 3.4 $\times$ 10$^9$ M$_\odot$    \\
      H$_2$ mass\tablefootmark{c} & \hfill 2.3 $\times$ 10$^9$ M$_\odot$    \\
      Stellar mass                & \hfill 2.8 $\times$ 10$^{10}$ M$_\odot$ \\
      \hline
   \end{tabular}
   \hfill
   \tablefoot{\\
      \tablefoottext{a}{Computed following Leroy et al. (2008) from the 24 $\rm{\mu} m$ SPITZER and far-ultraviolet GALEX data.} \\
      \tablefoottext{b}{Calculated from the VIVA \hi ~data cube (\citealt{2008ASPC..395..364C}) for a distance of 17 Mpc.} \\
      \tablefoottext{c}{Calculated using the modified \aco\ conversion factor presented in \sect{sect:xco}. } \\
   }
\end{table}
{\citet{2014PASJ...66...11C} conducted a study on NGC~4654 by combining the \hi\
VIVA data with the extragalactic CO CARMA Survey Toward
Infrared- bright Nearby Galaxies (STING, \citealt{2011AAS...21813001R}). They showed that the star-formation
efficiency with respect to the molecular gas ($\rm SFE_{H_2}$) reaches unusually high values in the
northwestern, high atomic gas surface density region. At the same time, they found that the 
R$_{\rm mol}$/P$_{\rm{tot}}$ ratio is unusually low in this same region. As the \sfe\ and \rmolp\ both depend on the surface density of the molecular gas,
it is essential to determine the CO-to-\h2 conversion factor as precisely as
possible. }

The purpose of this paper is to investigate the influence of gas compression and stellar feedback on the gas distribution {and SFR} within spiral galaxies through the study of NGC 4654 and its overdense northwestern region. We refer to the high atomic gas surface density region as "Hi$\Sigma_{\rm HI}$" for the remainder of the paper. This article is organized in the following way. In \sect{sect:obs} and 
\sect{sect:co}, the new CO(2-1) IRAM 30m data are presented, with ancillary data in introduced in
\sect{sect:ad}. The CO-to-H$_2$ conversion factor is estimated using the HERSCHEL 250$\rm \mu m$ data 
and direct metallicity measurements from \citet{1996ApJ...462..147S} in \sect{sect:xco}. 
In \sect{sect:mol}, we use this conversion factor to compute molecular and the total gas maps of NGC~4654, 
and use the resulting maps to investigate the relation between molecular fraction and total mid-plane pressure of the gas 
(\sect{sect:rmol}), the star-formation efficiency (\sect{sect:sfe}), and the Toomre stability criterion (\sect{sect:q}). 
An analytical model is used in \sect{sect:ana} to reproduce the observed radial profiles. In \sect{sect:dyn}, 
a dynamical model is used to reproduce the available observations. All the results are discussed in \sect{sect:dis}. 
We give our conclusions in \sect{sect:ccl}. The general physical properties used for NGC~4654 in this study are presented in \tab{tab:ngc}. 

\section{Observations} \label{sect:obs}
NGC~4654 CO(2-1) data were observed by François Nehlig with the IRAM 30m single-dish telescope at Pico Veleta, Spain. OTF maps have been done to
cover the entire galaxy using EMIR instrument in four different positions with a
scanning speed of 5''/s, using FTS backends. The velocity window goes from 770 $\rm km\ s^{-1}$ to 1410 $\rm km\ s^{-1}$
with spectral resolution channel of 10.4 $\rm km\ s^{-1}$. We reached an
average rms noise of 7 mK after a total of 25 hours of observation. We obtained a
data cube with a spatial resolution of 12'' using CLASS GILDAS xy\_map routine.
The pixel size is 3''. Compared to the interferometric data CARMA STING used by \citet{2014PASJ...66...11C} our data are four times deeper and are sensitive to extended large-scale emission.

We produced CO moment maps using the VIVA \hi\ data cube assuming that the CO line is located within the \hi\ line profile (see \citealt{2012A&A...547A..39V}). To do so, we resampled the channels of the \hi\ data cube to fit the CO data cube. A first 3D binary mask was produced by clipping the
\hi\ data cube at the 4$\sigma$ level. For the CO data cube, we calculated
the rms noise level for each spectrum at velocities devoid of an \hi\ signal and subtracted a constant baseline. A second 3D binary mask was produced based on the CO data cube. If the maximum intensity of a spectrum exceeded 5 times the rms, we fit a Gaussian profile to the CO spectrum and fixed the mask limit to $\pm$3.2$\sigma$. The \hi\ and CO 3D binary masks were added, applied to the CO data cube, and moment 0, 1, and 2 maps were created. 

\section{Results} \label{sect:co}

\begin{figure}[ht!]
   \centering
   \includegraphics[width=\hsize]{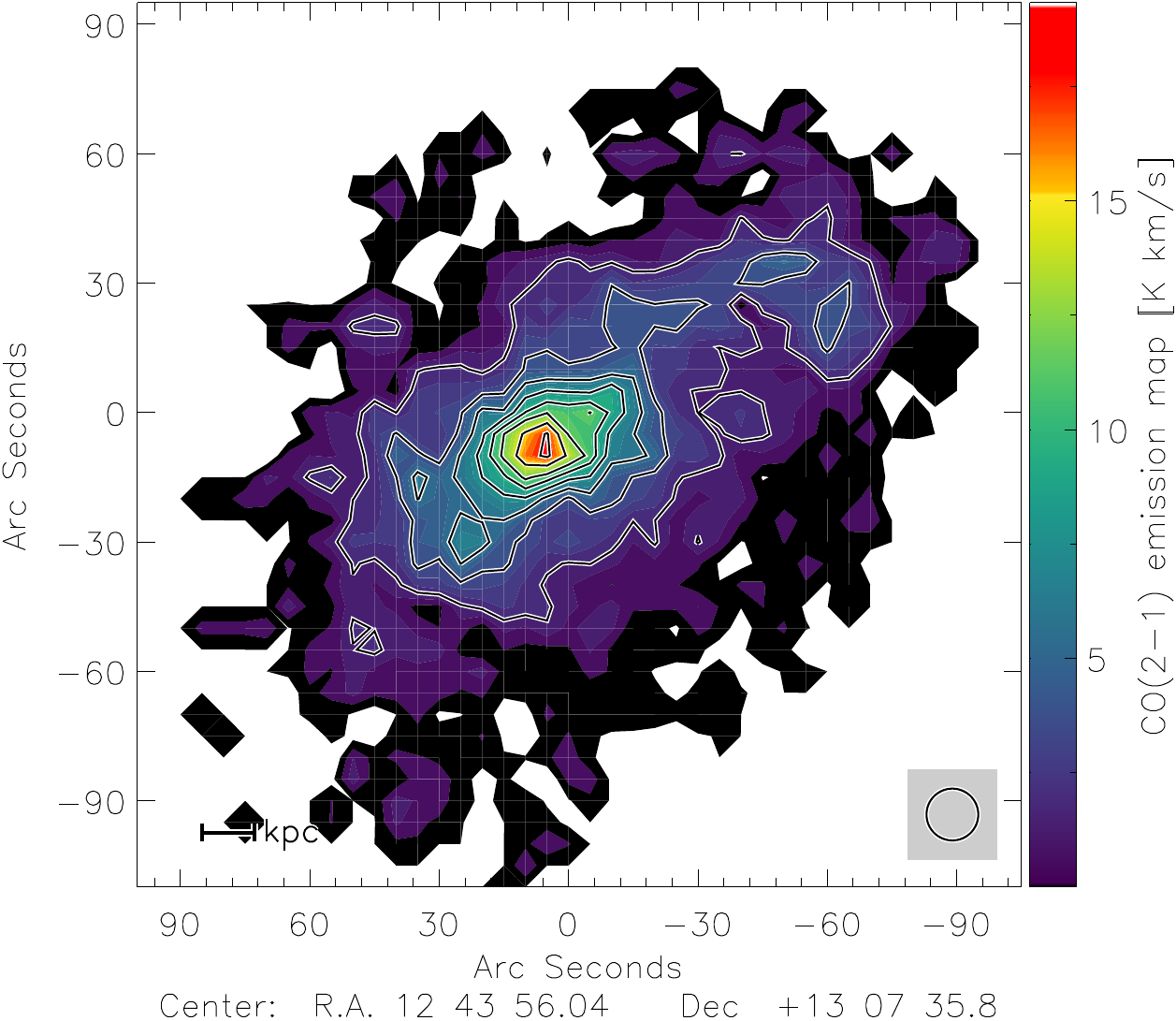}
   \caption{NGC~4654 CO(2-1) emission map. Contour levels are 2, 4, 6, 8, 10, 12, 15 and 18 K $\rm km\ s^{-1}$. The resolution is 12''.}
   \label{fig:m0co}
\end{figure}

The integrated CO(2-1) map was obtained from the reduced, windowed data cube
(\fig{fig:m0co}). The associated rms map is presented in \fig{fig:errco}. The
global emission distribution is asymmetric, the signal being detected
up to 8 kpc to the northwest and 7 kpc to the southeast. The maximum value of
19.3 K $\rm km\ s^{-1}$ is reached in the galaxy center. The emission map presents an
enhanced flux all along the galaxy spiral arm extending toward the northwest,
corresponding closely to the dense stellar arm distribution. Within the spiral
arm, the CO(2-1) flux is almost constant from 4 to 5 K $\rm km\ s^{-1}$. These values are 2
to 3 times higher than those in the inter-arm, and 4 to 5 times higher than the
southeast region at the same distance from the galaxy center.

\paragraph{}

\begin{figure}[ht!]
   \centering
   \includegraphics[width=\hsize]{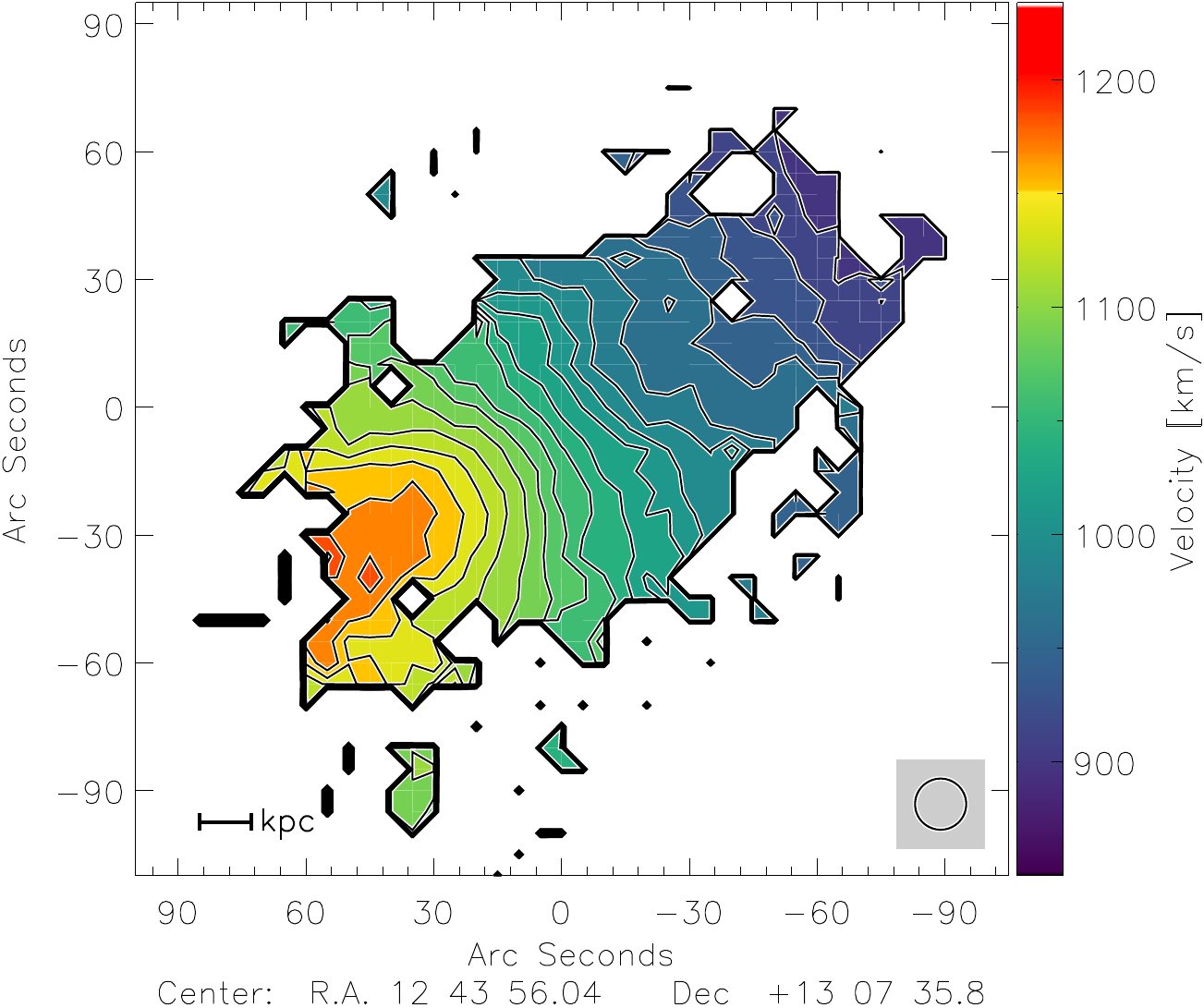}
   \caption{NGC~4654 CO(2-1) velocity field. Contours levels are from 900 to 1200 $\rm km\ s^{-1}$ in steps of 20 $\rm km\ s^{-1}$.}
   \label{fig:m1co}
\end{figure}

\begin{figure}[ht!]
   \centering
   \includegraphics[width=\hsize]{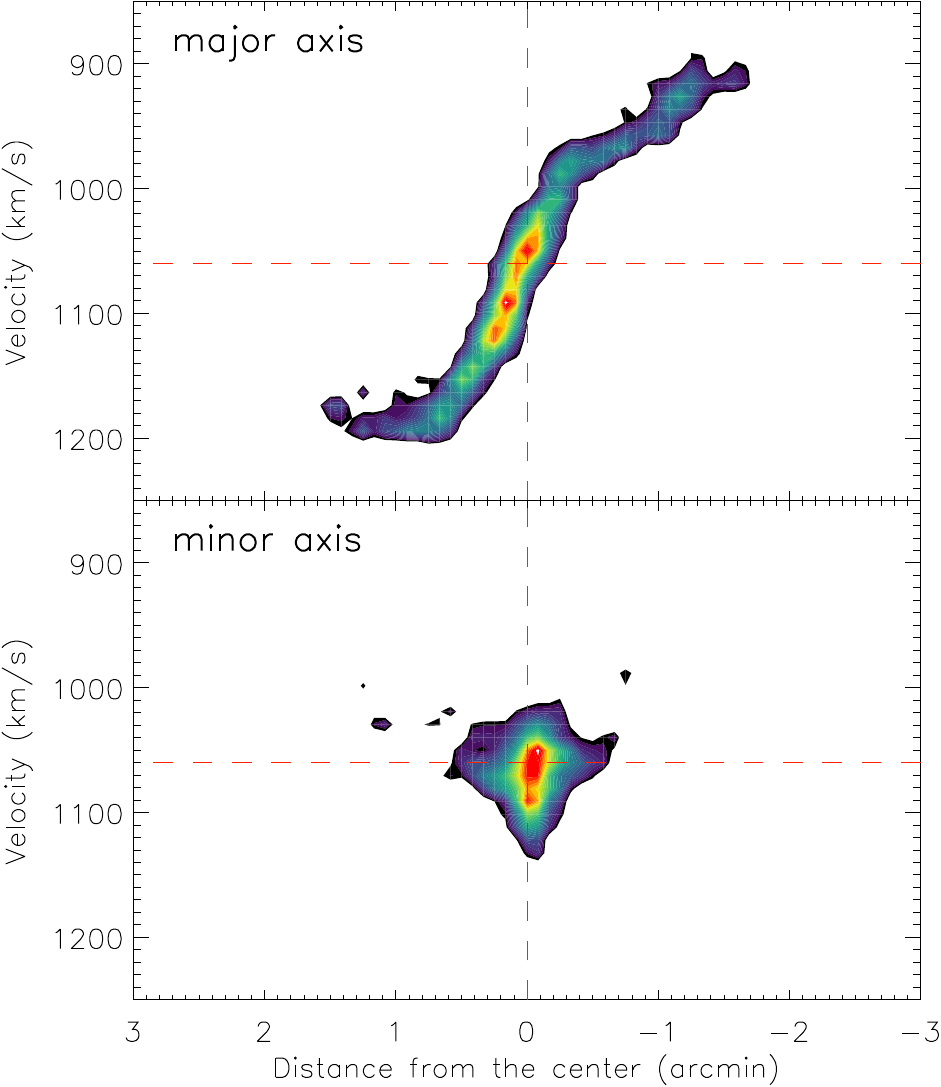}
   \caption{Position-velocity diagram of the CO(2-1) emission. \textit{Top panel}: Along the major axis. \textit{Bottom panel}: Along the minor axis. The dotted red line corresponds to the systemic velocity of 1060 $\rm km\ s^{-1}$.}
   \label{fig:pvco}
\end{figure}

As a consequence of the gravitational interaction between NGC~4654 and
NGC~4369, the CO velocity field (\fig{fig:m1co}) shows an asymmetric profile
along the major axis, with a constant velocity plateau reached in the southeast
that is absent in the northwest. We determine the systemic velocity at 1060
$\rm km\ s^{-1}$, measured in the optical center of the galaxy {using \ha\ data from
GOLDMINE (\citealt{2003A&A...400..451G}). This is consistent with the value of $1050\ \rm km\ s^{-1}$ given by the SDSS DR12 catalog (\citealt{Alam_2015})}. Considering 1060 $\rm km\ s^{-1}$, the maximum velocities reached in
the southeast and the northwest regions are +150 $\rm km\ s^{-1}$ and -170 $\rm km\ s^{-1}$,
respectively.
The position-velocity diagrams along the major and the minor axis of the galaxy
are presented in \fig{fig:pvco}. In the southeast, the behavior of the diagram
is standard: a gradient around the center of the galaxy ending with a velocity
plateau at 3 kpc. On the other hand, in the northwest, the plateau is reached at
2 kpc and is followed by a second gradient at 4 kpc. The position-velocity
diagram along the minor axis does not show any strong asymmetry.

\begin{figure}[ht!]
   \centering
   \includegraphics[width=\hsize]{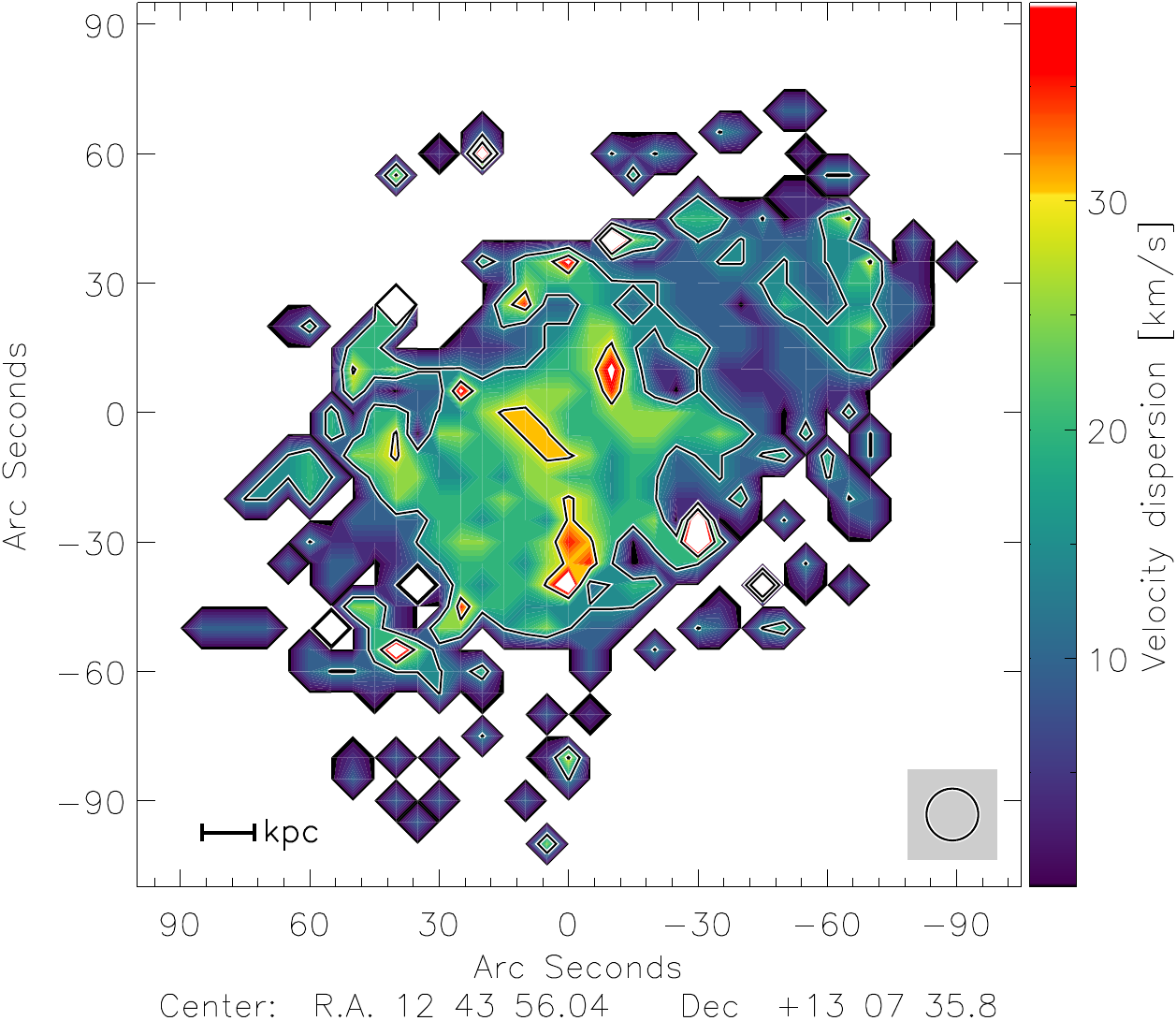}
   \caption{CO(2-1) moment 2 map. Contour levels correspond to 15 and 30 $\rm km\ s^{-1}$.}
   \label{fig:m2co}
\end{figure}

The moment 2 map for the CO(2-1) data cube is presented in \fig{fig:m2co}.
Because of galactic rotation, the observed velocity dispersion strongly increases in the region
around the center of the galaxy, with $\Delta \nu$~$\simeq$~40~$\rm km\ s^{-1}$. In the
northwestern high \hi\ surface density region, the CO(2-1) linewidth is 5-10
$\rm km\ s^{-1}$ broader than in the inner arm between the \hi\ region and the center of the
galaxy.

\section{Ancillary data} \label{sect:ad}

In order to compute the SFR, the total gas and the ISM pressure,
we used existing FUV 1528 \AA\ data from GALEX \citep{2005ApJ...619L...1M}, infrared 3.6$\rm{\mu m}$ and
24$\rm{\mu m}$ from SPITZER \citep{2004ApJS..154....1W}, \ha\ emission from GOLDMINE
\citet{2003A&A...400..451G} and \hi\ data from VIVA survey
\citet{2008ASPC..395..364C}. The above-mentioned data are presented in the
following subsections :

\subsection{Atomic gas}

The atomic gas surface density \sga\ is obtained using the VLA \hi\ data from
\citet{2008ASPC..395..364C}:
\begin{equation}
   \frac{\Sigma _{\rm{HI}}}{(\rm{M_\odot pc^{-2}})} = 0.020 \ \frac{{I_{21cm}}}{(\rm{K~\rm km\ s^{-1}})}~,
\end{equation}
with ${I_{21cm}}$ converted from Jy/beam to K $\rm km\ s^{-1}$. The expression contains a
coefficient of 1.36 to reflect the presence of Helium. The rms at 3~$\sigma$ is
about 0.5 \msunpc\ and the spatial resolution is 16''. The contour map of \sga\
is shown in \fig{fig:hi}. For the following sections, we will consider the \hsd\ as the northwestern area where \sga~>~28~\msunpc. 

\subsection{Star-formation rate} \label{sect:sfr}
{\citet{2014PASJ...66...11C} computed the SFR of NGC~4654 using the 1.4 GHz radio continuum data from the NRAO VLA Sky Survey (NVSS, \citealt{1998AJ....115.1693C}). However, the conversion between the radio continuum emission and SFR has a large scatter (\citealt{2020A&A...633A.144V}) and is
uncertain in galaxies whose ISM is affected by ram pressure stripping (\citealt{2006ApJ...638..157M}). We therefore use GALEX FUV all-sky survey (\citealt{2005ApJ...619L...1M}) and SPITZER $24 \rm{\mu}$m data (PI: J.D.P. Kenney) for the calculation of the star-formation map.} We computed the SFR following \citet{2008AJ....136.2782L}:
\begin{equation}
   \dot{\Sigma}_{\star}(FUV+24{\mu} m) = 8.1 \times 10^{-2} I_{FUV} + 3.2 \times 10^{-3}I_{24 {\mu} m}\ ,
\end{equation}
where $\dot{\Sigma}_{\star}(FUV+24~\rm{\mu} m)$ is the SFR in \msun~kpc$^{-2}$
yr$^{-1}$ and $I_{FUV}$ and $I_{24 \rm{\mu} m}$ are given in MJy/sr. The
maximum of 0.27 \msun~kpc$^{-2}$ yr$^{-1}$ is reached in the high atomic gas
surface density region at the outer edge of the stellar arm. This value is 1.5
times higher than that of the galaxy center. The total integrated SFR is 1.84
\msun yr$^{-1}$, {which places NGC~4654 on the star-forming main sequence, considering its stellar mass} (e.g,
\citealt{2004MNRAS.351.1151B}, \citealt{2007ApJ...660L..43N},
\citealt{2007ApJ...670..156D}). The \hsd\ concentrates 20\% of the total SFR of NGC~4654 (0.37~\msun kpc$^{-2}$yr$^{-1}$). The compact region of high star-formation, where \sfr~>~0.20~\msun kpc$^{-2}$yr$^{-1}$, accounts for 5\% of the total SFR of NGC~4654 (0.09~\msun kpc$^{-2}$yr$^{-1}$), which corresponds to $\sim$25\% of the SFR in the entire \hsd.  

\begin{figure}[ht!]
   \centering
   \includegraphics[width=\hsize]{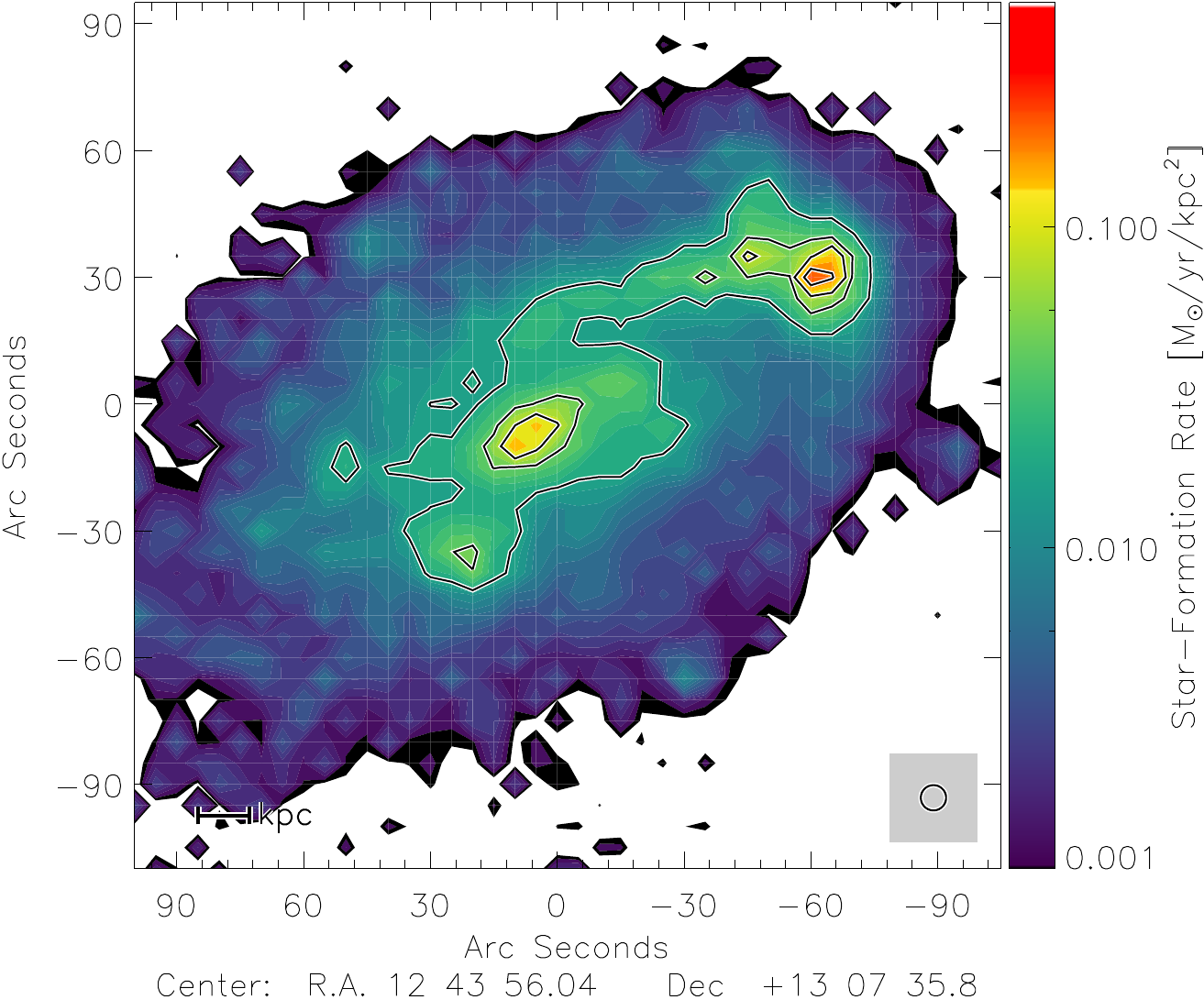}
   \caption{Star-formation rate based on GALEX FUV and SPITZER 24 $\rm{\mu} m$ data. Contour levels correspond to 0.02, 0.05, 0.10, and 0.20 \msun kpc$^{-2}$yr$^{-1}$. The resolution is 6''.}
   \label{fig:sfr}
\end{figure}

These results are consistent within 5\% with the SFR computed using the \ha\
emission from {GOLDMINE \citet{2003A&A...400..451G}}, following
\citet{2008AJ....136.2782L}. A third method was tested, combining the FUV data
with the TIR emission computed from the 3.6, 24 (SPITZER), 100 and 160$\rm{\mu}
m$ flux (HERSCHEL). The method is described in \citet{2011ApJ...741..124H} and
\citet{2013MNRAS.431.1956G}. {The SFR$_{FUV+TIR}$ shows slightly different results with a maximum of the SFR in the center of the galaxy. This is expected because the interstellar radiation field (ISRF), which heats the dust emitting in the infrared, contains a significant fraction of optical starlight}. In the \hsd,
$\dot{\Sigma}_{\star}(FUV+24 \rm{\mu} m)$ is about 2 times higher than
$\dot{\Sigma} _{\star}(FUV+TIR)$. The total integrated SFR using TIR as infrared
contribution gives 1.53 \msun yr$^{-1}$, which is 20\% lower than the total SFR
calculated from the FUV + 24 $\rm{\mu} m$ or \ha\ recipes.

\subsection{Stellar surface density} \label{sect:star}

The stellar surface density $\Sigma_{\star}$ is obtained from 3.6 $\rm{\mu} m$
SPITZER data following \citet{2008AJ....136.2782L}:
\begin{equation}
   \frac{\Sigma_\star}{\rm{(M_\odot pc^{-2})}} = 280\frac{I_{3.6~\rm{\mu} m}}{\rm{(MJy~sr^{-1})}}~.
\end{equation}
The resulting map is presented in \fig{fig:hi}. The stellar surface density
makes it possible to calculate the vertical stellar velocity dispersion,
$v_{\rm{disp}}^\star$, necessary to estimate the ISM pressure of the disk:
\begin{equation} \label{eq:vdisp}
   v_{\rm{disp}}^\star = \sqrt{\Sigma_{\star}\frac{2\pi G l_{\star}}{7.3}}\ ,
\end{equation}
where $l_{\star}$ is the stellar scale length computed from the radial profile
of the stellar surface density. {To do so, we averaged the stellar surface density within ellipsoidal annuli of 0.6 kpc width and determined the slope in lin-log space.} We obtained $l_{\star}$~=~2.3~kpc. The
exponential stellar scale height is assumed constant with
$l_{\star}$/$h_{\star}$~=~7.3~$\pm$~2.2 (\citealt{2002MNRAS.334..646K}), corresponding to $h_\star \sim 0.3$~kpc. 

\section{CO-to-\h2 conversion factor from dust emission} \label{sect:xco}

The CO-to-H$_2$ conversion factor is critical for our estimation of the
molecular surface density and for the subsequent physical quantities of this
study, the star-formation efficiency and the molecular fraction. For
non-starburst spiral galaxies at low redshift, {it is quite common to use the
Milky Way standard value of \xcomw\ = 2 $\times$ 10$^{20}$ \xcou\ (e.g, \citealt{2013ARA&A..51..207B}).} In the
following sections, we will use the equivalent expression of the conversion
factor expressed in solar masses, \acomw\ = 4.36 \acou\ that takes into account
a factor of 1.36 for the presence of Helium.

\begin{figure}[ht!]
   \centering
   \includegraphics[width=\hsize]{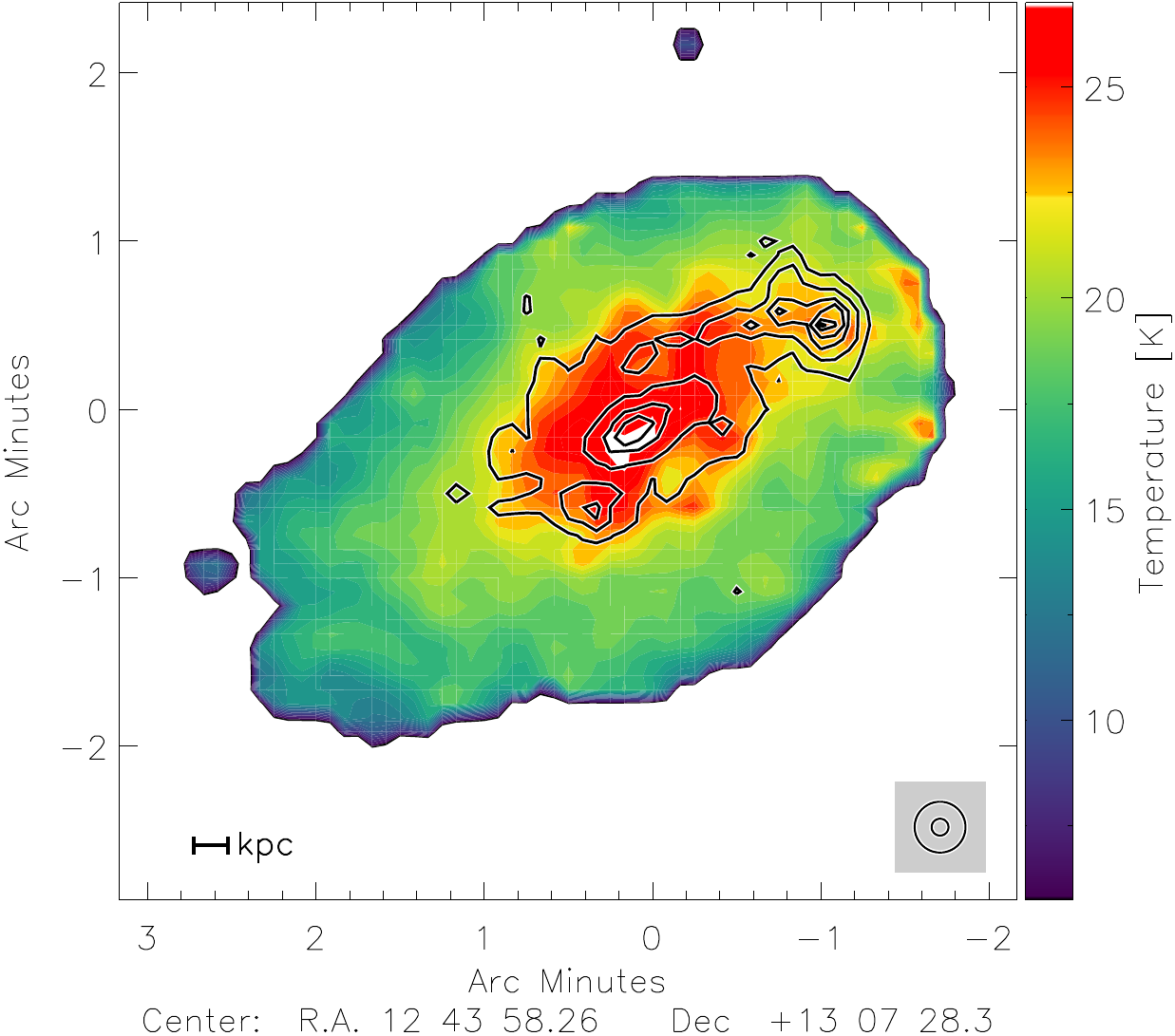}
   \caption{Dust temperature map based on HERSCHEL 250$\rm{\mu} m$ and 350$\rm{\mu} m$ data. Contour levels correspond to $\dot{\Sigma}_\star =$ 0.02, 0.05, 0.10, and 0.20 \msun kpc$^{-2}$yr$^{-1}$.}
   \label{fig:temp}
\end{figure}

\citet{2013ARA&A..51..207B} suggested that below $\sim$1/2-1/3 of the solar
metallicity, the conversion factor tends to increase significantly. Since the
high \hi\ surface density region is located close to the optical radius, where
generally the metallicity is Z~$\sim$~1/2-1/3~Z$_\odot$, a higher conversion
factor is not excluded in this region. \citet{2014PASJ...66...11C} also
suggested that the conversion factor in this region might be higher than the
Galactic standard value. We estimated the CO-to-H$_2$ conversion factor using
the hydrogen column density $N_H$ calculated from the far infrared HERSCHEL
bands at 250$\rm{\mu} m$ and 350$\rm{\mu} m$ and the dust-to-gas ratio, DGR, by
applying the method presented in \citet{2011ApJ...737...12L}. By combining the
dust emission with the CO and \hi\ data, the following equation can be
established:
\begin{equation} \label{eq:xco}
   \frac{N_{\rm{H}}^{\rm{dust}}}{\eta} = N_{\rm{H}}^{\rm{HI}} + 2~N_{\rm{H}}^{\rm{H_2}} = N_{\rm{H}}^{\rm{HI}} + 2~X_{\rm{CO}}~I_{\rm{CO}}(1-0)\ ,
\end{equation}
{where $I_{CO(1-0)}$ is the CO(1-0) emission computed from the CO(2-1) emission assuming a line ratio 
$\rm{I_{CO(1-0)} = I_{CO(2-1)} / 0.8}$ (\citealt{1997IAUS..170...39H}) and $\eta$~=~DGR/DGR$_\odot$ is the DGR normalized by the solar DGR (DGR$_\odot$~$\sim$~1/100, \citealt{2007ApJ...663..866D}).} The use of $\eta$ introduces a metallicity dependence in the calculations. The optical depth is given by:
\begin{equation} \label{eq:tau}
   \tau_\nu = \int \sigma_\nu~N_{\rm{H}}~ds\ ,
\end{equation}
where $\sigma_\nu$ is the absorption coefficient such that $\sigma_\nu$
$\propto$ $\nu ~^\beta$ with $\beta$~=~2. In order to calculate the column density, it is necessary to
determine the temperature of the dust at each point of the galaxy using:
\begin{equation}
   \frac{F_{250\rm{\mu} m}}{F_{350\rm{\mu} m}} = \frac{B_{250 \rm{\mu} m}(T)}{B_{350 \rm{\mu} m}(T)}~\left(\frac{\nu_{250 \rm{\mu} m}}{\nu_{350 \rm{\mu} m}}\right)^\beta \ ,
\end{equation}
where $\rm{F_{250\mu m}}$ and $\rm{F_{250\mu m}}$ are the HERSCHEL fluxes and
$\rm{B_{\nu}(T)}$ is a modified black-body function. The resulting map is shown
in \fig{fig:temp}. The maximum of the dust temperature, reached in the galaxy
center, is about 27~K. The dust is significantly warmer along the dense stellar
arm than in the inter-arm region with a local maximum of 22~K in the \hsd. The
temperature is 5~K higher in this region than in the rest of the disk at the
same radius.

\begin{figure}[ht!]
   \centering
   \includegraphics[width=\hsize]{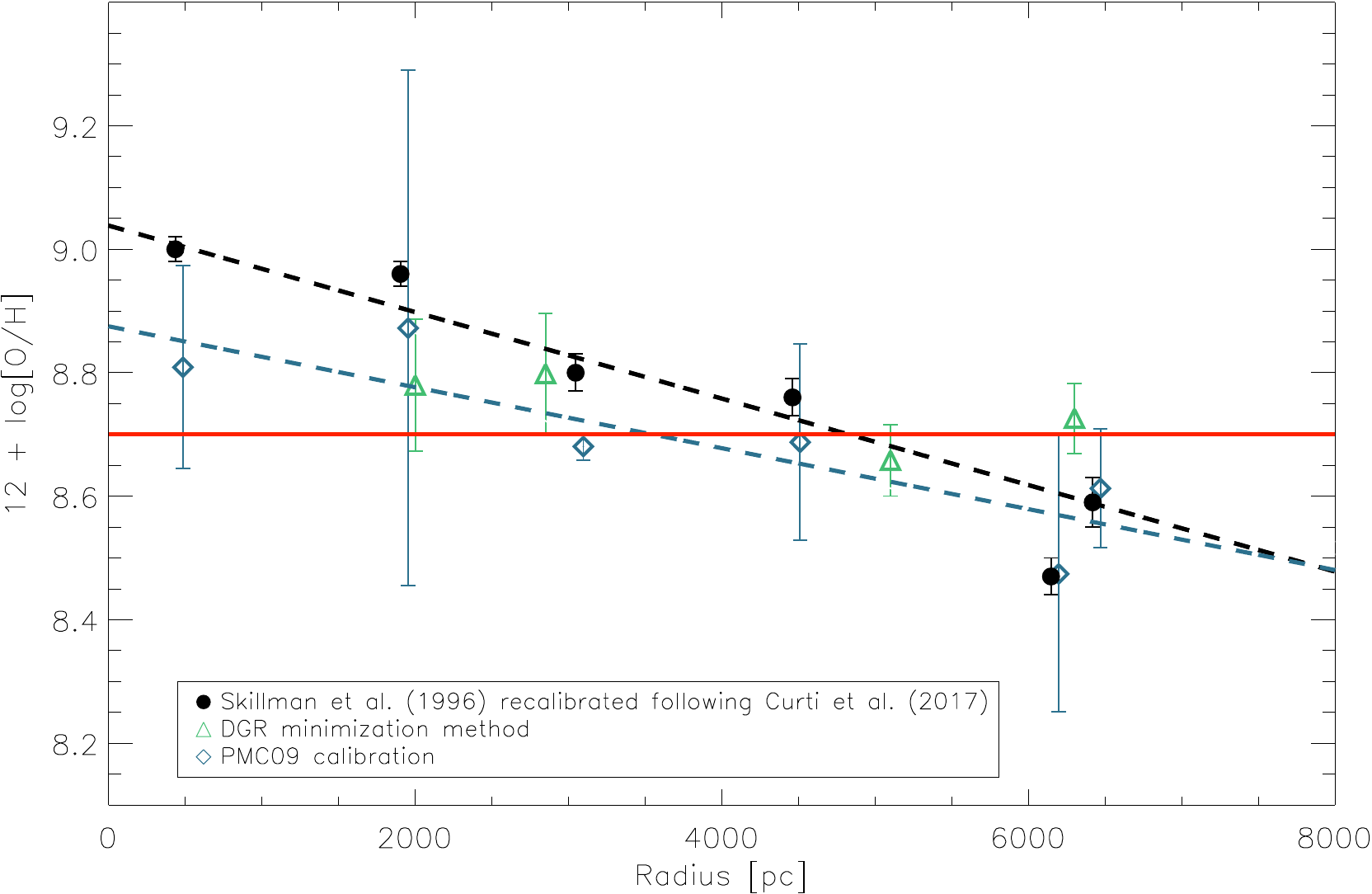}
   \caption{Radial profile of the metallicity (12+log[O/H]).
         The strong-line method measurements
from \citet{1996ApJ...462..147S} corrected following \citet{2020MNRAS.491..944C} are shown in black. The PMC09 recalibration is shown in blue. The minimization of the DGR variation is shown in green. The solar metallicity $\rm (12+log[O/H])_\odot = 8.7$ is shown in red. }
   \label{fig:sk1}
\end{figure}

The hydrogen column density is obtained using the 250 $\rm{\mu}$m band together with a dust absorption coefficient $\sigma_{250 \rm{\mu}m}$:
\begin{equation} \label{eq:nh}
   N_{\rm{H}}^{\rm{dust}} = \frac{ F_{250 \rm{\mu} m} }{ B_{250 \rm{\mu} m} (T)~\sigma _{250 \rm{\mu} m}}\ .
\end{equation} 
The dust absorption coefficient $\sigma _{250 \rm{\mu} m}$ is calibrated in a region where no CO is detected (i.e, I$_{\rm{CO}}$~=~0) by combining
\eq{eq:xco} and \eq{eq:nh}: 
\begin{equation} \label{eq:sig}
   \sigma _{250 \rm{\mu} m}~\eta = \frac{ F_{250 \rm{\mu} m} }{ B_{250 \rm{\mu} m}(T) N_{\rm{H}}^{\rm{HI}}}\ .
\end{equation}
At this point, we only determine the value of the product between $\eta$ and
$\sigma _{250\rm{\mu}m}$ but not their respective contributions. For a
comparison with the Galactic absorption coefficient, we assume that the DGR
evolves linearly with the metallicity of the gas. \citet{2014ApJ...797...85G}
calculated the grain absorption cross section per unit mass $\kappa _{\lambda}$
{using the Herschel HERITAGE survey (\citealt{2014AN....335..523M})}: 
\begin{equation} \label{eq:kap}
   \kappa_{\lambda} =  \kappa_{160 \mu m}\ \left(\frac{\lambda}{160}\right)^{-\beta}\  , 
\end{equation}
{where $\kappa_{160 \mu m}$~=~30.2~cm$^2$~g$^{-1}$ (SMBB model in Table 2 of \citealt{Gordon_2017}).}
With a coefficient $\beta$~=~2, the solar neighborhood value within the Milky
Way at 250 $\rm{\mu}m$ is $\kappa _{250
\rm{\mu}m}^{~\rm{MW}}$~=~12.36~cm$^2$g$^{-1}$. \eq{eq:sig} and \eq{eq:kap} lead
to the relation:    
\begin{equation} 
   \sigma_{250 \rm{\mu}m}\ \eta =  \left(\frac{Z}{Z_\odot}\right) \kappa_{250 \rm{\mu}m}^{~\rm{MW}}m_p\rm{DGR}_\odot \ .
\end{equation}
We calculated this ratio in regions without any CO detection and found
metallicity of $\sim$1/2~Z$_\odot$ at 9~kpc from the galaxy center. This
result is consistent with direct measurements of \citet{1996ApJ...462..147S} in
\fig{fig:sk1}. We obtained $\sigma _{\rm{dust}}$~=~8.0~$\times$~10$^{-26}$~cm$^2$, 
which is consistent with the standard value of
1.1~$\times$~10$^{-25}$~cm$^2$ (\citealt{1984ApJ...285...89D}). We used
this value in \eq{eq:nh}. and computed the hydrogen column density from dust
emission for the entire galactic disk.

\begin{figure}[ht!]
   \centering
   \includegraphics[width=\hsize]{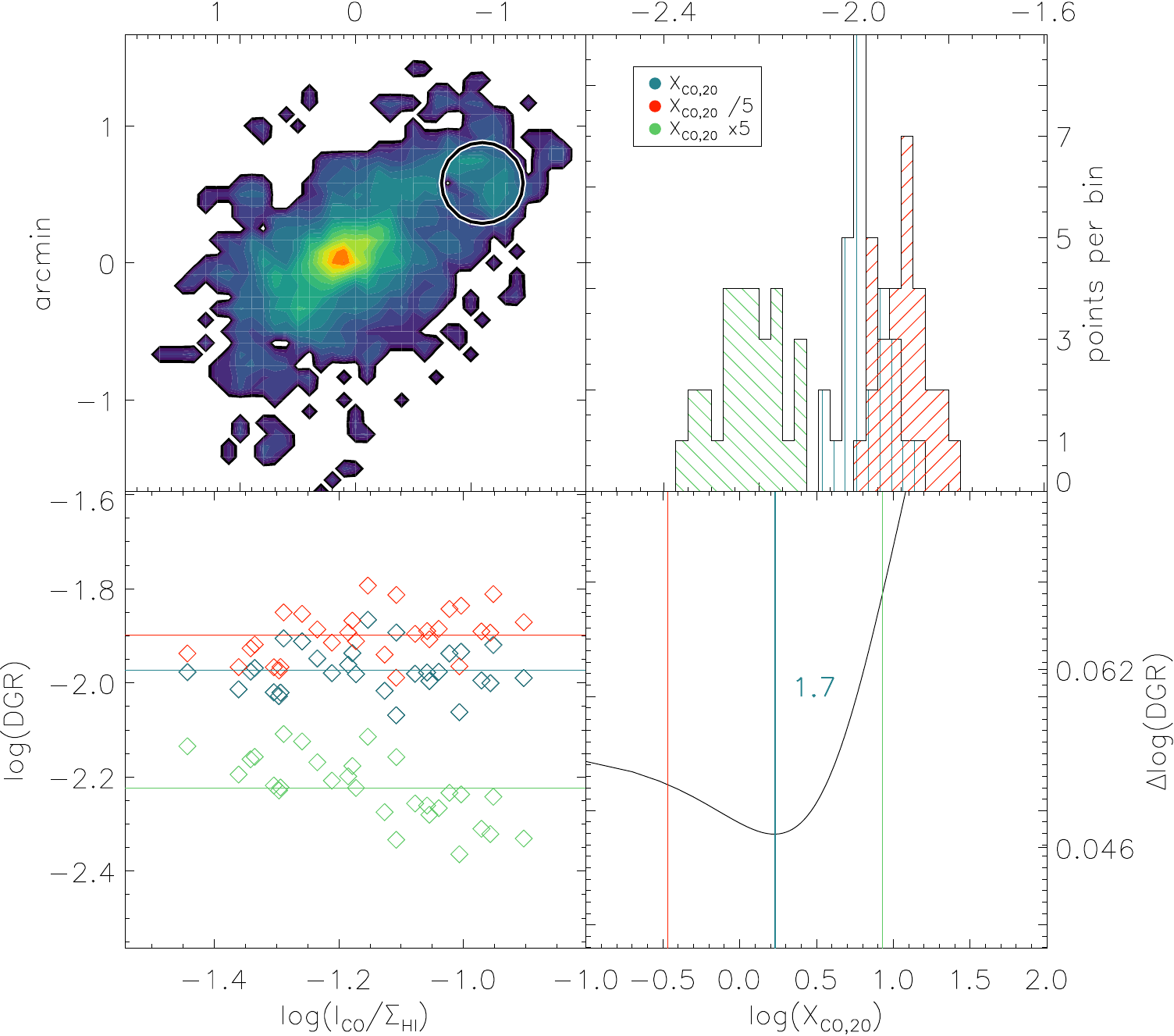}
   \caption{Simultaneous determination of the dust-to-gas ratio and the CO-to-\h2 conversion factor based on method described in \citet{2013ApJ...777....5S}. \textit{Top-left panel:} Selected region for the study on the CO(2-1) map. \textit{Top-right panel:} Scatter of the DGR calculated within resolution elements inside the selected area for different X$_{\rm{CO}}$. \textit{Bottom-left panel: }Standard deviation of the DGR as a function of the ration between the CO(2-1) flux and the H{\sc i} gas surface density. \textit{Bottom-right panel: } DGR dispersion as a function of X$_{\rm{CO}}$.}
   \label{fig:sd4}
\end{figure}

Based on the assumption that the DGR varies only slightly at kpc-scale within
galactic disks (\citealt{2011ApJ...737...12L}, \citealt{2013ApJ...777....5S}, ),
the DGR and \xco\ can be simultaneously determined using \eq{eq:xco}. For given
$\rm{I_{\rm{CO}}}$ and \sga, the process consists in searching for the conversion factor
that minimizes the dispersion of the DGR values in kiloparsec-size regions. We adopted
the method suggested by \citet{2013ApJ...777....5S}: the study have to be carried
out i) in areas where CO is detected; ii) within a range of conversion factors
from 0.1-100~$\times$~10$ ^{20}$~\xcou; {iii) by selecting regions containing 9 resolution elements. We defined four distinct regions in NGC~4654: a first ring around the galactic center at a distance of 2 kpc, a second ring from 2 to 3 kpc, a region that contain the northwest stellar arm and the \hsd. \fig{fig:sd4} presents the DGR minimization method for the \hsd. The other regions can be found in \app{sect:dgrbis}. }

{The results of the DGR variation method for the four regions are presented in \tab{tab:rst}. The uncertainties in the DGRs are estimated via bootstrapping (see \citealt{2013ApJ...777....5S}). All CO-to-H$_2$ conversion factors are consistent with the Galactic value within the errors.}

\begin{table*}[ht!]
   \caption{{Results of the minimization of the DGR variation method.}}
   \label{tab:rst}
   \begin{center}
    \renewcommand{\arraystretch}{1.5}
   \begin{tabular}{ccccc}
      \hline
         & Central ring & Outer ring & Inner arm & $\rm Hi\Sigma_{HI}$ \\
        \hline
        \hline
        D [kpc] & 2.0 & 2.9 & 5.1 & 6.3 \\
        \hline
        $\alpha_{\rm CO}$ & 3.7 $\pm$ 1.2 & 3.3 $\pm$ 1.1 & 6.5 $\pm$ 0.6 & 3.7 $\pm$ 0.6\\
        \hline
        DGR/DGR$_\odot$ & 1.2 $\pm$ 0.2 & 1.3 $\pm$ 0.2 & 0.9 $\pm$ 0.1 & 1.0 $\pm$ 0.1 \\ 
        \hline
        12+log[O/H] & 8.8 $\pm$ 0.1 & 8.8 $\pm$ 0.1 & 8.7  $\pm$ 0.1 & 8.7 $\pm$ 0.1 \\ 
      \hline
   \end{tabular}
      \end{center}
\end{table*}

As the DGR is expected to depend linearly on metallicity, we compare the
metallicity based on the DGR variations to direct measurements.
\citet{1996ApJ...462..147S} obtained 12+log[O/H] measurements using the
strong-line method. For a reliable detection of the electronic temperature,
the detection of faint auroral lines is necessary. However, \citet{1996ApJ...462..147S} carried
out the determination of the metallicity with a constant $\rm{T_e}$. By using a
spectrum stacking method in regions where auroral lines are
usually not detected, although necessary for metallicity measurements,
\citet{2017MNRAS.465.1384C} were able to define a new mass-metallicity
calibration (MZR) for star-forming galaxies. This new calibration reveals a
systematic and significant offset for high mass galaxies. At the typical mass
of NGC~4654, this offset is about 0.3 dex below previous calibrations (see
\citealt{2020MNRAS.491..944C}). We therefore recalibrated the data from
\citet{1996ApJ...462..147S} by applying this offset.
{A second recalibration was performed based on the work of \citet{2009MNRAS.398..949P} (PMC09). This recalibration consists in the estimation of the metallicity
\begin{equation}
   \rm 12 + log[O/H] = 8.74 - 0.31 \times O3N2
\end{equation}
with the O3N2 ratio 
\begin{equation}
    \rm O3N2 = log \left[ \frac{I[OIII]\ (5007 \AA)}{I[{H\beta}]} \times \frac{I[H\alpha]}{I[NII] (6584 \AA)} \right]
\end{equation} 
derived from the data of \citet{1996ApJ...462..147S}.}

{The \citet{1996ApJ...462..147S} metallicity
profiles recalibrated following \citet{2020MNRAS.491..944C} and PMC09 are compared to the metallicity profile determined by the DGR method in \fig{fig:sk1}. These methods lead to metallicities close to the solar value (see \tab{tab:rst}). The \citet{2020MNRAS.491..944C} and PMC09 recalibrations resulted in metallicity gradients of $-0.07$ dex/kpc and $-0.05$ dex/kpc, respectively (\fig{fig:sk1}). Both profiles lead to a solar metallicity at the distance of 4.5 kpc.}

\begin{figure}[ht!]
   \centering
   \includegraphics[width=\hsize]{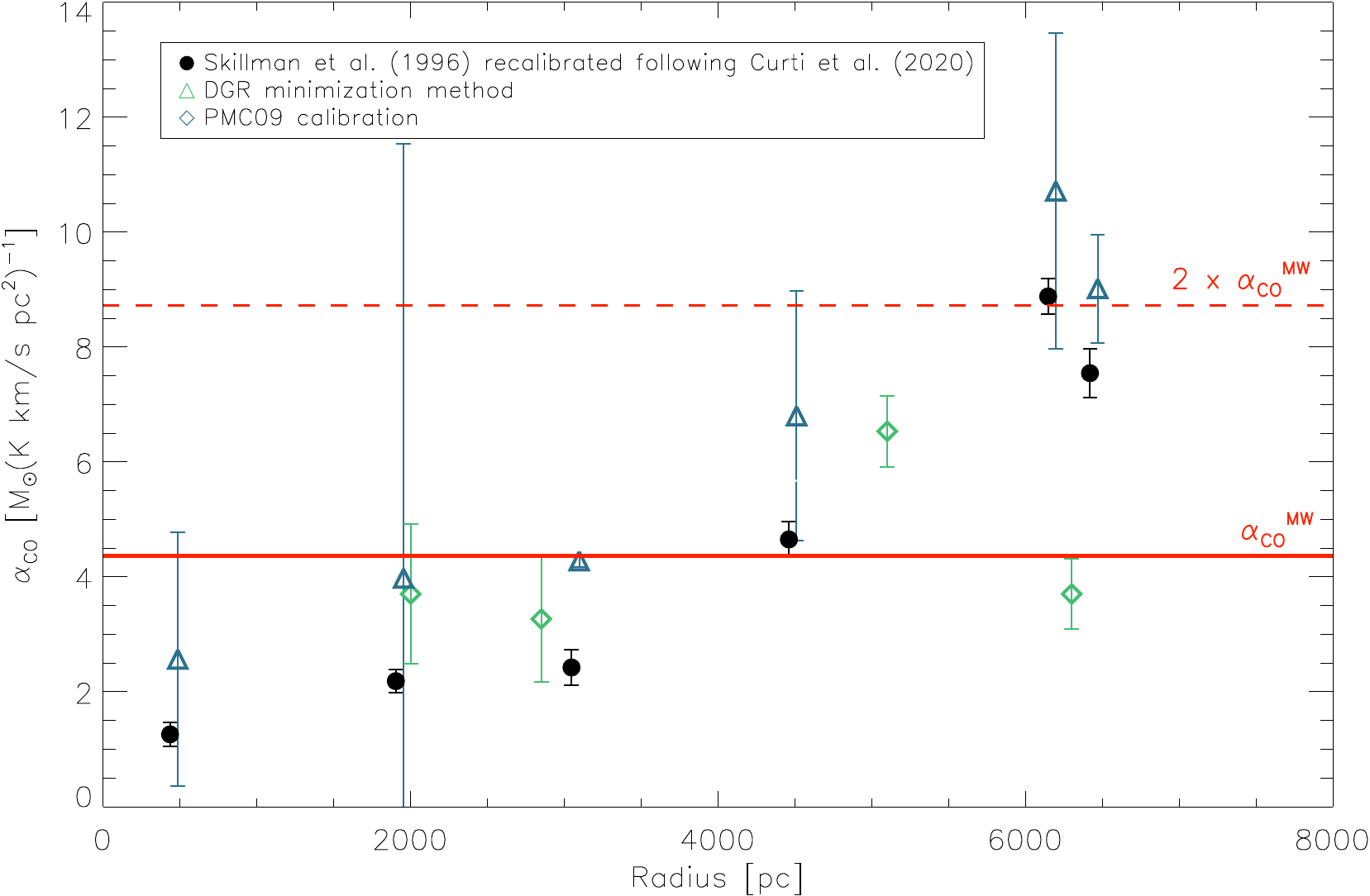}
   \caption{Radial profile of the CO-to-H$_2$ conversion factor. The red line corresponds to the Galactic conversion factor $\alpha _{\rm{CO}} ^{\rm{MW}}$ = 4.36 M$_\odot$(K $\rm km\ s^{-1}$ pc$^2$)$^{-1}$. Black dots correspond to the \aco\ estimation using \citet{1996ApJ...462..147S} metallicities. Black triangles correspond to the conversion factors estimated using the DGR dispersion minimization method.}
   \label{fig:sk2}
\end{figure}

{An alternative CO-to-H$_2$ conversion factor can be obtained by converting the
recalibrated metallicities to DGRs.} The \xco\
derived from \eq{eq:xco} are presented in \fig{fig:sk2}. {Between 0 and 5 kpc the
mean CO-to-H$_2$ conversion factors are \aco~=~3.9~\acou\ for the \citet{2020MNRAS.491..944C} recalibration and \aco~=~5.7~\acou\ for the PMC09 recalibration. These values are close to the Galactic conversion factor. The conversion factors in the \hsd\ are \aco~$\sim$~8.2~\acou\ and \aco~$\sim$~8.7~\acou, about twice the Galactic value.} We therefore set \aco~=~2~\acomw\ in the high \hi\ surface
density region and \aco~=~\acomw\ otherwise. The maps presented in the following
sections are based on this modified conversion factor. The same maps for a Galactic
conversion factor, \acomw\, in the \hsd\ are presented in \app{cte}.

\section{Molecular and total gas} \label{sect:mol}

\begin{figure}[ht!]
   \centering
   \includegraphics[width=\hsize]{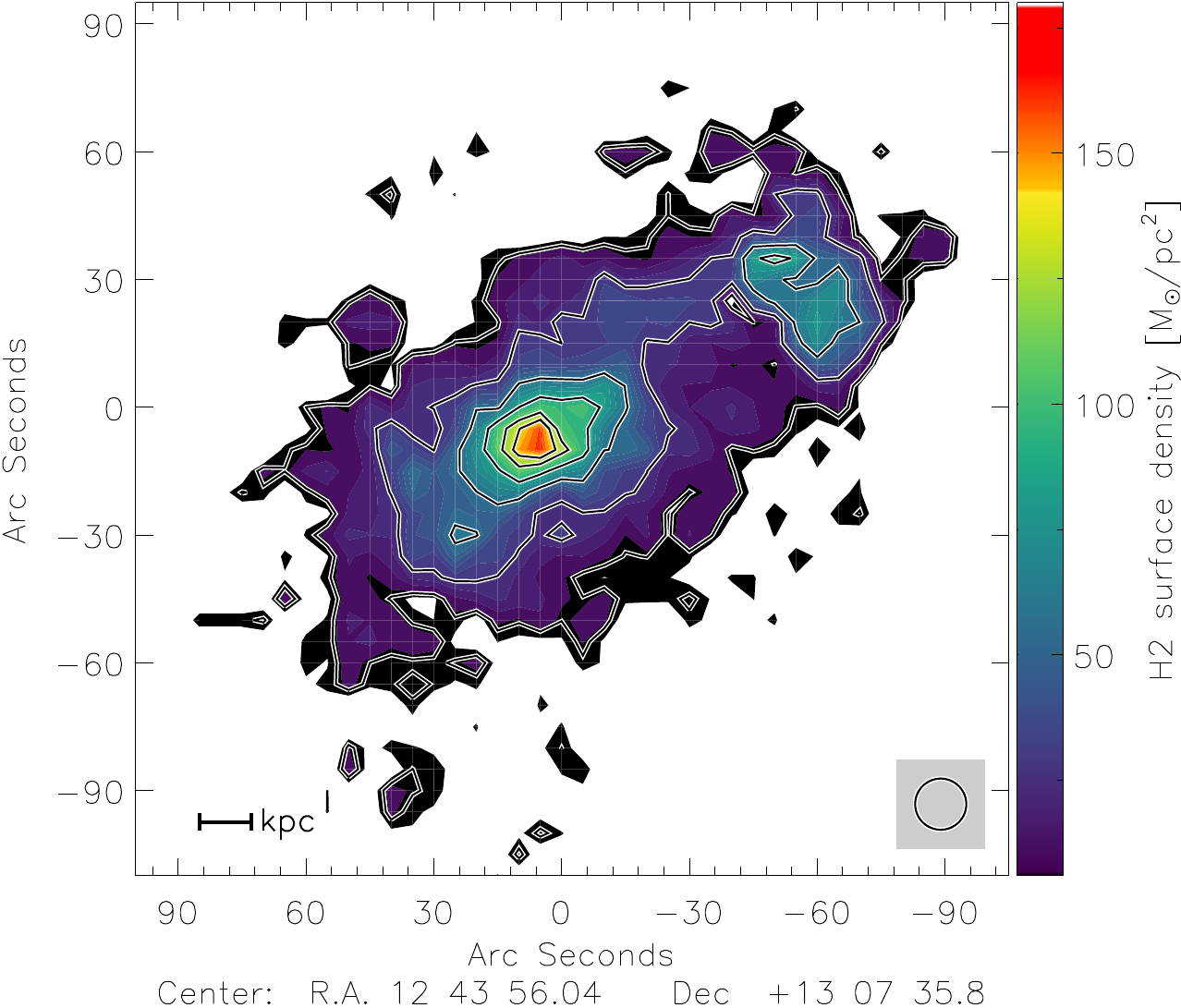}
   \caption{H$_2$ surface density. Contour levels are $10$, $30$, $60$, $90$, $120$ and $140$ \msunpc. The resolution is 12''.}
   \label{fig:h2}
\end{figure}

The molecular gas surface density is computed using the CO-to-H$_2$ conversion
factor presented in \sect{sect:xco}. and is given by the following equation:
\begin{equation}
   \Sigma _{\rm{H_2}} = \alpha _{\rm{CO}}\ \frac{I_{\rm{CO}}(2-1)}{0.8}\ .
\end{equation}
A factor of 1.36 reflecting the presence of Helium is included in \aco. The 3$\sigma$ detection limit of the CO data is $0.5$~\msunpc\ for a constant channel width of 10.4 $\rm km\ s^{-1}$. The resulting map is presented in \fig{fig:h2}. using the modified
$\alpha _{\rm{CO}}$ conversion factor and in \fig{fig:h2b} with a constant
\acomw. The maximum of the molecular gas surface density is reached in the
galaxy center, \sgm~=~173~\msunpc. The distribution of the molecular gas is
strongly asymmetric along the major axis, extended to $\sim$1.6' (7 kpc) to the
southeast and $\sim$1.8' (8 kpc) to the northwest. Toward the northwest, a high
molecular gas surface density arm at 40-50 \msunpc\ is detected, following the
stellar arm (\fig{fig:hi}.). The surface density values along the arm are 2
 to 3 times higher than in the inter-arm region. Within the \hsd, where the conversion factor is assumed to be 2 $\times$ \acomw,
the maximum value is \sgm~=~99~\msunpc. The total molecular gas mass is
$\rm{M_{\rm{H_2}}}$~=~2.1~$\times$~10$^9$~\msun, which is about 10\% higher than the
value using a constant \acomw.

\begin{figure}[ht!]
   \centering
   \includegraphics[width=\hsize]{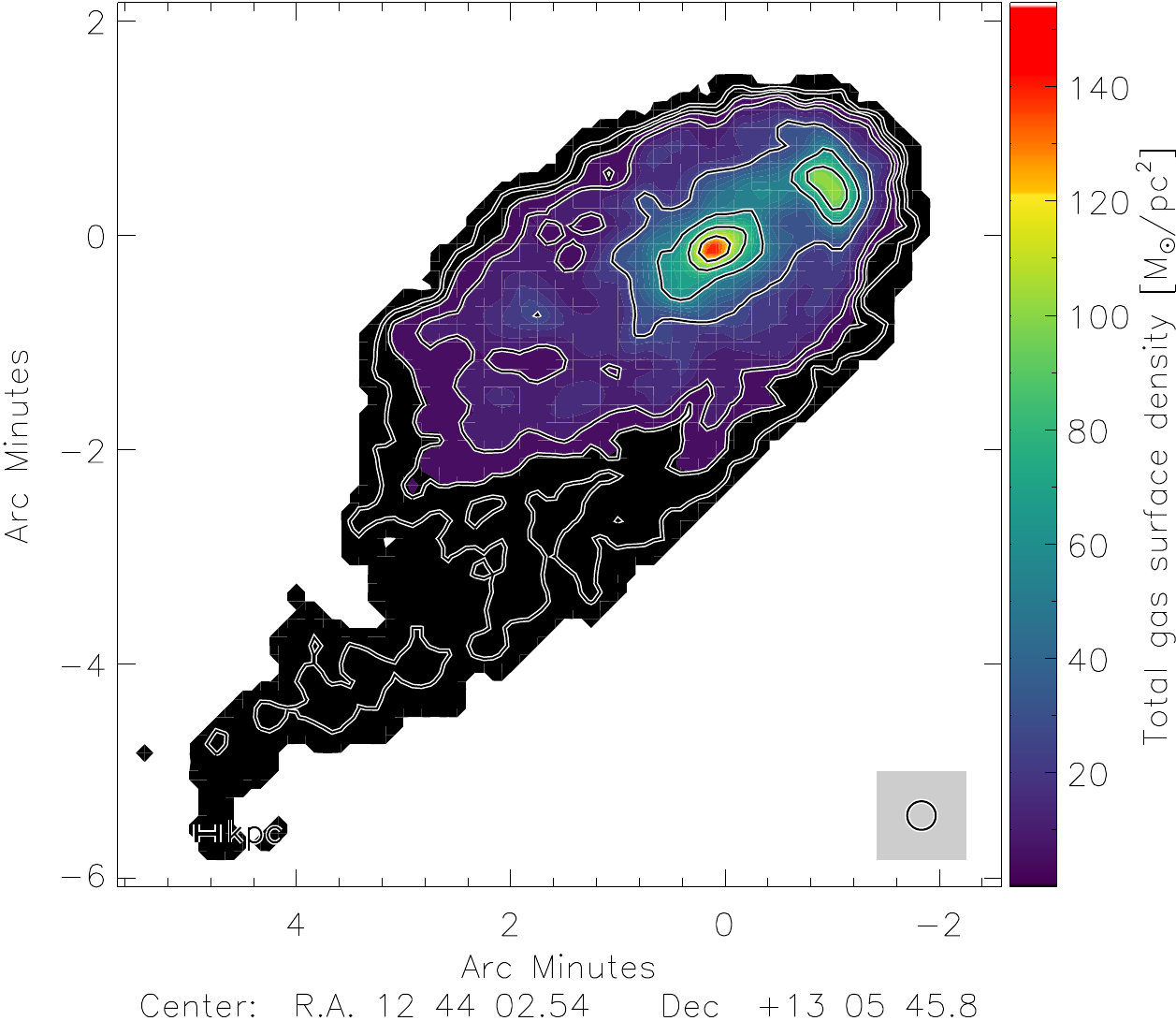}
   \caption{Total gas surface density $\Sigma _{\rm gas} = \Sigma _{\rm{HI}} + \Sigma _{\rm{H_2}}$. Contour levels are 2, 5, 10, 30, 60, 90 and 120 \msunpc. The resolution is 16''.}
   \label{fig:g}
\end{figure}

Convolved to the spatial resolution of 16'', the molecular surface density can be
added to the atomic surface density to compute the total gas map
(\fig{fig:g}). The maximum gas surface density is reached in the galaxy center
with $\rm{\Sigma_{\rm{gas}} = 154\ M_\odot pc^{-2}}$, which is lower than the value
presented for \h2\ due to the convolution of the data. Along the spiral arm
toward the northwest, the density is approximately constant, $\rm{\Sigma_{\rm{gas}}
= 60-70\ M_\odot pc^{-2}}$. Within the \hsd , a local
maximum of 117~\msunpc\ is observed.

\section{Molecular fraction and pressure}\label{sect:rmol}

\begin{figure}[ht!]
   \centering
   \includegraphics[width=\hsize]{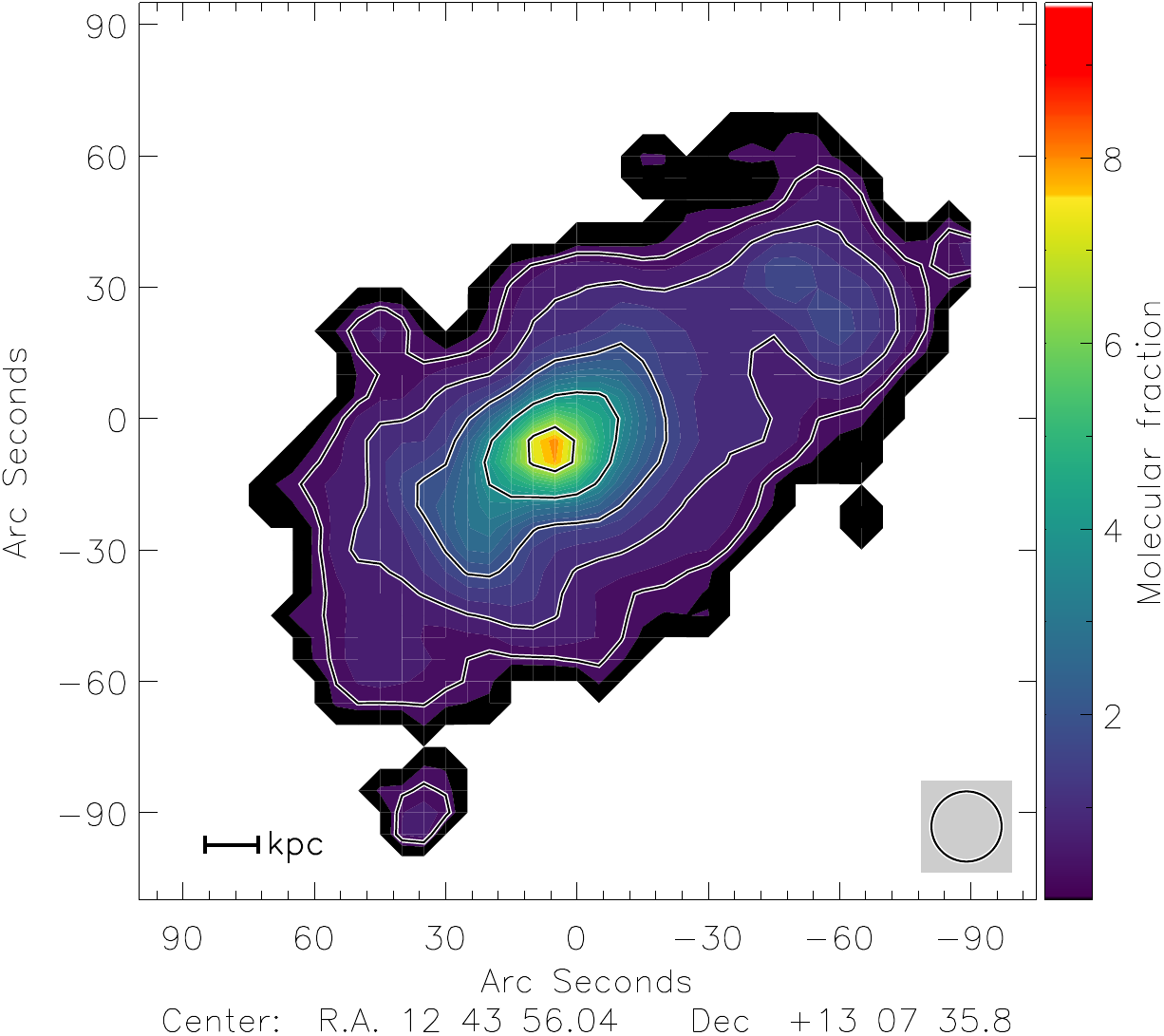}
   \caption{Molecular fraction $R_{\rm{mol}}$ = $\Sigma _{\rm{H_2}}$ /  $\Sigma _{\rm{HI}}$. Contours levels are 0.5, 1, 2, 4 and 7. The resolution is 16''.}
   \label{fig:rmol}
\end{figure}

The molecular fraction corresponds to the ratio of the molecular surface density
divided by the atomic surface density, $\rm{R_{\rm{mol}} =
\Sigma_{\rm{H_2}}/\Sigma_{\rm{HI}}}$ (\fig{fig:rmol}). The molecular fraction
reaches its maximum in the galaxy center, where \rmol~=~8-9. Within the disk of
NGC~4654, \rmol\ decreases with increasing radius. In the \hsd, the molecular fraction is 2-3 times higher than in the disk at
the same radius, reaching a local maximum, $\rm{R_{\rm{mol}} \sim 2}$. 

\begin{figure}[ht!]
   \centering
   \includegraphics[width=\hsize]{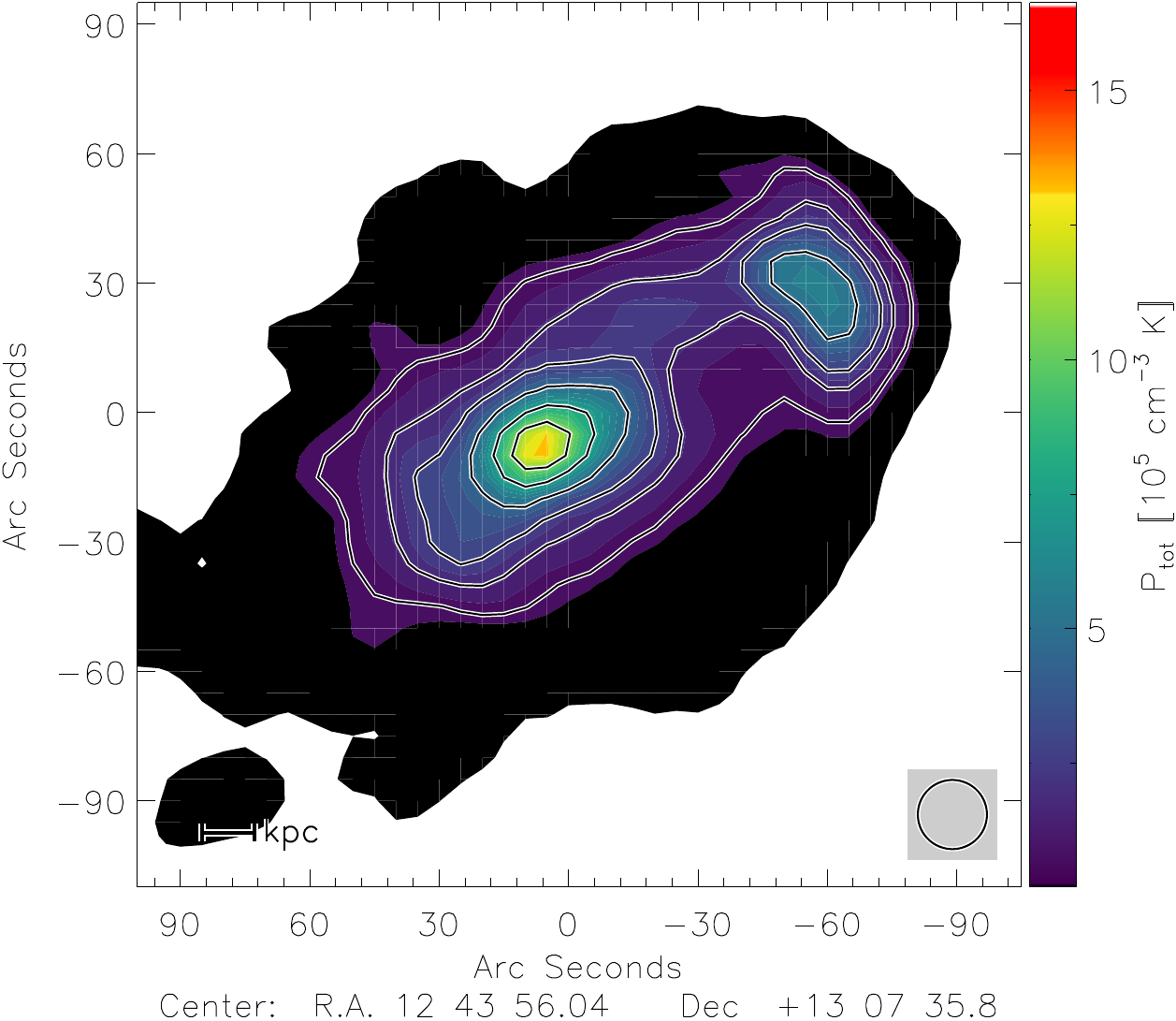}
   \caption{Total ISM mid-plane pressure. Contours levels are 1, 2, 3, 5, 8 and 11 $\times$ 10$^{5}$ cm$^{-3}$ K. The resolution is 16''.}
   \label{fig:pmod}
\end{figure}

\begin{figure}[ht!]
   \centering
   \includegraphics[width=\hsize]{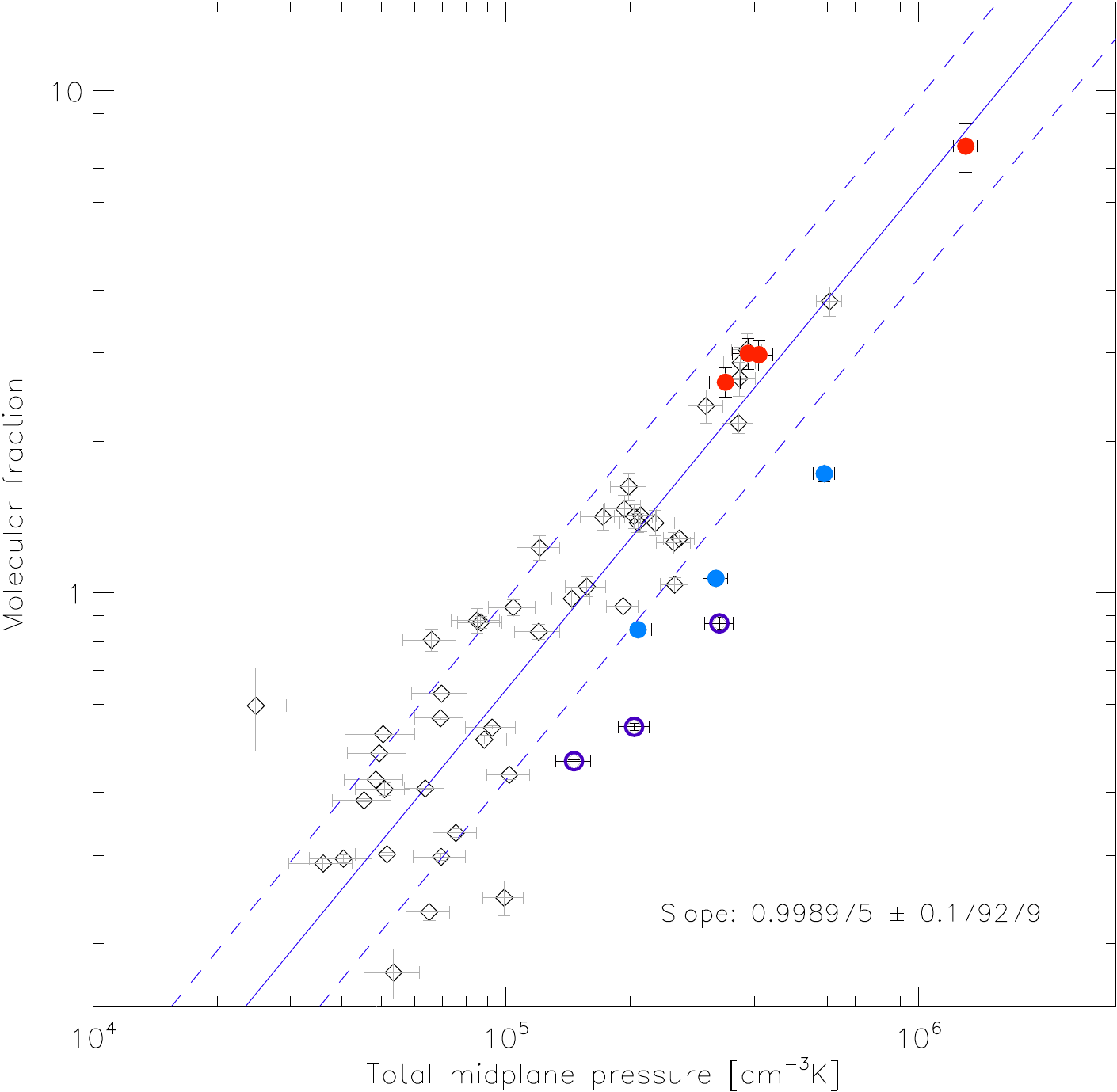}
   \caption{Molecular fraction $R_{\rm{mol}}$ as a function of the ISM pressure $P_{\rm{tot}}$.  Light blue points correspond to the high H$\textsc{I}$ surface density region with \aco = 2 $\times$ \acomw. Dark blue circles correspond \aco = \acomw. Red points correspond to the galaxy center. The dashed lines correspond to $\pm$ 1 $\sigma$.}
   \label{fig:relation1}
\end{figure}

\citet{2006ApJ...650..933B} found a close correlation between the molecular
fraction and the total gas mid-plane pressure in the star-forming galaxies, such
that \rmol~=~$P^{0.92\pm 0.07}_{\rm{tot}}$. Assuming hydrostatic equilibrium within
the disk, the total mid-plane pressure is computed following
\citet{1989ApJ...338..178E}:
\begin{equation} \label{eq:ptot}
P_{\rm{tot}} = P_{\rm{gas}} + P_{\star} = \left( \frac{\pi}{2} G \Sigma_{\rm{gas}}^2 \right) +  \left( \frac{\pi}{2} G \Sigma_{\rm{gas}}\Sigma_{\star} \frac{v_{\rm{disp}}}{v_{\rm{disp}}^\star} \right)\ ,
\end{equation}
{where G is the gravitational constant and $v_{\rm{disp}}$ the gas velocity dispersion $v_{\rm disp}=10(\pm 2)$~km\,s$^{-1}$ (\citealt{2009AJ....137.4424T})}. The resulting map is shown in \fig{fig:pmod}, the ratio between $P_{\rm gas}$ and $P_\star$ is presented in \fig{fig:pp}. The general
distribution of the pressure within the disk is similar to that of the molecular
fraction, with a global maximum at the galaxy center and a local maximum in the
\hsd. The relation between the molecular fraction and the
ISM pressure is presented in \fig{fig:relation1}. 

The slope of the correlation between \rmol\ and $\rm{P_{\rm{tot}}}$ is
(1.00~$\pm$~0.18). This result is consistent with the the value reported by
\citet{2006ApJ...650..933B} cited above. However, the points corresponding to the \hsd\
deviate from the correlation. These values are about 2-3~$\sigma$ lower than
expected by the linear correlation. For a constant \acomw, this offset is still
present because both \rmol\ and $\rm{P_{\rm{tot}}}$ depend on \sgm. By decreasing the
conversion factor the points corresponding the high HI surface density move
parallel to the correlation. We conclude that the lower
$\rm{R_{\rm{mol}}/P_{\rm{tot}}}$ ratio in the \hsd\ is independent of the choice of
\aco.  

\section{Star-formation efficiency}\label{sect:sfe}

\begin{figure}[ht!]
   \centering
   \includegraphics[width=\hsize]{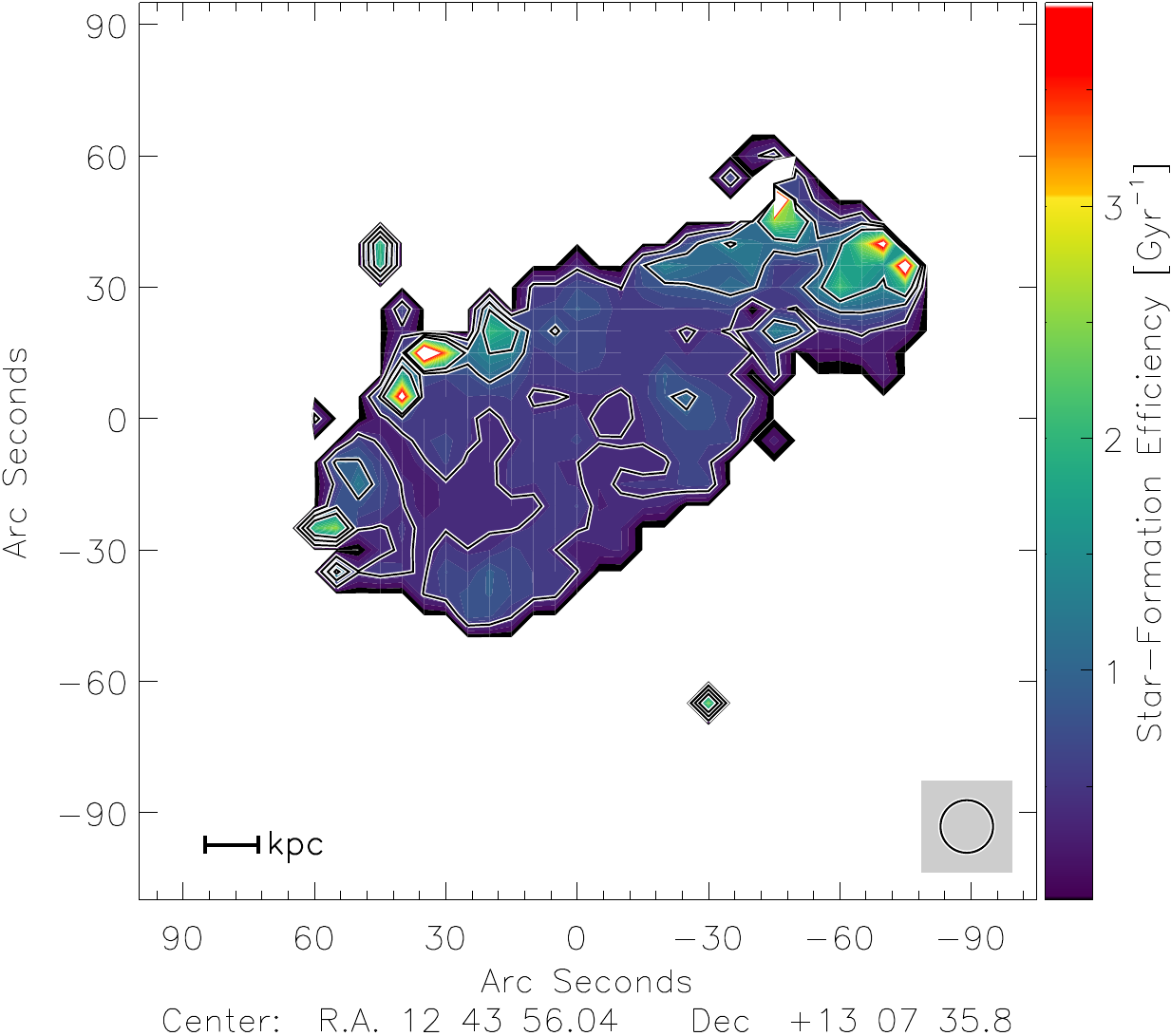}
   \caption{Star-formation efficiency with respect to the molecular gas. Contours levels are 0.5, 1 and 1.5 Gyr$^{-1}$. The resolution is 12''.}
   \label{fig:sfe}
\end{figure}

Since molecular gas is closely correlated with star-formation, the study of
cases where the $\rm{SFE_{\rm{H_2}} = \dot{\Sigma}_\star/\Sigma_{\rm{H_2}}}$ is not constant
provide valuable information on the physics of the ISM.
\citet{2008AJ....136.2846B} showed that variations of the star-formation efficiency within a galaxy are lower than those between individual galaxies. \citet{2014PASJ...66...11C}
suggested that the \sfe\ is significantly higher in the \hsd\ than in the rest
of the disk. We convolved the maps presented in Figs. \ref{fig:sfr} and \ref{fig:h2}
to a resolution of 12'' to obtain the $\rm{SFE_{\rm{H_2}}}$ map shown in
\fig{fig:sfe}.

The molecular gas depletion timescale is defined as $\rm{\tau_{depl}^{\rm{H_2}} = SFE_{\rm{H_2}}^{-1}}$. The mean value within the galactic disk is $\rm{\tau_{depl}^{\rm{H_2}}} = \rm 1.5\ Gyr$, and $\rm{\tau_{depl}^{\rm{H_2}}} = \rm 2\ Gyr$ if we exclude the \hsd\ in the calculations. The maximum of the star-formation
efficiency is reached in the \hsd , where
$\tau_{\rm depl}^{\rm{H_2}}$~$\sim$~500~Myr, which is three times shorter than the mean
value within the disk. 

\begin{figure}[ht!]
   \centering
   \includegraphics[width=\hsize]{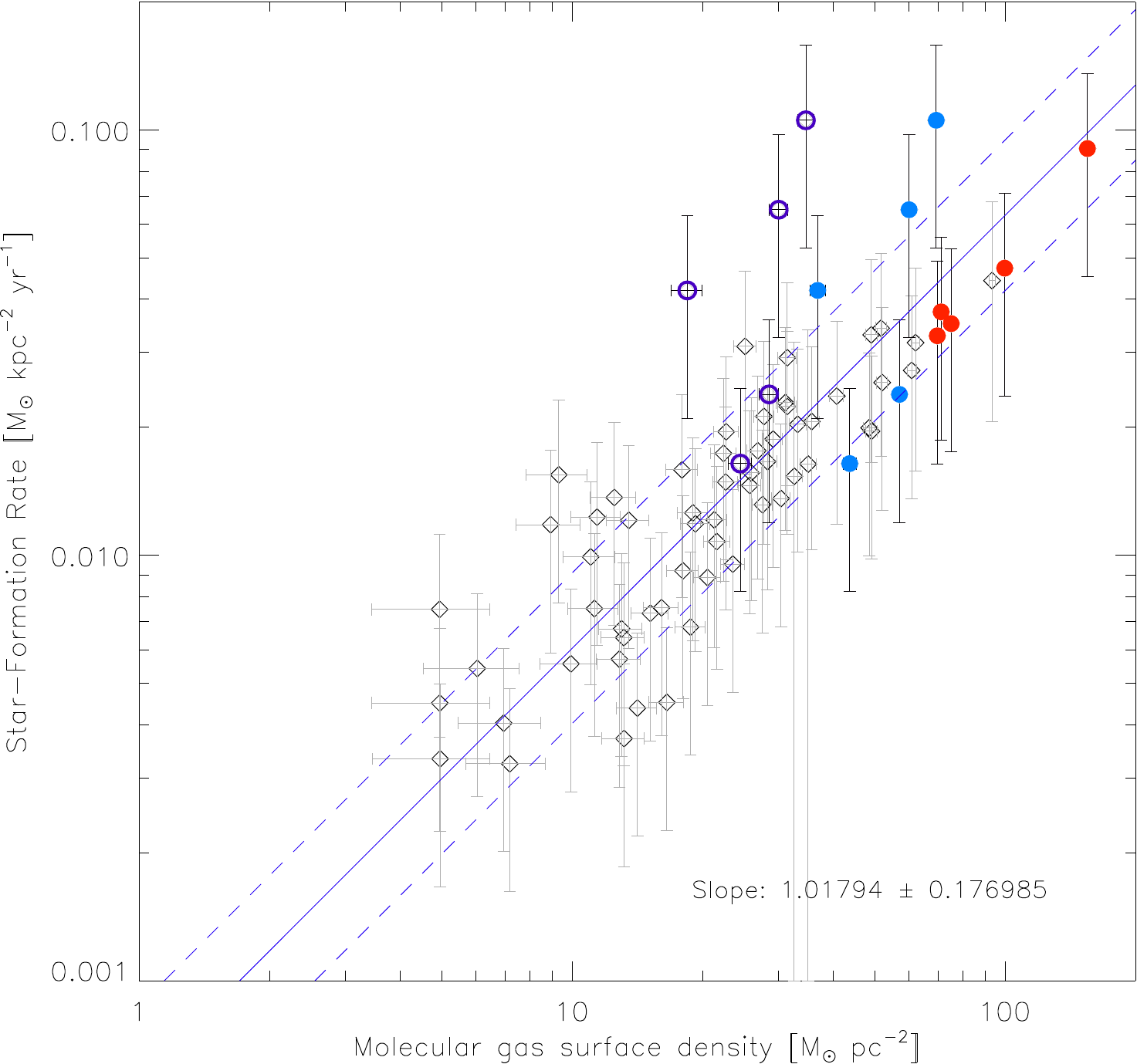}
   \caption{Star-formation rate $\rm{\dot{\Sigma}_\star}$ as a function of the molecular gas surface density $\rm{\Sigma _{\rm{H_2}}}$.  Light blue points correspond to the high $\rm{H\textsc{I}}$ surface density region with \aco = 2 $\times$ \acomw. Dark blue circles correspond to \aco = \acomw. Red points corresponds to the galaxy center. The dashed lines correspond to $\pm$ 1 $\sigma$.}
   \label{fig:relation2}
\end{figure}

The SFR as a function of the molecular gas surface density is
presented in \fig{fig:relation2}. A linear power law is observed between the two
quantities for NGC~4654, with a slope of 1.02~$\pm$~0.18 in agreement with
\citet{2008AJ....136.2846B} (\sfr~$\propto \Sigma^{1.0 \pm 0.2}_{\rm{H_2}}$). The star
formation efficiency of the galaxy center slightly is marginally lower by approximately 0.5~$\sigma$. {For the modified \aco, the $\rm{SFE_{\rm{H_2}}}$
of three resolution elements within the \hsd\ are 1-2~$\sigma$ higher than those of the rest of the disk. With a
constant \acomw, the \sfe\ of the same resolution elements are 2-3~$\sigma$ higher than their values expected from the linear correlation. Two other resolution elements within the \hsd\ show values consistent with the linear correlation. These resolution elements are located at the northern edge of the \hsd. }

\section{The Toomre stability criterion} \label{sect:q}

\begin{figure}[ht!]
      \centering
      \includegraphics[width=\hsize]{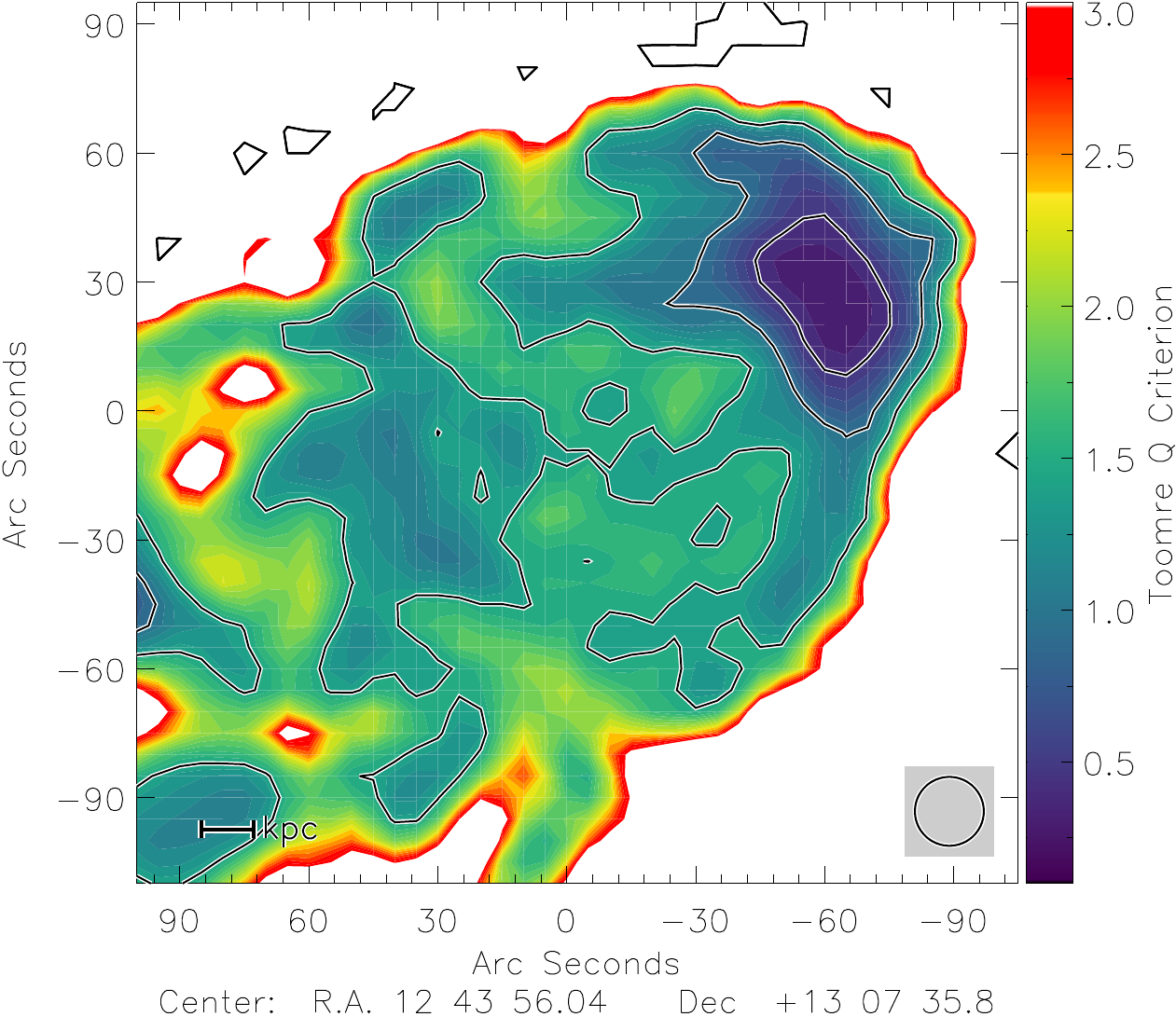}
      \caption{The Toomre $Q$ parameter. Contour levels are 0.5, 1, and 1.5.}
      \label{fig:qana}
   \end{figure}

   The Toomre stability criterion (\citealt{1964ApJ...139.1217T}) describes the
   stability of a gas disk against fragmentation. This criterion depends on the surface
   density and the velocity dispersion of both gas and stars. For our purpose we
   only evaluate the Toomre $Q$ parameter of the gas disk. Since the combined
   Toomre $Q$ parameter is always lower than the individual $Q$ parameters, the
   $Q$ parameter based on the gas can be regarded as an upper limit of the total
   $Q$ parameter. We use the Toomre criterion to evaluate if the northwestern
   region remains stable (Q~$\geq$~1) with an enhanced \aco. The Toomre $Q$ for
   the gas is calculated from the following equation
   (\citealt{1964ApJ...139.1217T}):
   \begin{equation} \label{eq:q}
      Q_{\rm{gas}} = \frac{\kappa v_{\rm{disp}}}{\pi G \Sigma_{\rm{gas}}}\ ,
   \end{equation}
   where $\kappa$ is the epicyclic frequency:
   \begin{equation} \label{eq:K}
      \kappa = \sqrt{2\ \frac{\Omega(R)}{R}\ \frac{d(R^2\Omega)}{dR}}\ .
   \end{equation}
   The angular velocity is $\Omega$(R)~=~$\rm{v_{\rm{rot}}(R) / R}$ and
   $v_{\rm{rot}}$ is the rotation velocity of the galactic disk.
   We assume {a constant gas velocity dispersion of $v_{\rm disp}=10(\pm 2)$~km\,s$^{-1}$ (\citealt{2009AJ....137.4424T}).}
   In Sect.~\ref{sect:ana} this assumption is dropped and a radial profile of the 
   velocity dispersion is calculated based on the analytical model of \citet{2003A&A...398..525V}. 
   We generated two different Toomre $Q$ maps. The first one is based on an approximation of the
   rotation curve following \citet{2003MNRAS.346.1215B}:
   \begin{equation}
      v_{\rm{rot}}=v_{\rm{flat}} \left( 1 - \rm{exp}\left(\frac{ -R }{ l_{\rm{flat}} }\right) \right)\ ,
   \end{equation}
   where $l_{\rm{flat}}$ and $v_{\rm{flat}}$ represent the length scale and velocity at
   which the rotation curve becomes flat. We estimated both values from the
   position-velocity diagram presented in \fig{fig:pvco}. The resulting map
   is shown in \fig{fig:qana}. The second Toomre $Q$ map is based on rotation
   velocities calculated from the deprojected CO velocity field
   (\fig{fig:qobs}). 
   The Toomre $Q$ parameter is lower than one in the whole disk except in the \hsd:
   the minimum is Q~=~$0.5$ for \acomw\ and Q~=~$0.3$ for the modified \aco.

\section{Modeling NGC~4654}

{Our modeling effort is based on the combination of a small-scale analytical model together with a large-scale dynamical model, to handle the properties of a turbulent ISM in a simplified way. The analytical model takes into account gas pressure equilibrium (\eq{eq:ptot}), molecule formation, the influence of self-gravity of giant molecular clouds, and stellar feedback. The large-scale dynamical model includes the gravitational interaction and ram pressure stripping. It gives access to the gas distribution and dynamics on scales of about $1$~kpc. The large-scale model does not include stellar feedback. The combination of the two models gives insight into the physics of the compressed ISM and its ability to form stars.}

\subsection{Analytical model} \label{sect:ana}

The analytical model used for this study is presented in detail in \citet{2011AJ....141...24V}.
   The analytical model describes a star-forming turbulent
   clumpy gas disk with a given Toomre $Q$, where the energy flux produced by
   supernova explosions is dissipated by the turbulence of the gas. This model
   generates radial profiles of the main quantities of this study
   ($\Sigma_{\rm{gas}}$, $\Sigma_{\rm{H_2}}$, $\Sigma_{\rm{HI}}$ and
   $\dot{\Sigma}_\star$) that can be compared with the observations to constrain
   its free parameters. The observational input parameters are the total stellar
   mass, the scale length of the stellar disk, and the rotation curve of the
   galaxy. In this model the ISM is considered as a single turbulent gas in
   vertical hydrostatic equilibrium:
    \begin{equation} \label{eq:3}
      p_{\rm{turb}}=\rho v_{\rm{turb}}^2 = \frac{\pi}{2}G\Sigma_{\rm{gas}}\left( \Sigma_{\rm{gas}} + \Sigma_\star \frac{v_{\rm{turb}}}{v_{\rm{disp}}^\star} \right)\ .
   \end{equation}
   
   Turbulence is driven by
   supernova explosions, which inject their energy into the ISM.
   The model considers turbulence as eddies with the largest eddies defined by a characteristic turbulent driving scale length $l_{\rm{driv}}$ and the associated velocity dispersion $v_{\rm{turb}}$. 
   Assuming a constant initial mass
    function independent of environment, the following equation can be written: 
       \begin{equation} \label{eq:4}
      \xi \dot{\Sigma}_\star = \Sigma_{\rm{gas}} \frac{v_{\rm{turb}}^3}{l_{\rm{driv}}} = \Sigma_{\rm{gas}} \nu \frac{v_{\rm{turb}}^2}{l_{\rm{driv}}^2}\  ,
   \end{equation}
      where $\rm{\xi = 4.6 \times 10^{-8} (pc/yr)^2}$ is the constant relating the
   SNe energy input into star-formation (\citealt{2003A&A...404...21V}). Turbulence gives also rise to the viscosity $\nu$ such that $\nu=v_{\rm{turb}}l_{\rm{driv}}$. In the model, the turbulent crossing time of a single cloud is compared to the gravitational free fall time in order to derive an expression of the volume filling factor $\phi_{\rm{v}}$. Self-gravitation can be assumed for clouds if the condition $\tau_{\rm{turb}}^{\rm{cl}} = \tau_{\rm{ff}}^{\rm{cl}}$ is verified. In such a case, the volume filling factor $\phi_{\rm{v}}$ links the average density of the disk $\rho$ to the
   density of individual clouds $\rho_{\rm{cl}}$ such that $\rho_{\rm{cl}} = \phi_{\rm{v}}^{-1} \rho$.
   The density of a single cloud $\rho_{\rm{cl}}$ refers to the density of the largest self-gravitating structures of a size $l_{\rm{cl}}$. The latter size is smaller than the driving length scale $l_{\rm{driv}}$ by a factor $\delta$, such that $l_{\rm{cl}}=l_{\rm driv}/\delta$. For self-gravitating clouds, the turbulent crossing time and the free-fall timescale can be written: 
   \begin{equation}
      \tau_{\rm{turb}}^{\rm{cl}} = \delta ^{-1} \frac{l_{\rm{driv}}}{ v_{\rm{turb}}}\ ,
   \end{equation}
   \begin{equation} \label{eq:tff}
      \tau_{\rm{ff}}^{\rm{cl}} = \sqrt{\frac{3\pi}{32G\rho_{\rm{cl}}}}\ .
   \end{equation}
   These timescales are used to control both the balance between the atomic and molecular gas phases and
   between turbulence and star-formation. Considering clouds self-gravitation, the star-formation can be finally defined as:
    \begin{equation} \label{eq:5}
      \dot{\Sigma}_\star = \phi_{\rm{v}} \frac{\rho}{\tau_{\rm{ff}}} l_{\rm{driv}} = \delta \phi_{\rm{v}} \rho v_{\rm{turb}}\ .
   \end{equation}
   
In the model, the turbulent motion is expected to redistribute angular momentum in the gas disk like an effective viscosity would do. With this consideration, accretion toward the center is allowed and one can treat the galaxy disk as an accretion disk. Assuming a continuous and nonzero external gas mass accretion rate $\dot{\Sigma}_{\rm{ext}}$, the global viscous evolution can thus be written as:
   \begin{equation}
       \frac{\partial \Sigma_{\rm{gas}}}{\partial t} \sim \frac{\nu \Sigma_{\rm{gas}}}{R^2} - \dot{\Sigma}_\star + \dot{\Sigma}_{\rm{ext}}\ .
   \end{equation}
   The mass accretion rate at a given radius in a galactic disk with a constant rotation curve is given by
     \begin{equation}
       \label{eq:mdotn}
       \dot{M}(R)=-2\,\pi R \Sigma v_{\rm r}=4 \pi R^{\frac{1}{2}} \frac{\nabla}{\nabla\,R}(\nu \Sigma R^{\frac{1}{2}})\ ,
     \end{equation}
    where $v_{r}$ is the radial velocity.
    If a stable total Toomre criterion Q$_{tot}$~$\sim$~1 is held over a few rotation periods by the balance between the mass accretion rate and the gas loss due to star-formation, the galaxy disk can be considered as stationary, such that $\partial \Sigma$/$\partial t$~$\sim$~0. For such a stationary gas disk, the local mass and momentum conservation yield:
    \begin{equation} \label{eq:1}
        \nu \Sigma_{\rm{gas}} = \frac{\rm{\dot{M}}}{2\pi}\ .
    \end{equation}
    Since NGC~4654 underwent a tidal interaction about $500$~Myr ago
    and is now undergoing ram pressure stripping, it cannot be considered as a stable disk.
    However, the relatively unperturbed southeastern half of the
    disk of NGC~4654 is actually not far from such an equilibrium with a constant mass accretion rate
    (left panels of Fig.~\ref{fig:models}).
    On the other hand, the mass accretion rate clearly has to vary in the perturbed northwestern half of the disk.
    We decided to keep Eq.~\ref{eq:1} for convenience and to radially vary $\dot{M}$.
    For each model, the real mass accretion rate can be calculated via Eq.~\ref{eq:mdotn}.  

   The separation between the atomic and molecular phase of the ISM is defined
   by the molecular fraction, estimated by the ratio of the free-fall
   timescale to the molecule formation timescale (\eq{eq:tff} and
   \eq{eq:tmol}):
   \begin{equation} \label{eq:rmol}
      R_{ \rm{mol} } = \frac{ \Sigma_{\rm{ H_2 }} }{ \Sigma_{ \rm{HI} } } \simeq \frac{ \tau_{ \rm{ff} }^{ \rm{cl} }}{ \tau_{ \rm{mol} }^{ \rm{cl} } } \ ,
   \end{equation}
   where the molecule formation timescale is:
   \begin{equation} \label{eq:tmol}
      \tau_{\rm{mol}}^{\rm{cl}} = \frac{\gamma}{\phi_{\rm{v}}^{-1} \rho}\ .
   \end{equation}
   The factor $\gamma$ corresponds to the coefficient of molecule formation timescale (\citealt{1985ApJ...291..722T}). If we
   assume no exchange between the ISM and the environment of the galaxy, we
   can estimate $\gamma$ based on a closed box model following
   \citet{2011AJ....141...24V}: 
   \begin{equation} \label{eq:gam}
      \gamma = \gamma_0 \ \left(\ln\left( \frac{ \Sigma_{\rm{gas}} + \Sigma_\star }{\Sigma_{\rm{gas}}}\right)\right)^{-1} ,
   \end{equation}
   where $\gamma_0$~=~7.2~$\times$~10$^7$~$\rm{yr\ M_\odot\ pc^{-3}}$. The
   metallicity $Z$ can be estimated using the solar value of the constant of
   molecule formation $\gamma_\odot = 4.7 \times 10^7 \ \rm{yr\ M_\odot\
   pc^{-3}}$ as $\rm{Z / Z_\odot = (\gamma / \gamma_\odot)^{-1}}$. In order to
   verify whether this approach is compatible with the results previously
   obtained using the strong-line method and the DGR variation method presented
   in \sect{sect:xco}, we used the radial profiles of the stellar and total gas
   surface densities to calculate the profile of the gas metallicity. We
   separated the galactic disk into two halves: (i) the unperturbed eastern half
   and (ii) the western half containing the \hsd. For the northwestern region,
   we separated two distinct cases: a constant conversion factor \acomw\ and the
   modified \aco\ in the \hsd . The three resulting
   metallicity profiles are shown in \fig{fig:sk1bis}. As the results obtained
   for both conversion factor assumptions are consistent with the previous
   metallicity estimates based on both, the recalibrated
   \citet{1996ApJ...462..147S} observations and the DGR variation method, we
   conclude that the closed-box model is consistent with observations. Using the molecular fraction computed from \eq{eq:rmol}, we obtained: 
   \begin{equation} \label{eq:fmol}
      \Sigma_{\rm{H_2}} = \Sigma_{\rm{gas}}\ f_{\rm{mol}} = \Sigma_{\rm{gas}} \left( \frac{\Sigma_{\rm{H_2}}}{\Sigma_{\rm{H_2}}+\Sigma_{\rm{HI}}}\right) = \Sigma_{\rm{gas}} \left( \frac{R_{\rm{mol}}}{1+R_{\rm{mol}}}\right)\ ,
   \end{equation} 
   \begin{equation} \label{eq:fmol2}
      \Sigma_{\rm{HI}} = \Sigma_{\rm{gas}}\ (1 - f_{\rm{mol}}) = \Sigma_{\rm{gas}} \left( \frac{1}{1+R_{\rm{mol}}}\right)\ .
   \end{equation} 
   
   \subsubsection{Additional free parameters} \label{sect:10.1}
   
   The three free parameters of the analytical model that can be varied to
   reproduce the observational profiles are: {the scaling factor between driving and dissipation length scale $\delta$, the Toomre stability parameter $Q$, and the accretion rate $\dot{M}$.}
   The modification of these parameters induces significant changes in the resulting model
   profiles: (i) an increase in $\rm{\dot{M}}$ leads to an increase in the gas
   velocity dispersion; (ii) a decrease in the Toomre $Q$ parameter leads to an
   increase in the gas surface density; (iii) an increasing $\delta$ leads to an
   increase in both, \rmol\ and \sfr.

   In addition to the free parameters $\delta$, $Q$ and $\dot{M}$, we decided to vary three other 
   parameters that have a significant impact on the radial profiles of the model: {the constant relating the supernova energy injection rate to the SFR $\xi$, the stellar vertical velocity dispersion $v_{\rm{disp}}^\star$, and the constant of molecule formation timescale $\gamma_0$}. Varying these parameters has several consequences on 
   the radial profiles generated by the model: (i) an increase in $\xi$ induces a decrease in the SFR (see \eq{eq:4}); (ii) a higher $v_{\rm{disp}}^\star$ leads to a lower SFR profile; (iii) an increase in $\gamma_0$ strongly 
   decreases the molecular fraction, which means that the \h2\ radial profile is lowered and the \hi\ radial 
   profile is augmented.

    \subsubsection{Degeneracies between parameters} \label{sect:10.2}
    
The study of the analytical model revealed degeneracies between the free parameters. A decrease in $Q$ or an increase in $\dot{M}$ both result in an higher gas surface density. To decrease the value of \rmol , $\delta$ or $\gamma_0$ have to be increased. Finally, four of the six free parameters have a significant impact on the SFR profile: $\dot{M}$ positively; $\delta$, $\xi$ and $v_{disp}$ negatively. 

      \subsubsection{Results}

   To determine which set of parameters best fits the observational data, we carried
   out two successive reduced-$\upchi^2$ minimizations: the first stage consists
   in the reproduction of the radial profiles averaged over the disk, excluding
   the northwest region, which differs from the rest. Once these parameters have
   been defined, a second minimization was carried out to adjust the values of $Q$
   and \mdot\ in the \hsd. We selected a portion equivalent to 30\% to the disk toward the northwest to include the entire \hsd\ and avoid bias caused by stochastic increases in the SFR. The results of the two successive $\upchi^2$
   minimizations are presented in Table~\hyperref[tab:ana]{3}. The
   best-fit models are models 1, 2 and 3, described in detail in the following section. 
5   {Model 1 has the default parameters. Model 2 has a two times higher vertical stellar velocity dispersion, model 3 a two times higher constant relating the supernova energy injection rate to the SFR than model 1.}
   We rejected models 4, 5, 6 {and 7} due to the too high $\upchi^2$ found to reproduce the \hsd. 
      \begin{table*}
      \label{tab:ana}
      \begin{center}
         \renewcommand{\arraystretch}{1.25}
         \addtolength{\tabcolsep}{-2.5pt}
         \begin{tabular}{c c c c c c c c c c c c c} 
            \hline
            \# & \xco & $\delta$ & $Q$ & $\dot{M}$ & $\gamma_0$ & $\xi$ & $v_{\rm{disp}}^\star$ & $v_{\rm{disp}}^{\rm{DISK}}\tablefootmark{(1)}$ & $v_{\rm{disp}}^{\rm{NW}}\tablefootmark{(2)}$ & $Q^{\rm{min}}\tablefootmark{(3)} $ &
            \begin{tabular}{ll} \begin{tabular}{cccc} $\upchi^2_{\rm{tot}}$\ & $\upchi^2_{\rm{HI}}$ &
             $\upchi^2_{\rm{H2}}$ & $\upchi^2_{\rm{SFR}}$ \end{tabular} \\ \hspace{1.1cm} DISK\tablefootmark{(4)}\end{tabular} &
             \begin{tabular}{ll} \begin{tabular}{cccc} $\upchi^2_{\rm{tot}}$\ & $\upchi^2_{\rm{HI}}$ &
             $\upchi^2_{\rm{H2}}$ & $\upchi^2_{\rm{SFR}}$ \end{tabular} \\ \hspace{1.1cm} NW\tablefootmark{(5)}\end{tabular}\\
            \hline
            \hline
            \begin{tabular}{cc} {1} & \begin{tabular}{ll} {(a)} \\ {(b)} \end{tabular} \end{tabular} & 
            \begin{tabular}{c} {-} \\ {$\times$2} \end{tabular}  &  {6.6}  &  {1.7}   &  {0.17}  &  {-}  &  {-}   &  {-}   &  {10.6}  &
            \begin{tabular}{c} {19.5}\\ {13.5}\end{tabular}  & \begin{tabular}{c} {1.26}\\{0.64}\end{tabular}  &
            \begin{tabular}{cccc}{3.3} & {1.0} & {1.0} & {1.3} \end{tabular} & \begin{tabular}{cccc}{5.6} & {3.7} & {1.0} & {0.9} \\ {4.7} & {0.8} & {1.4} & {2.4}\end{tabular}  \\ [1ex]
            \hline
            \begin{tabular}{cc} \textbf{2} & \begin{tabular}{ll} \textbf{(a)} \\ (b) \end{tabular} \end{tabular} &
            \begin{tabular}{c}\textbf{-} \\$\times$2 \end{tabular}  & \textbf{ 5.7}  &  \textbf{1.5}   &  \textbf{0.10}  &  \textbf{-}  & \textbf{-}   &   \textbf{$\times$2}   &  \textbf{8.7}  &
            \begin{tabular}{c}\textbf{ 18.0} \\ 12.0 \end{tabular}  & \begin{tabular}{c}\textbf{ 1.20} \\ 0.58 \end{tabular}  &
            \begin{tabular}{cccc} \textbf{3.1} & \textbf{0.9} & \textbf{1.1} & \textbf{1.1} \end{tabular} & \begin{tabular}{cccc} \textbf{3.2} &\textbf{ 0.8} & \textbf{0.8} & \textbf{1.6} \\ 3.7 & 1.1 & 1.7 & 1.0\end{tabular}  \\ [1ex]    
            \hline
            \begin{tabular}{cc} \textbf{{3}} & \begin{tabular}{ll} (a) \\ \textbf{{(b)}} \end{tabular} \end{tabular} &            
            \begin{tabular}{c}- \\ \textbf{{$\times$2}} \end{tabular}  &  \textbf{{4.8}}  &  \textbf{{1.6}}   &  \textbf{{0.11}}  & \textbf{-}  &  \textbf{ {$\times$2}}   &  \textbf{{-}}   &  \textbf{{10.0}}  & 
            \begin{tabular}{c}19.8\\ \textbf{{12.0}}\end{tabular}  &  \begin{tabular}{c}1.23 \\ \textbf{{0.50}}\end{tabular}  & 
            \begin{tabular}{cccc}\textbf{ {3.0}} & \textbf{{0.9}} & \textbf{{1.0}} & \textbf{{1.0}} \end{tabular} & \begin{tabular}{cccc} 4.1 & 0.9 & 1.1 & 2.1 \\ \textbf{{2.6}} &\textbf{ {1.1}} &  \textbf{{0.8}} & \textbf{{0.8}}\end{tabular}  \\ [1ex]
            \hline
            \hline
            \begin{tabular}{cc} 4 & \begin{tabular}{ll} (a) \\ (b) \end{tabular} \end{tabular} &
            \begin{tabular}{c}- \\$\times$2 \end{tabular}   &  11.1  &  1.6   &  0.13  &  $\times$2  &  -   &  -   &  9.4  &
            \begin{tabular}{c}18.8\\11.1\end{tabular}  & \begin{tabular}{c}1.25\\0.42\end{tabular}  &
            \begin{tabular}{cccc}3.1 & 1.0 & 1.1 & 1.0 \end{tabular} & \begin{tabular}{cccc}8.0 & 1.0 & 0.9 & 6.0 \\ 4.8 & 3.2 & 0.8 & 0.8 \end{tabular}  \\ [1ex]                  
            \hline
            \begin{tabular}{cc} 5 & \begin{tabular}{ll} (a) \\ (b) \end{tabular} \end{tabular} &
            \begin{tabular}{c}- \\$\times$2 \end{tabular}  &  5.7 &  1.8   &  0.17  &  -  &  $\times$2   &  $\times$2   &  11.2  &
            \begin{tabular}{c}20.5\\14.1\end{tabular}  & \begin{tabular}{c}1.28\\0.57\end{tabular}  &
            \begin{tabular}{cccc}2.9 & 0.9  & 1.1 & 0.9 \end{tabular} & \begin{tabular}{cccc} 7.9 & 0.8 & 0.9 & 6.2 \\ 4.9 & 2.8 & 0.8 & 1.3\end{tabular}  \\ [1ex]  
            \hline
            \begin{tabular}{cc} 6 & \begin{tabular}{ll} (a) \\ (b) \end{tabular} \end{tabular} &
            \begin{tabular}{c}- \\$\times$2 \end{tabular}  &  10.2  &  1.9   &  0.22  &  $\times$2  &  $\times$2  &  -   &  12.1  &
            \begin{tabular}{c}20.1\\14.2\end{tabular}  & \begin{tabular}{c}1.29\\0.44\end{tabular}  &
            \begin{tabular}{cccc}4.0 & 1.5 & 1.1 & 1.3 \end{tabular} & \begin{tabular}{cccc}12.9 & 0.8 & 0.8 & 11.2 \\ 7.2 & 5.5 & 0.9 & 0.8 \end{tabular}  \\ [1ex]  
            \hline
            \begin{tabular}{cc} 7 & \begin{tabular}{ll} (a) \\ (b) \end{tabular} \end{tabular} &
            \begin{tabular}{c}- \\$\times$2 \end{tabular}  &  11.1 &  1.5   &  0.13  &  $\times$2  &  -  &  $\times$2  &  9.1  &
            \begin{tabular}{c}18.0\\10.7\end{tabular}  & \begin{tabular}{c}1.30\\0.40\end{tabular}  &
            \begin{tabular}{cccc}3.8 & 1.5 & 1.1 & 1.1 \end{tabular} & \begin{tabular}{cccc}11.9 & 1.1 & 1.6 & 9.3 \\ 6.7 & 4.4 & 0.8 & 1.4 \end{tabular}  \\ [1ex]  
            \hline
         \end{tabular}
      \end{center}
      \caption{Sets of parameters obtained from $ \upchi^2$-minimization to fit the observation profiles with the analytical model.}
      \tablefoot{\\
{      \tablefoottext{1}{Mean velocity dispersion in the unperturbed disk. } \\
      \tablefoottext{2}{Maximum velocity dispersion reached in the \hsd.}
      \\
      \tablefoottext{3}{Minimum Toomre Q reached in the \hsd.}
        \\
      \tablefoottext{4}{$\upchi^2$ obtained for the southeastern unperturbed disk}
        \\
      \tablefoottext{5}{$\upchi^2$ obtained for the northwestern perturbed disk}}
      } 
   \end{table*}
   
    \subsubsection*{The unperturbed eastern disk}

{All three models reproduce the observational profiles for the unperturbed disk in a satisfactory way. The total $\upchi^2$ are all about three. We found values of $\delta$ between $4.8$ and $6.6$, a Toomre Q parameter between $1.5$ and $1.7$, and a mass accretion rate $\dot{M}$ between $0.11$ and $0.17$ $\rm M_\odot yr^{-1}$. These parameters lead to a mean velocity dispersion between $8.7$ and $10.6$ km~$ \rm s^{-1}$ within the unperturbed disk, which corresponds to common values for undisturbed local spiral galaxies (e.g, \citealt{2009AJ....137.4424T}). }
        
    \subsubsection*{The northwestern region with a constant \acomw}

\begin{figure*}[ht!]
      \centering
        \includegraphics[width=0.75\hsize]{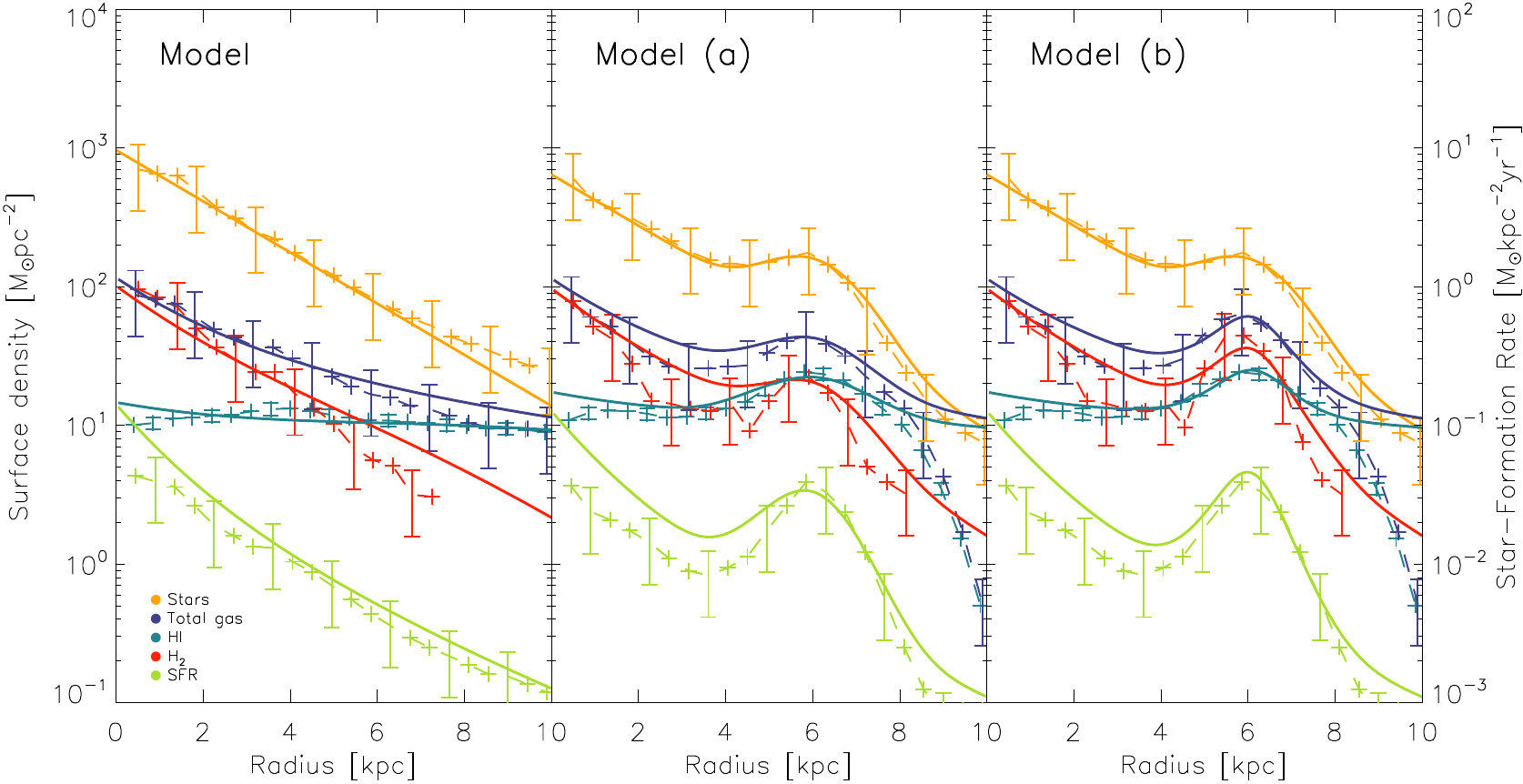}\label{fig:model1}
        \hfill
        \includegraphics[width=0.75\hsize]{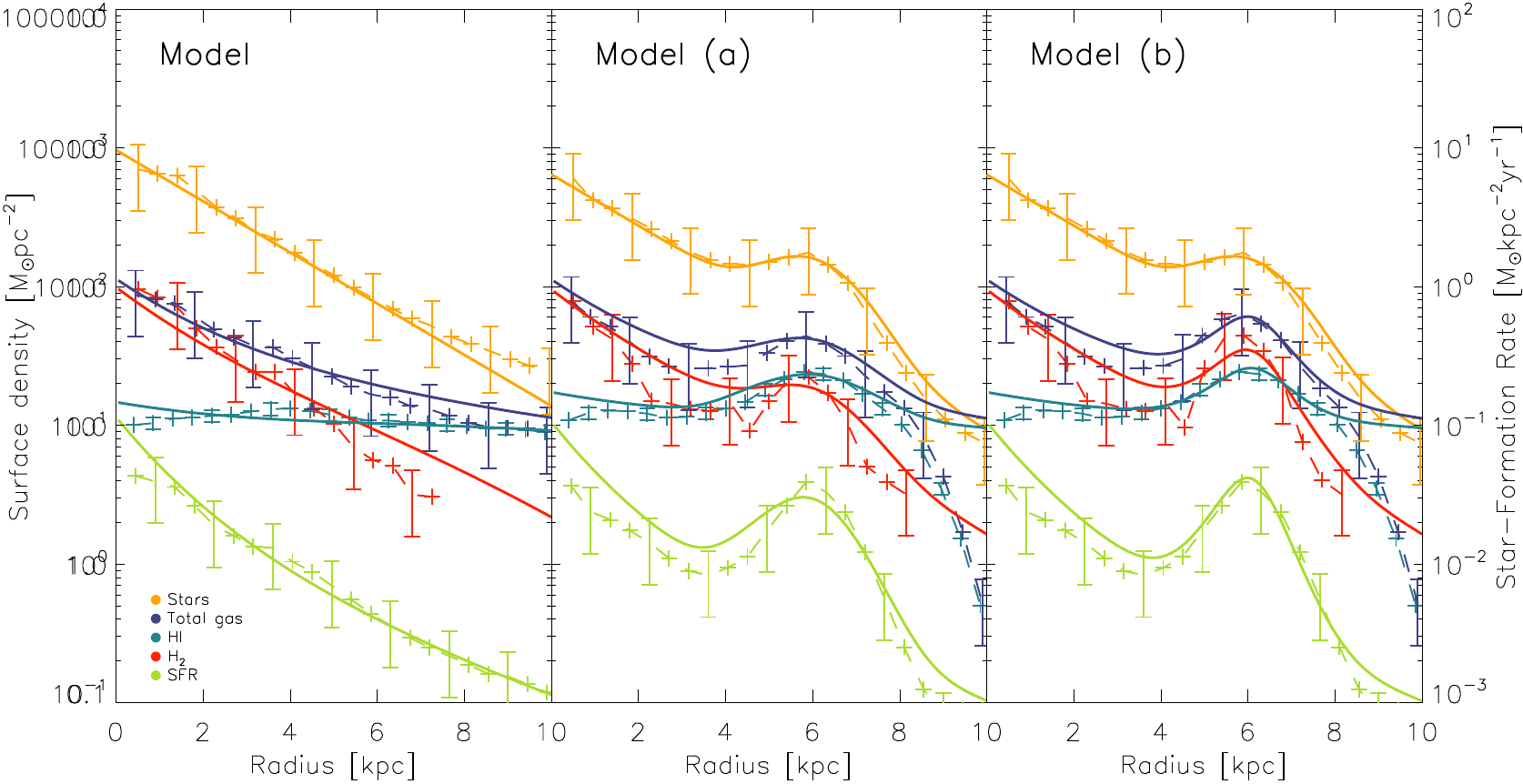}\label{fig:model2}
        \hfill
        \includegraphics[width=0.75\hsize]{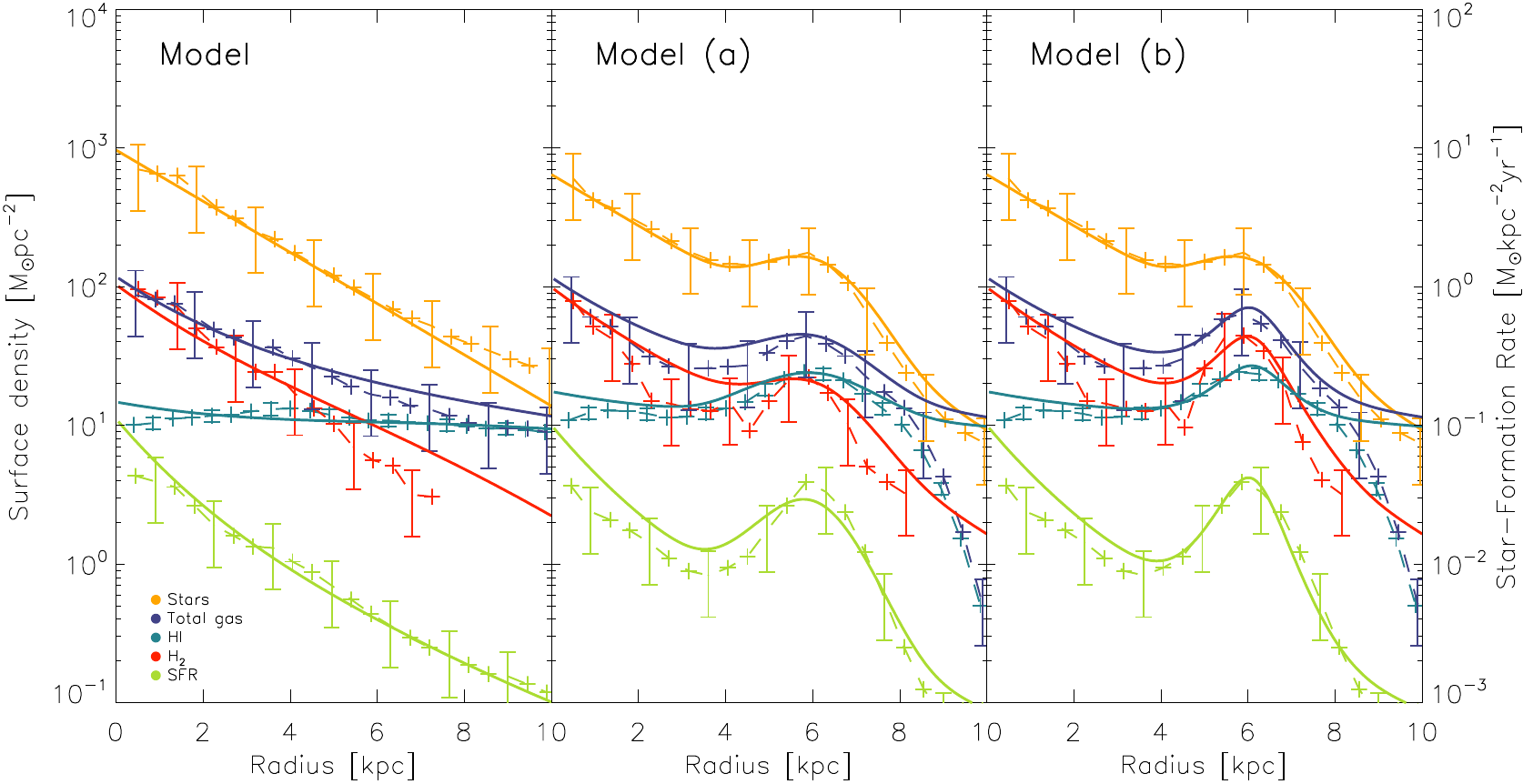}\label{fig:model3}
      \caption{{Radial profiles derived from observations (dashed lines) and from the analytical model (solid lines). The models correspond to model 1, 2, and 3. The models are presented in Table~\hyperref[tab:ana]{3}. The \h2\ profiles are limited by the detection threshold of 3~$\sigma$ within three velocity channels, corresponding to 2.8 \msunpc.} }
      \label{fig:models}
   \end{figure*}

{Model~1(a) is the solution with the highest $\chi^2_{\rm{tot}}$ of the three selected models. This model reproduces almost perfectly the \h2\ and SFR radial profiles within the \hsd\ but is the worst to reproduce the \hi\ in the same region (see \fig{fig:models}). The drop of the Toomre Q parameter is half that of the observations while the increase in the velocity dispersion is two times higher. However, the advantage of this model is that it does not need any modification of the additional parameters to reproduce the observations. }
 
{Model~2(a) is the best model to fit the \hsd\ with a constant conversion factor. The star-formation profile within the \hsd\ is reproduced in an acceptable way (see \fig{fig:models}). However the model requires a nearly 10 km~$s^{-1}$ increase in velocity dispersion to reproduce the observations, while a fairly small decrease in the Toomre Q compared to the observations is found. }

{Model~3(a) is a decent but imperfect solution. Although the two gas phases in the \hsd\ are reproduced quite accurately compared to observations (see \fig{fig:models}). The deviation of the SFR from
observations makes this model an acceptable but unfavorable
choice for this study. The increase in the gas velocity dispersion and the drop in the Toomre Q criterion are comparable with the previous models. }

{Given its high $\chi^2_{\rm{tot}}$, we first reject model~1(a). We prefer model~2(a) to model~3(a) because it is the model with the lowest $\chi^2_{\rm{tot}}$ of all considered solutions for the northwestern region with a constant \aco.} 

    \subsubsection*{The northwestern region with a modified \aco}
    
{Model~1(b) has the highest $\chi^2_{\rm{tot}}$ of the three considered models. Contrary to model~1(a), the \hi\ is well reproduced but the SFR deviates from observations (see \fig{fig:models}). The increase in the velocity dispersion is small ($\Delta v$~$\sim$~2.9~$\rm km\ s^{-1}$) and the Toomre Q parameter drops significantly below the critical value of 1, with $Q$~=~$0.64$. }

{Model~2(b) is an acceptable solution. As for Model~1(b), the increase in the velocity dispersion is small with $\Delta v$~$\sim$~3.3~$\rm km\ s^{-1}$ and the Toomre Q drops to $Q$~=~$0.58$ in the \hsd. }

{Model~3(b) is the best-fit model for a modified conversion factor. This model presents the lowest $\chi^2_{\rm{tot}}$ of all models presented in Table~\hyperref[tab:ana]{3}. Model~3(b) closely reproduces all observational profiles in the \hsd\ (see \fig{fig:models}). The increase in the velocity dispersion is only $\Delta v$~$\sim$~2.0~$\rm km\ s^{-1}$ and the Toomre Q parameter is $Q$~=~$0.5$.}

{Model~3(b) is thus our preferred model for the modified conversion factor. All three models show a rather small increase in the velocity dispersion and a significant drop of the Toomre parameter in the \hsd. }

\subsubsection*{Conclusion for the analytical model}

{The unperturbed disk is reproduced by model 1, 2, and model 3.  the unperturbed disk. We found mean values of $\delta$~$\sim$~5.6, $Q$~$\sim$~1.6, and $\dot{M}$~$\sim$~$0.13\ \rm{M}_\odot \rm{yr}^{-1}$. }
    
For all the solutions and regardless of the choice of \aco, an increase in
the velocity dispersion combined with a decrease in the Toomre criterion is mandatory to
reproduce the \hsd. {Models using a constant conversion factor lead to a significant increase in the velocity dispersion and a slight drop in the Toomre Q parameters in the \hsd\ compared to the unperturbed disk. On the other hand, models using a modified \aco~=~2~\acomw\ lead to a significant drop in the Toomre Q parameter with only a slight increase in the velocity dispersion. Overall, the increase in the velocity dispersion in the \hsd\ is $\Delta v$~$\sim$~2-10~$\rm km\ s^{-1}$ and the Toomre Q parameters ranges between $Q$~=~$0.5-1.3$. Since we think that the CO-to-H$_2$ conversion factor lies between one and two times the Galactic value, we expect a Toomre Q parameter of $Q$~$\sim$~$0.8$ and $\Delta v$~$\sim$~5~$\rm km\ s^{-1}$. }

\subsection{The dynamical model} \label{sect:dyn}

The dynamical model is based on the N-body sticky particle code
described in \citet{2001ApJ...561..708V}. The particles
are separated in two distinct phases: a non-collisional phase that reproduces
the dark matter and stellar component of NGC~4654 and a collisional component for the ISM. The
model takes into account both the gravitational interaction with the neighboring galaxy
NGC~4369 and the influence of ram pressure stripping over $1$~Gyr. The ram pressure
time profile is of Lorentzian form, which is consistent with
highly eccentric orbits of galaxies within the Virgo cluster
(\citealt{2001ApJ...561..708V}). The effect of ram pressure stripping is
simulated by an additional pressure on the clouds in the wind direction, defined as $p_{\rm{ram}} = \rho v_{\rm{turb}}^2 \sim$~200~cm$^{-3}\rm km\ s^{-1}$. Only
clouds that are not shielded by other clouds are affected by ram pressure.
The star-formation is assumed to be proportional to the cloud collision rate. Stars are formed by cloud-cloud collision and added to the total number of particles as zero-mass points with the position and the velocity of the colliding clouds. The information about the time of creation is attached to each new star particle, making it possible to model the \ha\ emission for each snapshot using stars created less than 10 Myr ago. The UV emission is modeled by the UV flux from single stellar population models from STARBURST99 (\citealt{1999ApJS..123....3L}). The total UV flux corresponds to the extinction-free distribution of the UV emission from newly created star particles. The cloud masses range between 3~$\times$~10$^5$ and
3~$\times$~10$^6$~$\rm{M_\odot}$. The atomic and molecular phases of the ISM are
separated by computing the molecular fraction $f_{\rm{mol}}$ using
(\eq{eq:fmol} and \eq{eq:fmol2}). Observational input parameters are presented in \tab{tab:ngc}. 
The dynamical model produces CO, \hi, FUV, stellar mass and \ha\ data cubes at the spectral and spatial resolution of the observations. This first set of modeled data is used to produce derived quantities such as total mid-plane pressure, SFR, star-formation efficiency and Toomre $Q$ criterion in order to carry out a complete comparative study between the model and the observations. 

Contrary to the analytical model, the dynamical model gives direct access to two
quantities that are difficult to observe although fundamental: the volume
density $\rho$ of the gas and its intrinsic 3D velocity dispersion $v_{\rm{disp}}$. The model neglects supernova feedback, 
which may be the origin of an increased gas velocity dispersion and an enhancement of the SFR within dense gas regions. 

Three
different versions of the model were produced: (i) constant pressure of $p=$~200~cm$^{-3}$$\rm km\ s^{-1}$; (ii) constant ram pressure of $p=$~100~cm$^{-3}$$\rm km\ s^{-1}$; (iii) no ram pressure. 
Only the model with a strong ram pressure wind was able to reproduce both the extended HI gas tail and the \hsd, so we chose to focus our study on this version. The maps obtained from the other versions are presented in \app{others}. We search by eye for the relevant timestep that reproduces the available observations best. A summary table of the comparison between the model and observations is presented in 
Table~\hyperref[tab:model]{4}. 

\subsubsection{Stellar and atomic gas surface densities}

\begin{figure}[ht!]
   \centering
   \includegraphics[width=\hsize]{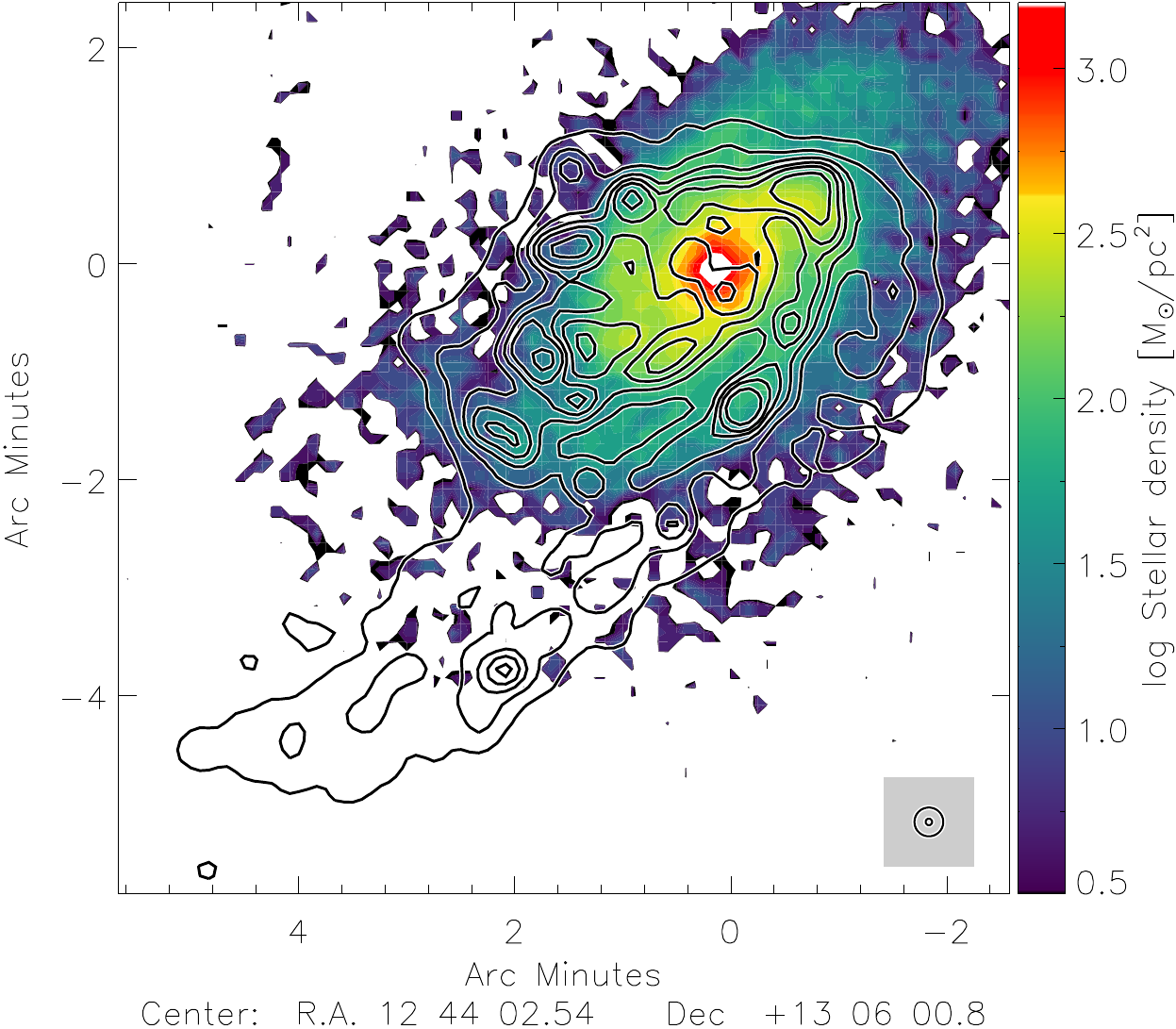}
   \caption{NGC~4654 dynamical model. The colors correspond to the stellar surface density. The contours correspond to the \hi\ surface density. Contour levels are 1, 5, 10, 15, 20, and 30 \msunpc. The model data are convolved to the same spatial resolutions as the observations.}
   \label{fig:simuhi}
\end{figure}

Maps of the surface density of atomic gas and stars are shown in \fig{fig:simuhi}. 
The stellar distribution of the dynamical model is highly asymmetric. The diffuse stellar disk extends 9 kpc to the southeast compared to 12 kpc to the northwest. Toward the northwest an overdense stellar arm is formed. At the end of this stellar arm, a region of high \hi\ surface density appears in the model, with total gas surface densities on the order of 50~\msunpc. Beyond the southeastern edge of the optical disk an extended \hi\ gas tail is formed with surface densities of 5-10~\msunpc. Another high \hi\ surface density region is observed south of the galaxy center with \sga~>~40 \msunpc. By studying previous timesteps of the model, we identified this region as an overdensity with a lifetime of few tens million years created accidentally at this position. We do not consider this overdensity as relevant for analysis. 

Qualitatively, the overall model stellar and gaseous distributions are consistent with observations (\fig{fig:hi}). The overdense stellar arm is well reproduced, with a comparable \hsd\ at its end. The \hi\ gas tail in the southeast direction is also well reproduced by the dynamical model. However, the northwestern half of the diffuse stellar disk is much more extended in the model than in the observations. Quantitatively, 
the stellar arm presents similar surface densities, with $\Sigma_\star = 200-500$~\msunpc.
The model \hsd\ presents comparable \hi\ surface densities, with a maximum of \sga~$\sim$~50~\msunpc\ compared to the observed maximum of 35~\msunpc. However, the \hi\ surface density in the southeastern gas tail is 2 to 3 times higher than in the observations.

The dynamical model is therefore able to reproduce the dense stellar arm, the \hsd\ in the northwest and the \hi\ tail. However, it overestimates the influence of the ram pressure stripping and does not reproduce properly the diffuse stellar component. 

\subsubsection{Molecular and total gas surface densities}

\begin{figure}[ht!]
   \centering
   \includegraphics[width=\hsize]{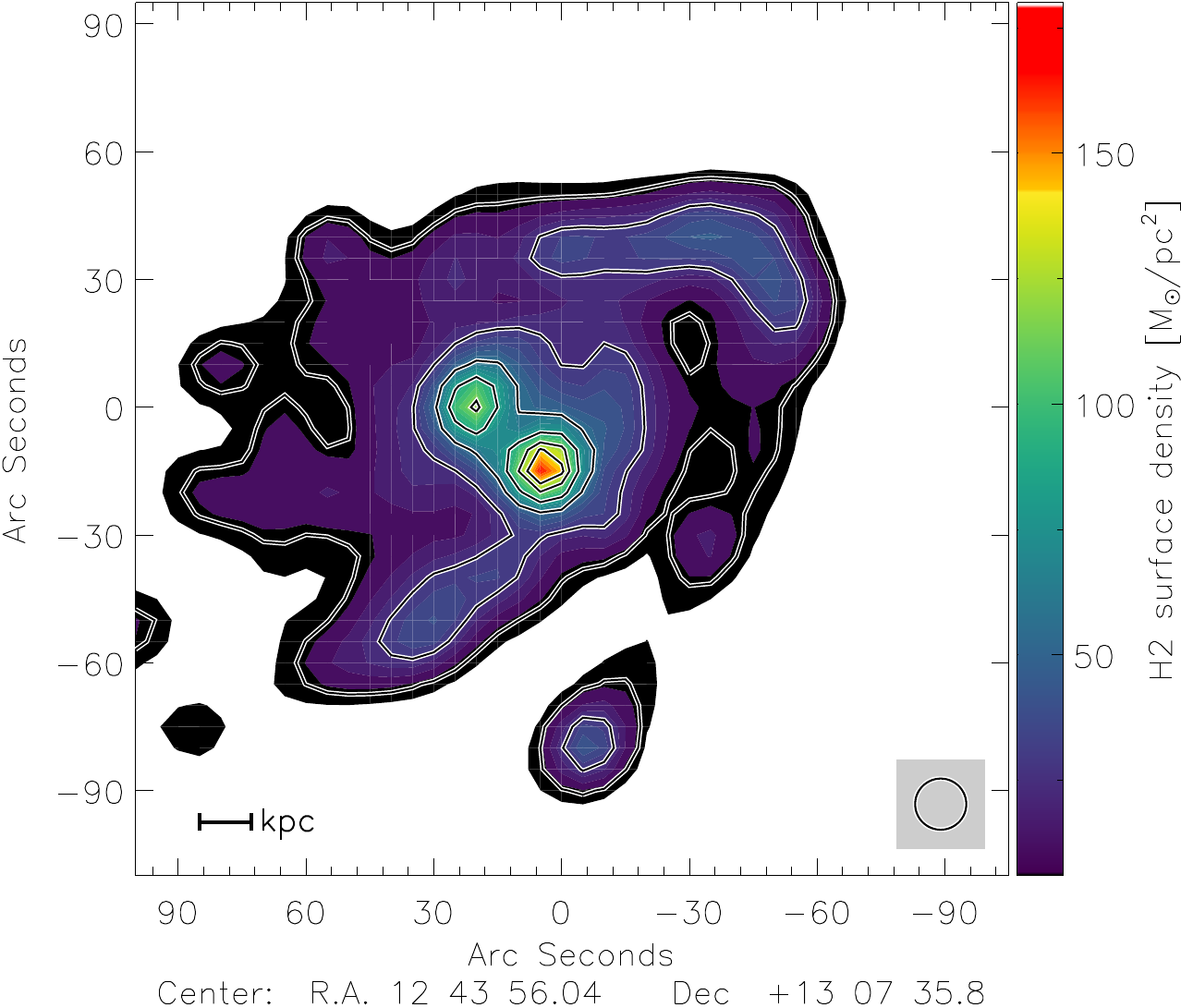}
   \caption{Molecular gas surface density of NGC~4654 from the dynamical model. Contour levels are 2, 5, 10, 30, 60, 90, and 120 \msunpc.}
   \label{fig:simuh2}
\end{figure}

The molecular gas map of the model is presented in \fig{fig:simuh2}. Its maximum is reached in
the galaxy center with \sgm~=~170~\msunpc. The model forms 
an overdense arm in the northwest direction with \sgm~$\sim$~40~\msunpc\, which follows the stellar arm. On the opposite side, a southeastern spiral arm with lower surface densities is present. Moreover, an external gas arm is created without any correlation with the stellar distribution in the southern edge of the disk. This gas arm is identified as a consequence of shear motions induced by the combined effect of galaxy rotation and ram pressure stripping.

As for the \hi\ gas, the general distribution of \h2\ gas is broadly
consistent with observations (\fig{fig:h2}). The average surface density within the northwestern arm is almost equivalent to the observations. The local maximum within the \hsd\ is about 40 \msunpc, which is almost two times lower than the observed value with a modified \aco\ and equivalent to observations with a constant conversion factor, \acomw. The galaxy center also presents comparable surface density as observations. The differences between the model and the observations are essentially: (i) the presence of the southeastern spiral arm that we did not intend to reproduce with the model; (ii) the overdense external gas arm formed in southern edge of the galactic disk, suggesting that the model slightly overestimates the strength of the ram pressure stripping. 

\begin{figure}[ht!]
   \centering
   \includegraphics[width=\hsize]{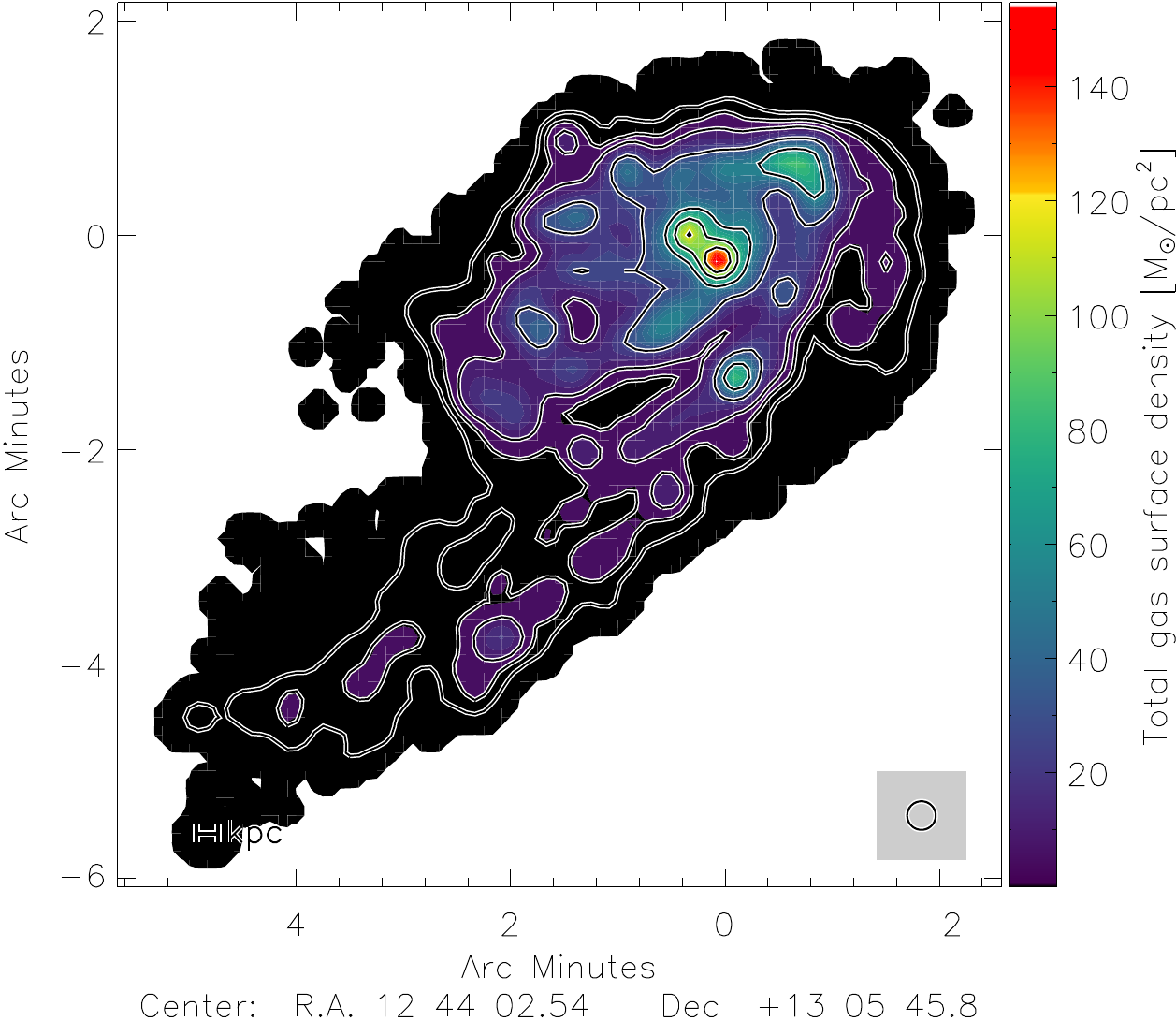}
   \caption{Total gas surface density of NGC~4654 from the dynamical model. Contour levels are 2, 5, 10, 30, 60, 90 and 120 $\rm{M_{\odot}pc^{-2}}$}
   \label{fig:simugas}
\end{figure}

We added the atomic gas to the molecular gas to study the distribution of the total gas of the model (\fig{fig:simugas}). Since on one hand the molecular gas surface density is consistent with observations with a constant conversion factor but on the other hand the \hi\ is slightly overestimated, the total gas map of the model provides an intermediate solution between observations with \acomw\ and a modified \aco. The maximum of $\Sigma_{\rm gas}$ reached in the \hsd\ is 90 \msunpc, which is 10 \msunpc\ above the observations with a constant conversion factor and 30 \msunpc\ below the observations with a modified conversion factor. To conclude, the model seems to be able to reproduce quite accurately the general distribution of the total gas of NGC~4654. It is, however, slightly more similar to observations with a constant factor than with a modified factor \aco. 

\subsubsection{Star-formation rate} 

\begin{figure}[ht!]
   \centering
   \includegraphics[width=\hsize]{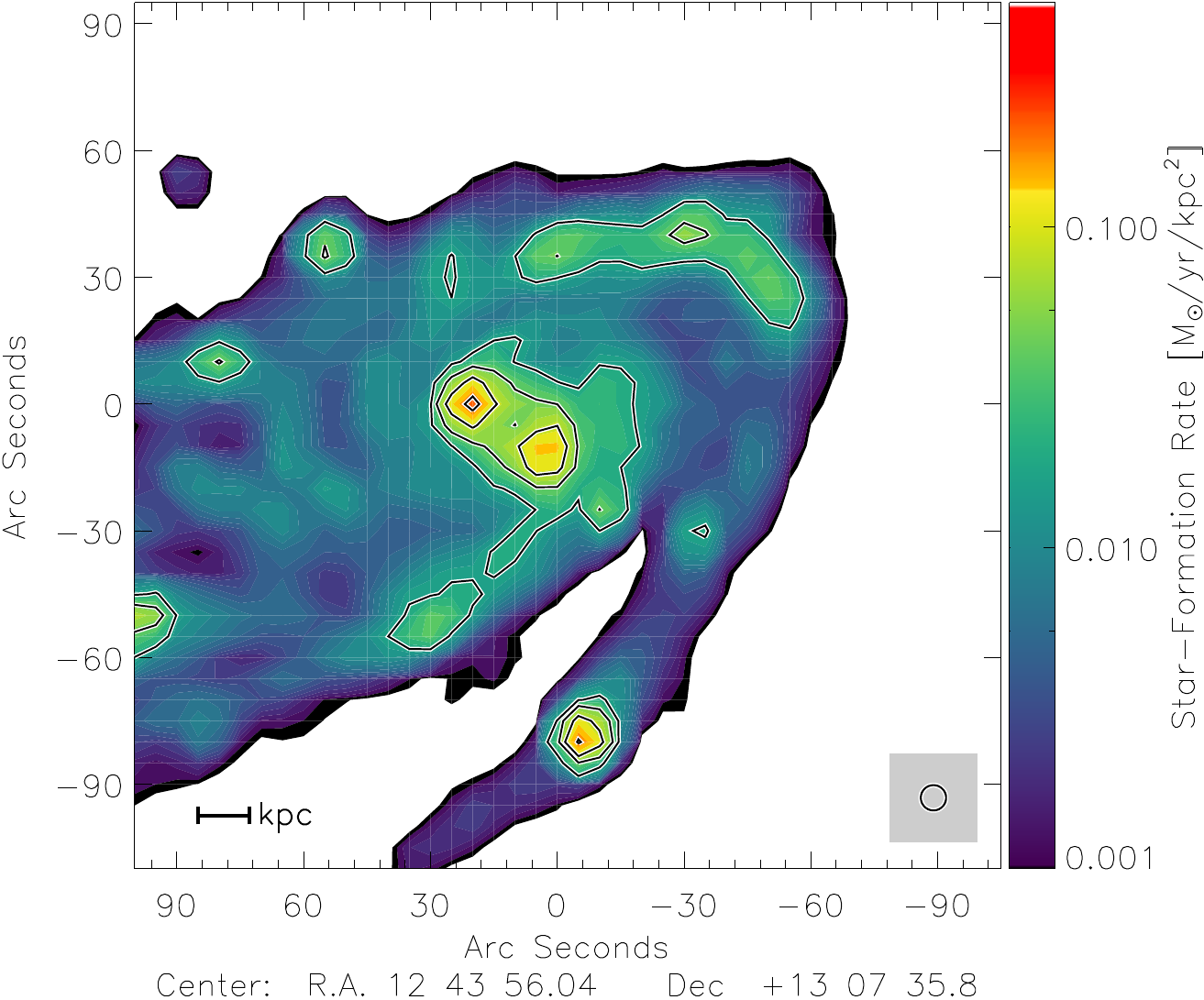}
   \caption{Star-formation rate of NGC~4654 from the dynamical model. Contours levels correspond to 0.02, 0.05, 0.10 and 0.20 \msun kpc$^{-2}$yr$^{-1}$.}
   \label{fig:simusfr}
\end{figure}

To compute the model SFR, we used the FUV map generated by the dynamical model 
normalized to the total observed SFR. The model SFR map is presented in \fig{fig:simusfr}. The maximum of the SFR is reached in the galaxy center with \sfr~=~0.2~M$_\odot$kpc$^{-2}$yr$^{-1}$. Toward the northwest, a higher SFR surface density arm is observed, corresponding to the dense stellar and molecular gas surface density arm. We also note the presence of a slightly enhanced \sfr\ arm following the southeastern spiral arm on the opposite side of the galaxy.

The general morphology of the SFR in the model is consistent with observations, except for the higher SFR arm along the southeast spiral arm, which is not observed (\fig{fig:sfr}). This could be however linked to the small enhancement of the \ha\ emission observed in \fig{fig:ha}. The maximum of the SFR is reached in the galaxy center, contrary to observations where the maximum is located in the \hsd. The model SFR is 2-3 times lower in the northwestern region and 2-3 times higher in the galaxy center than the corresponding observed SFRs. 

To conclude, the model seems to be able to reproduce qualitatively the SFR distribution of NGC~4654 but not quantitatively. As mentioned in the introduction of this section, we can presume that this is due to the absence of supernova feedback in the model.

\subsubsection{\rmol-$P_{\rm{tot}}$ and \sgm-\sfr\ correlations}

\begin{figure}[ht!]
   \centering
   \includegraphics[width=\hsize]{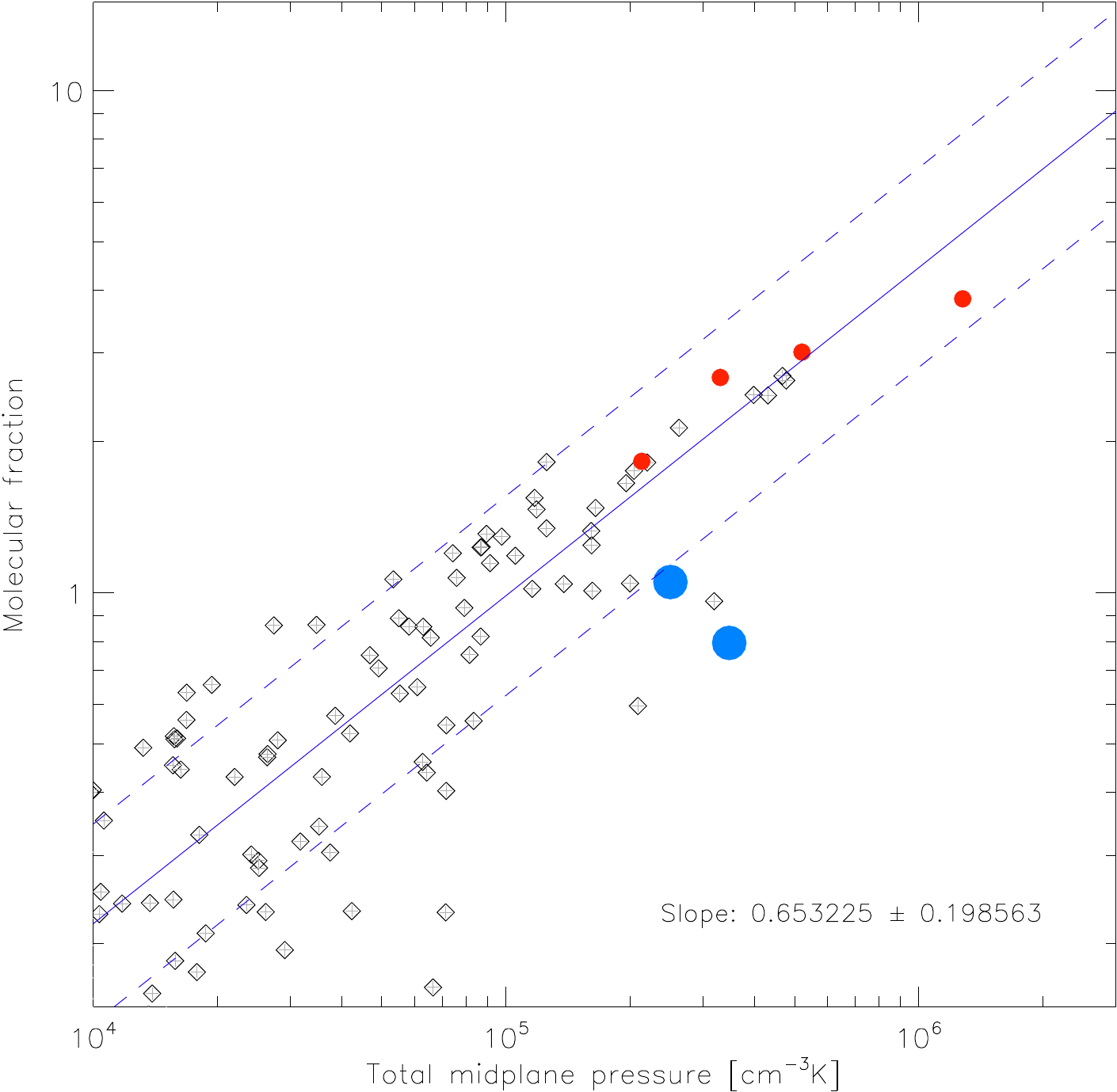}
   \caption{Molecular fraction $R_{\rm{mol}}$ as a function of the ISM pressure $P_{\rm{tot}}$ of NGC~4654 from the dynamical model. Blue points correspond to the high $\rm{H\textsc{I}}$ surface density region. Red points correspond to the galaxy center. The dashed lines correspond to $\pm$ 1 $\sigma$.}
   \label{fig:1dsimu1}
\end{figure}

The model correlation between the molecular fraction of the ISM and the total gas pressure is presented in \fig{fig:1dsimu1}. The \rmol-$P_{\rm{tot}}$ slope is 0.65~$\pm$~0.19, which is significantly flatter than the
predictions of \citet{2006ApJ...650..933B}. The slope is consistent within the errors bars with the results of \citealt{2008AJ....136.2782L} ($0.73$), measured for local spiral galaxies of the THINGS survey. As for the observations, the model molecular fraction within the \hsd\ deviates by 1 to 2 $\sigma$ from the overall correlation.

\begin{figure}[ht!]
   \centering
   \includegraphics[width=\hsize]{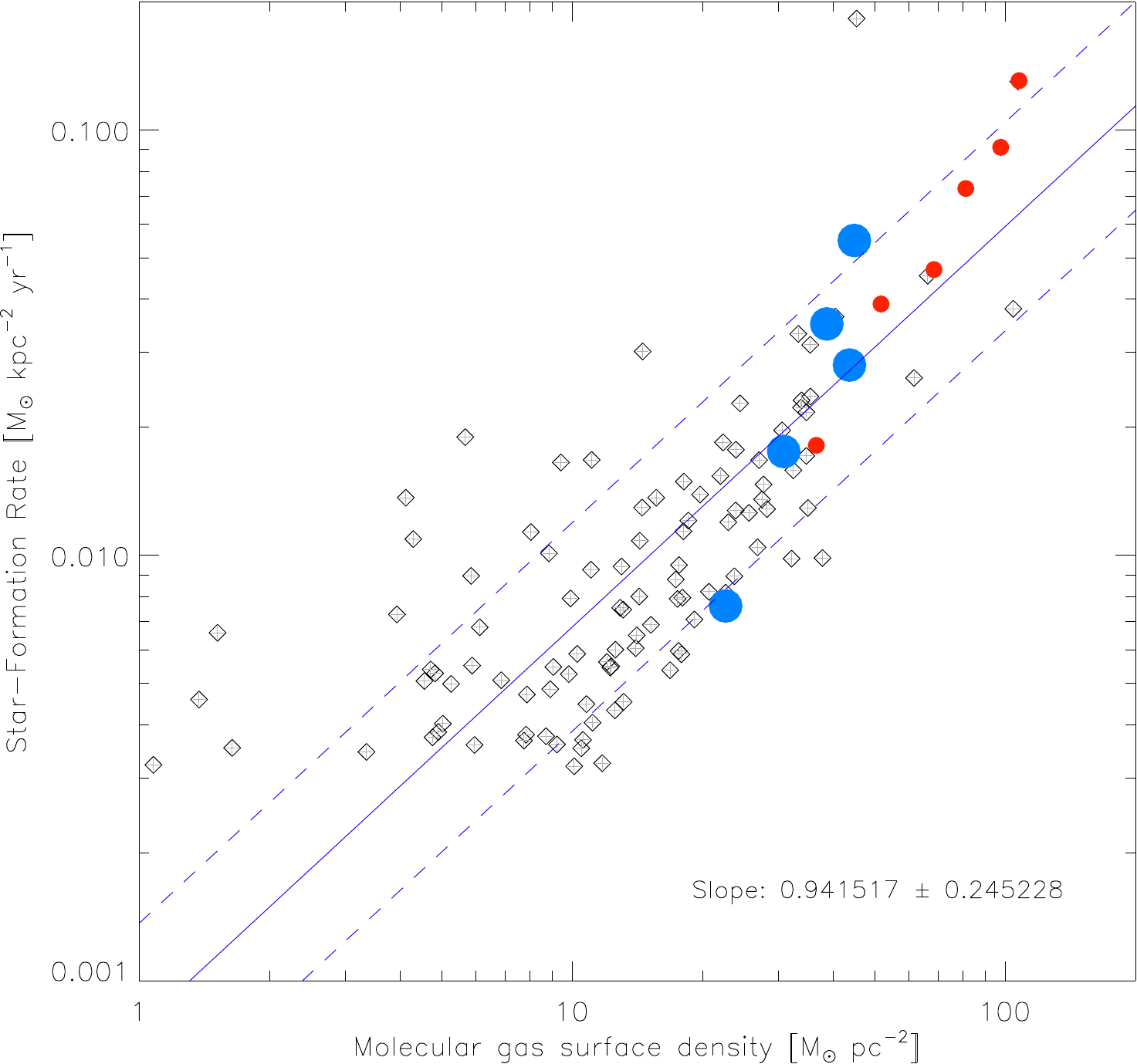}
   \caption{Star-formation Rate $\rm{\dot{\Sigma}_\star}$ as a function of molecular gas surface density $\rm{\Sigma _{\rm{H_2}}}$ of NGC~4654 from the dynamical model. Blue points correspond to the high $\rm{H\textsc{I}}$ surface density region. Red points correspond to the galaxy center. The dashed lines correspond to $\pm$ 1 $\sigma$.}
   \label{fig:1dsimu2}
\end{figure}

The slope of the model \sgm-\sfr\ relation is 0.94~$\pm$~0.25, which is consistent within the error bars with the results of \citet{2008AJ....136.2846B}
and the observed relation (\fig{fig:1dsimu2}). The SFE$_{\rm{H_2}}$ varies little within the model disk, with a scatter of 0.25 dex. In contrast to our observations, the \sfe\ of the \hsd\ does not deviate from the general correlation, with all the corresponding points included in the $\pm$~1~$\sigma$ error bars.

The study of the \rmol-$P_{\rm{tot}}$ and the \sgm-\sfr\ correlations revealed that the dynamical model reproduces faithfully the decrease in \rmolp\ in the \hsd. However, the model is not able to reproduce the observed enhancement of the \sfe\ in this region. 

\subsubsection{Velocity field, linewidths, and velocity dispersion}

\begin{figure}[!tbp]
  \centering
  \includegraphics[width=\hsize]{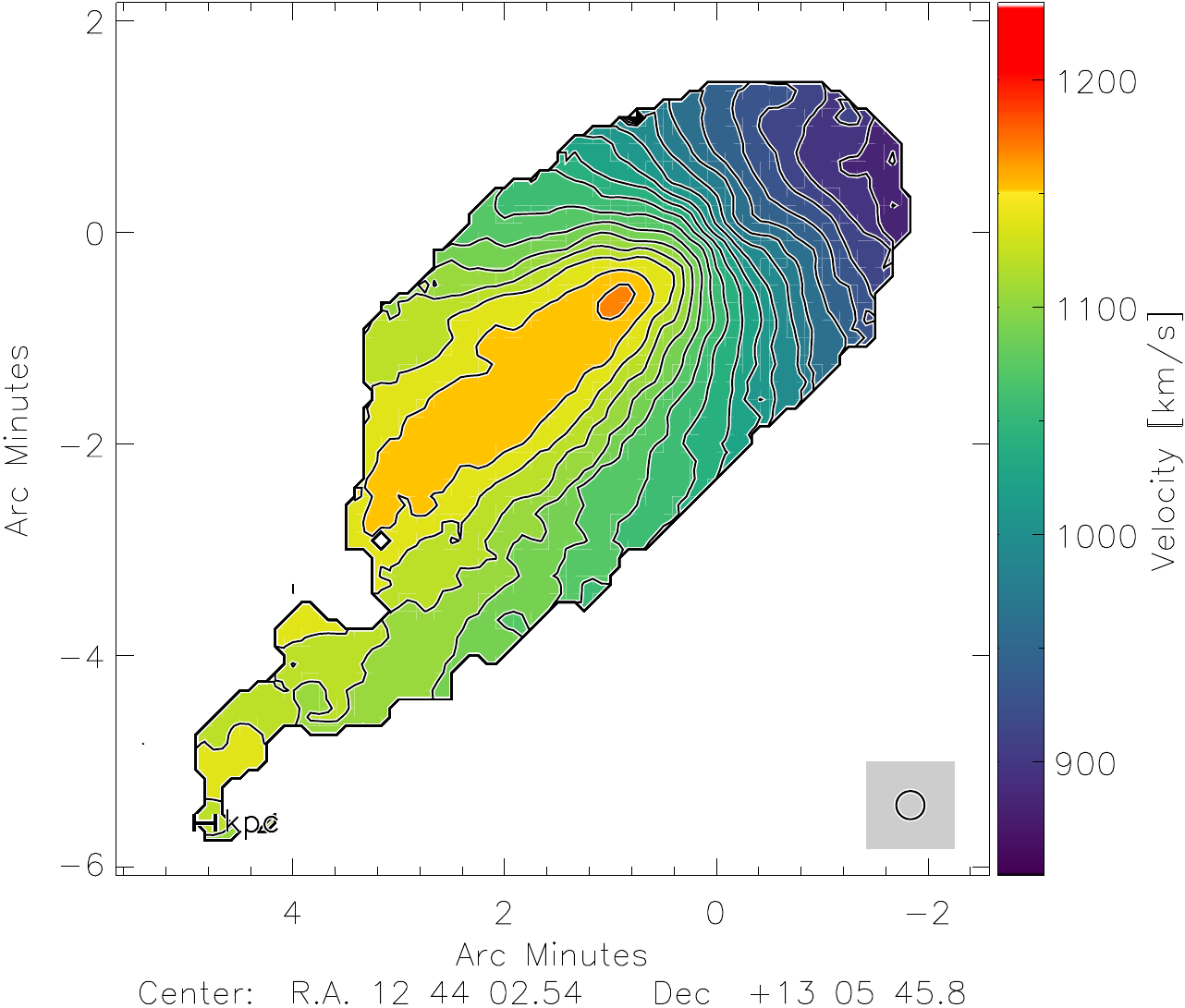}\label{fig:ff1}
  \hfill
  \includegraphics[width=\hsize]{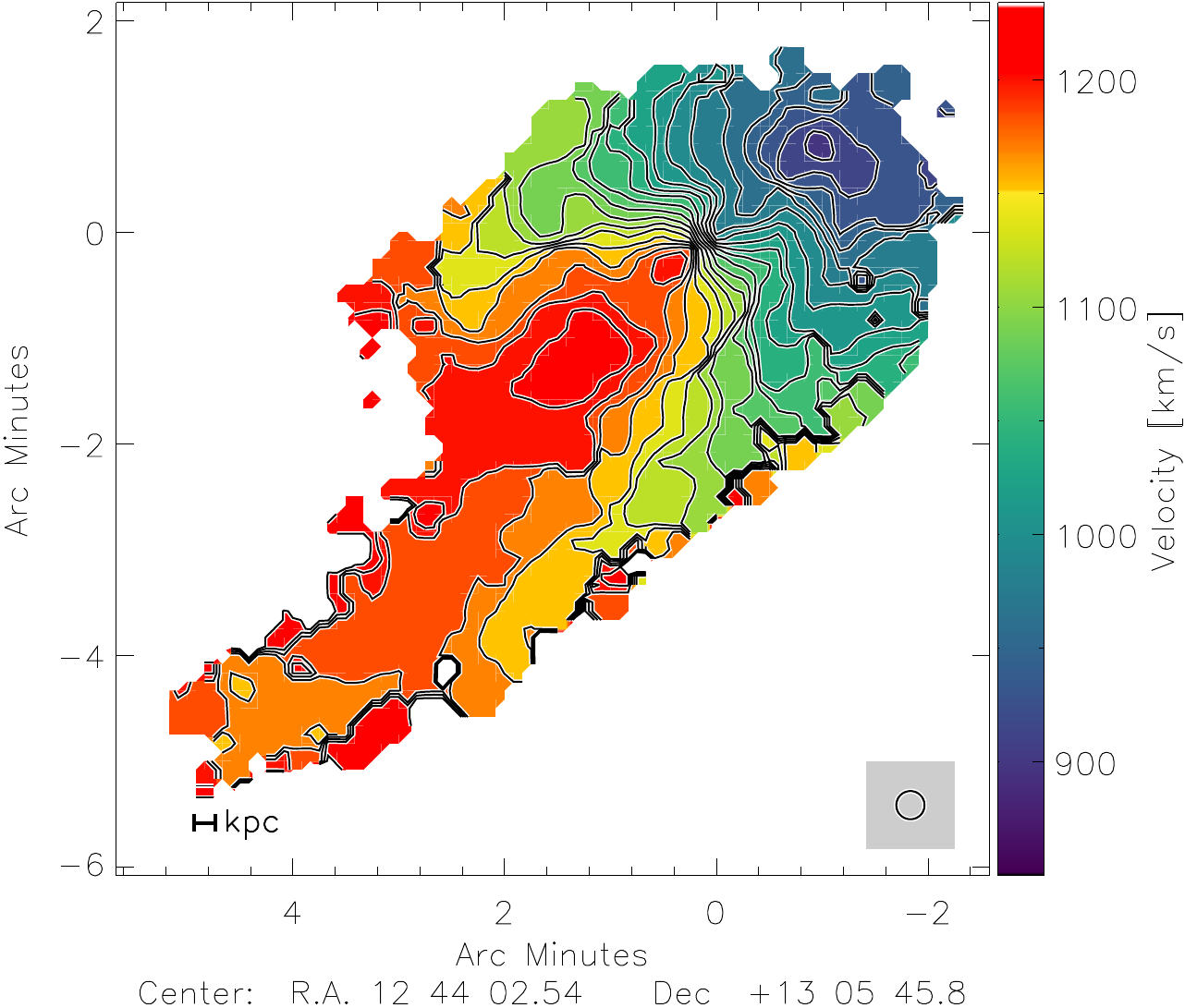}\label{fig:ff2}
  \caption{\hi\ velocity field. \textit{Top panel:} From VIVA data. \textit{Bottom panel:} From the dynamical model.}
  \label{fig:simuv}
\end{figure}

The velocity field of the
dynamical model (\fig{fig:simuv}) presents straight iso-contours
along the major axis in the northwest direction. These results reveal a well-defined velocity gradient toward the northwest along the major axis, that is also observed. A velocity plateau is reached in the southeast side of the disk along the major axis. The iso-contours along the minor axis are curved to the southeast, which is also consistent with the observations. The velocity field in the \hi\ tail increases slightly therein, while in the observations, the region is decoupled from the galaxy rotation. All these results are consistent with previous studies of \citet{2003A&A...398..525V}.
 
\begin{figure}[!tbp]
  \centering
  \includegraphics[width=0.9\hsize]{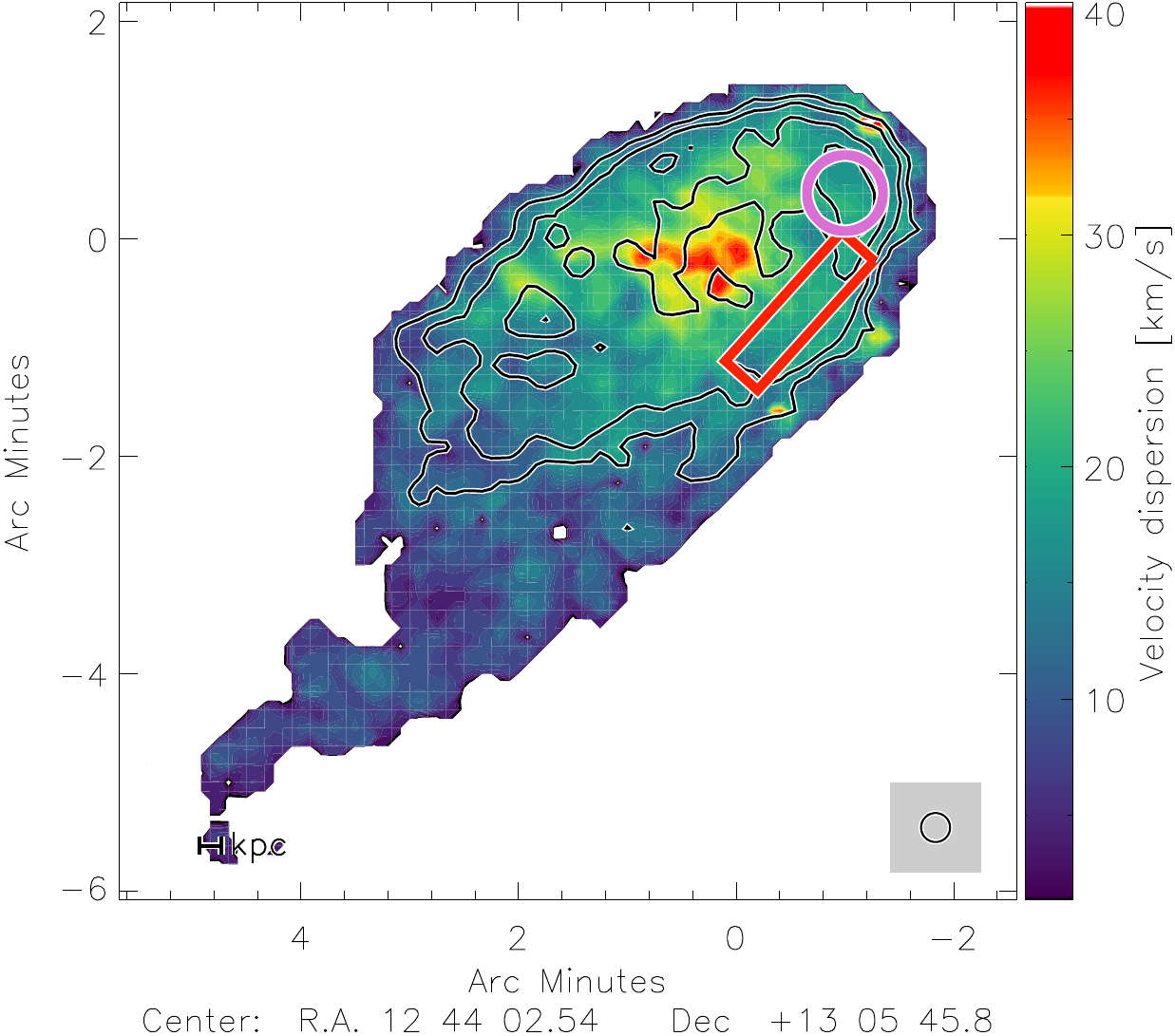}
  \hfill
  \includegraphics[width=0.9\hsize]{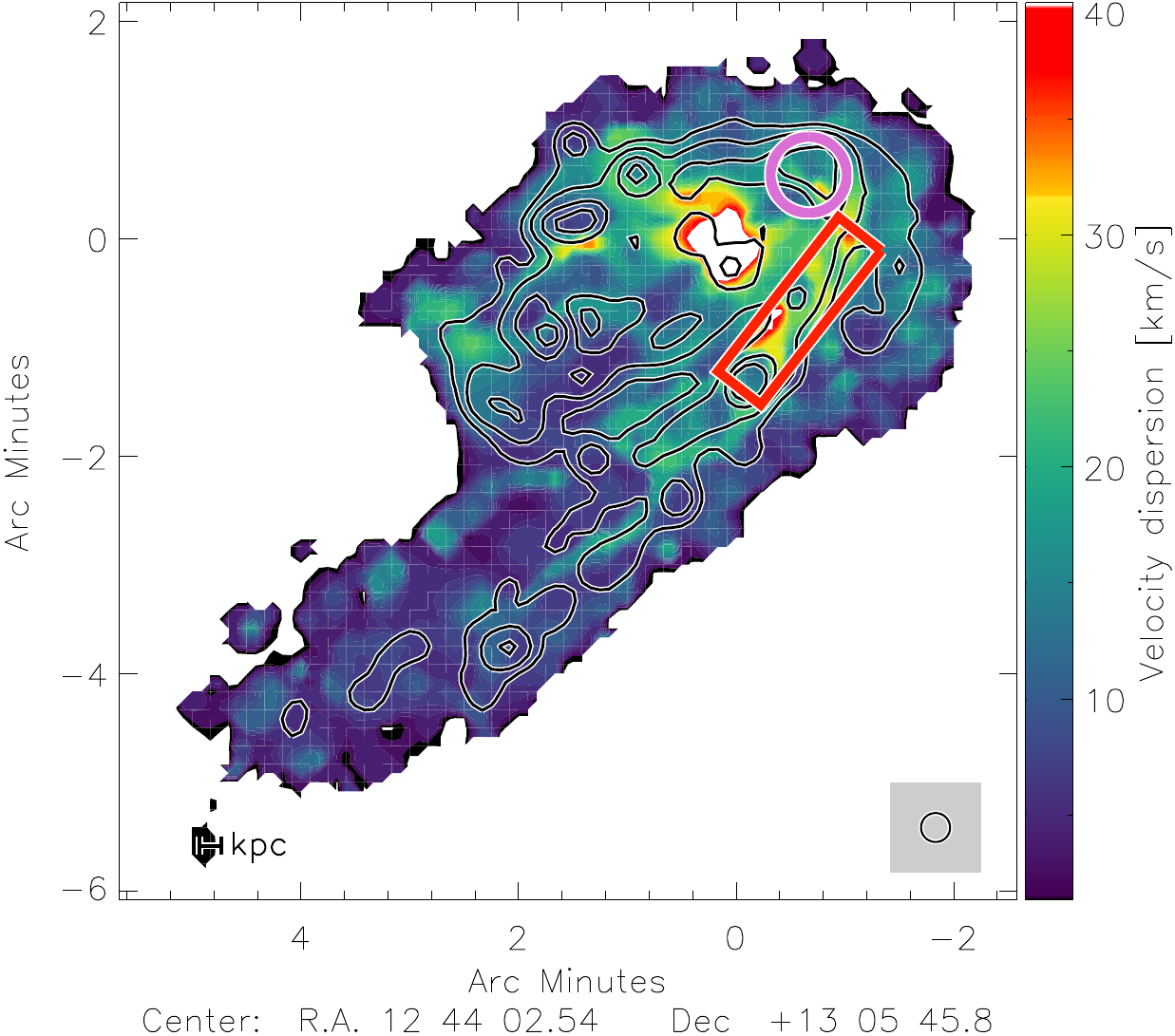}
  \hfill
  \includegraphics[width=0.9\hsize]{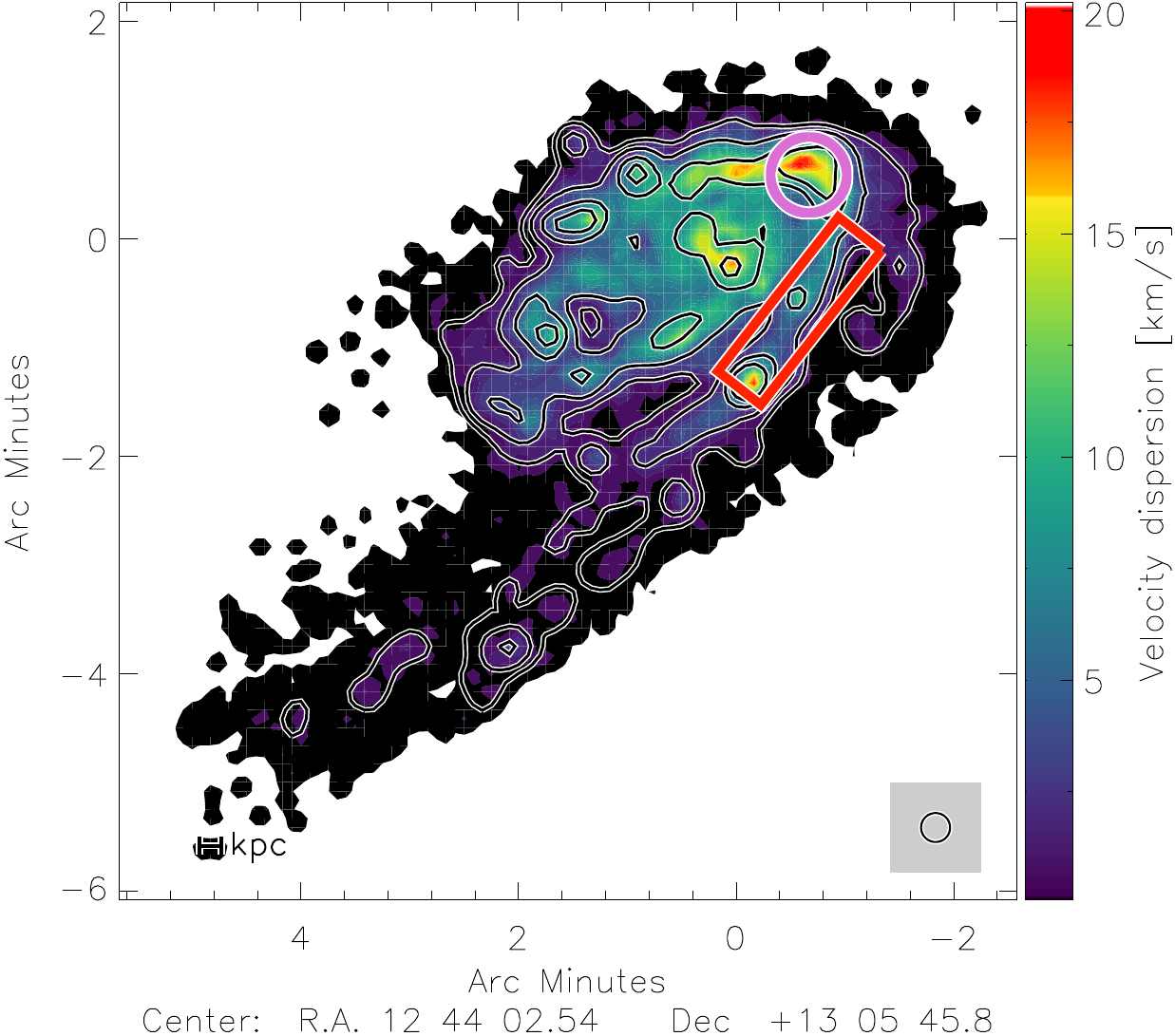}
  \caption{Gas velocity dispersion. \textit{Top panel:} observed VIVA \hi\ velocity dispersion (moment 2). \textit{Middle panel:} velocity dispersion (moment 2) of the dynamical model. \textit{Bottom panel:} intrinsic 3D velocity dispersion. The pink circle corresponds to the \hsd. The red rectangle corresponds to the external gas arm. Contours correspond to \hi\ surface densities of \sga~=~5, 10, 20 and 50 \msunpc.}
  \label{fig:simuv2}
\end{figure}

As mentioned in the introduction of this section, the 3D dynamical model allows us to investigate the velocity dispersion of the ISM. An increase in the \hi\ linewidth within the disk can be explained by: (i) the presence of a velocity gradient within a resolution element or (ii) an increase in the intrinsic velocity dispersion of the gas caused by local phenomena and physical conditions. In order to separate the broadening of the linewidth generated by these two scenarios, we computed two different maps: (i) the \hi\ moment 2 map of the dynamical model; (ii) the 3D velocity dispersion map based on the velocity dispersion obtained for each particle using its 50 closest neighbors. The two resulting maps are presented in \fig{fig:simuv2}.

The moment 2 map of the model reveals that the broadest velocity dispersion is created in the galaxy center with $\Delta$v~$\sim$~50~km~s$^{-1}$. In the \hsd, the velocity dispersion is only 5~km~s$^{-1}$ above the values at the same radius on the opposite side of the disk, with a mean $\Delta$v~$\sim$~20~km~s$^{-1}$. Within the external gas arm, the velocity dispersion is broader than in the rest of the disk, with $\Delta$v~=~20-40~km~s$^{-1}$. This result is consistent with the assumption that the ram pressure strength in the model is too strong because the band disappears in models without any ram pressure stripping. The velocity dispersion map shows an enhancement of the dispersion within the \hsd, reaching a maximum of 18 $\rm km\ s^{-1}$. Within the external gas arm, no increase is measured. The mean velocity dispersion is $\overline{v}_{\rm disp} = 5~\rm km~s^{-1}$ therein. We therefore suggest that the velocity dispersion within the northwest region is unrelated to the velocity gradient, while the rise in the external gas arm seems to be its direct outcome.  

\subsubsection{The model Toomre stability criterion}

The Toomre stability criterion of the dynamical model is computed using \eq{eq:q}. To compute $\kappa$, we used direct measurements of velocities and distances to the center for every mass point within the model. The velocity dispersion $v_{\rm disp}$ corresponds to the real dispersion presented in \fig{fig:simuv2}. A local minimum is reached in the \hsd, with $Q < 0.5$ while in the rest of the disk the Toomre criterion is around $Q$~=~1. This result is consistent with observations, suggesting that even in the dynamical model, the region is unstable with respect to fragmentation. 

\begin{figure}[ht!]
   \centering
   \includegraphics[width=\hsize]{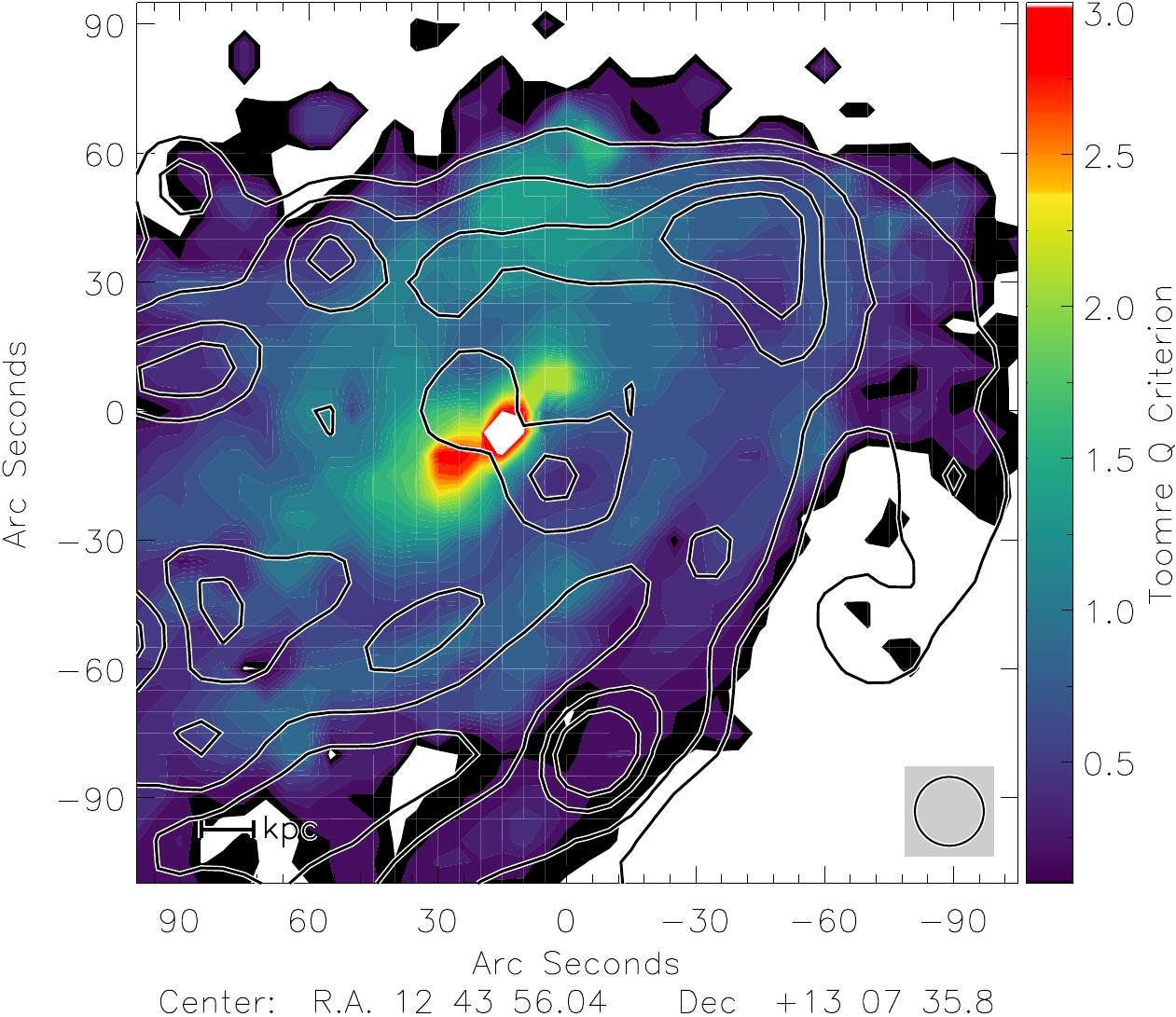}
   \caption{Toomre $Q$ parameter of NGC~4654 from the dynamical model. Contours levels are 0.5, 1 and 1.5.} 
   \label{fig:qsimu}
\end{figure}

\subsubsection{Summary of the dynamical model} \label{sect:11.7}

Table~\hyperref[tab:model]{4} gathers all results from the study of the dynamical model. The dynamical model appears to be closer to the observations with a constant conversion factor \acomw. The model reproduces \hi\ observations of NGC~4654: the \hi\ tail and the \hsd\ being well reproduced with quite comparable surface densities. Although the stellar arm is well reproduced by the model, the diffuse stellar disk is much more extended toward the northwest than the observations. The spatial distribution of the molecular gas matches the observations but is underestimated by a factor of 2 along the dense stellar arm. The observed enhancement of the SFR in the \hsd\ is not reproduced by the model, suggesting that the inclusion of local physical phenomenon such as supernova feedback is mandatory to reproduce such high star-formation. The deviation of the \hsd\ from the galaxy correlation \rmol-$P_{\rm{tot}}$ is also present in the model. However, the model \sfe\ does not decrease in the \hsd\ as it is observed. The Toomre stability parameter of the dynamical model decreases below one within the \hsd, suggesting that the region is unstable with respect to gas fragmentation. Excluding the formation of an external gas arm caused by an overestimation of the ram pressure stripping in the model, the reproduction of the velocity field of the dynamical model is also robust with the observations. A velocity gradient in the northwestern direction is measured with comparable strength. Separate studies of the observed velocity dispersion and the intrinsic velocity dispersion within the model suggests that the broader linewidth measured in the \hsd\ are produced by a real increase in the intrinsic velocity dispersion and not by a sole consequence of the velocity gradient caused by the ram pressure stripping. 

\begin{table*}[ht!]
   \begin{center}
   \label{tab:model}
   \renewcommand{\arraystretch}{1.5}
   \addtolength{\tabcolsep}{-1pt}
   \caption{{Comparison between the observations and the dynamical model.}}
      \begin{tabular}{c c c c} 

        \hline 

         & \begin{tabular}{c} Observations \\ \begin{tabular}{ccc} ( \acomw\ & / & 2\acomw ) \end{tabular}\end{tabular}  & Model  \\
         \hline
         \hline
        
        Mean \sga\ in the gas tail & 1.86 \msunpc & 2.96 \msunpc \\
        
        Max \sga\ in the \hsd\ & 40.6 \msunpc & 43.6 \msunpc \\
        
        Mean \sgm\ in the northwestern dense stellar arm & 30-40 \msunpc\ & 30-40 \msunpc \\
        
        Max \sgm\ in the galaxy center & 173 \msunpc\ & 170 \msunpc \\
        
        Max \sgm\ in the \hsd\ & \begin{tabular}{ccc}
  49.6 \msunpc\ & / & 99.3 \msunpc\ 
        \end{tabular} & 49.5 \msunpc \\
        
        Max \sfr\ in the \hsd\ & 0.27 \msun~kpc$^{-2}$
yr$^{-1}$ & 0.07 \msun~kpc$^{-2}$
yr$^{-1}$ \\
        Slope of \sgm-\sfr correlation & 1.02 ($\pm$0.18) & 0.94 ($\pm$0.25) \\
        \hsd\ deviation from the correlation & 0.1-0.2 dex & - \\
        
        Slope of \rmol-$\rm{P_{\rm{tot}}}$ correlation & 1.00 ($\pm$0.18) & 0.65 ($\pm$0.20)\\
         \hsd\ deviation from the correlation & 0.2-0.3 dex & 0.2 dex \\

         \hline
      \end{tabular}
   \end{center}
\end{table*}

\section{Discussion} \label{sect:dis}

In this section we investigate which kind of physical environment is required to maintain the \hsd\ in its current state. We first compare the region of NGC~4654 with a similar region of enhanced H{\sc i} surface density in another Virgo galaxy, NGC~4501. Then, we carry out a comparative study between the analytical and dynamical models to highlight what these regions have in common and what distinguishes them. Finally, we conclude the discussion on the suggested youth of the northwestern region.  

\subsection{Comparison with NGC~4501 \label{sect:12.1}}

\begin{figure}[ht!]
   \centering
   \includegraphics[width=\hsize]{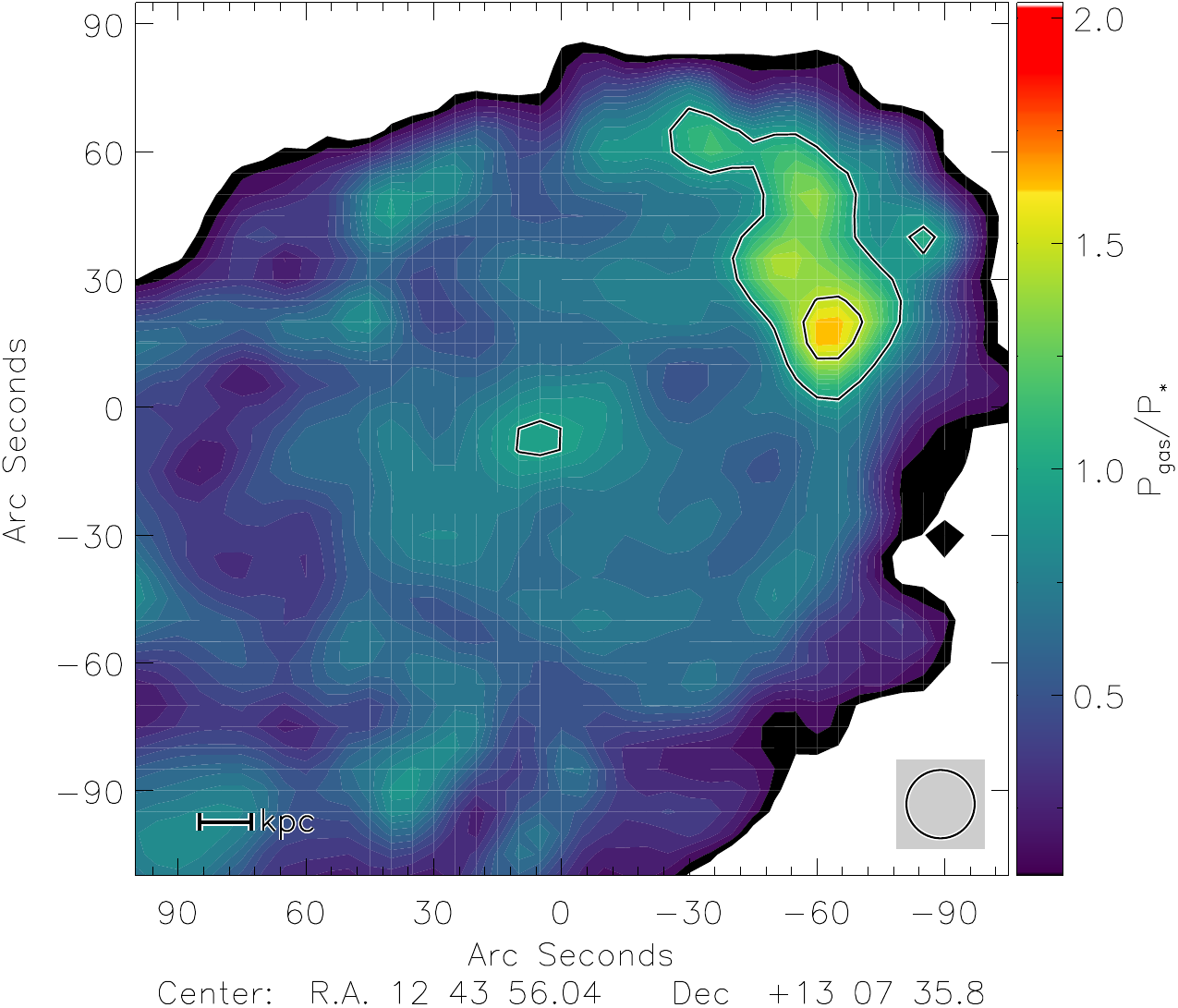}
   \caption{NGC~4654 observed mid-plane pressure counterparts ratio $P_{\rm gas}$/$P_\star$. Contour levels correspond to 1 and 1.5.}
   \label{fig:pp}
\end{figure}

NGC~4501 is another Virgo galaxy studied in detail in \citet{2016A&A...587A.108N} that is undergoing active ram pressure stripping. The interaction is nearly edge-on, leading to a well-defined compression front on the western side of the disk. NGC~4501 also presents a region where the \hi\ surface density is particularly high, located in this compressed front. \citet{2016A&A...587A.108N} studied the variation of the \rmolp\ and \sfe\ correlations within the galaxy. They showed that the \hsd\ of NGC~4501 presents: (i) an excess in the \hi\ surface density up to 27~\msunpc; (ii) a slight increase in the \sfe\ of 0.1~dex; (iii) a significant decrease in \rmolp\ of 0.3-0.4~dex; (iv) a drop of the Toomre $Q$ parameter close to the value of 1; (v) an increase in the $P_{\rm gas} / P_\star$ ratio up to 0.7. In the following, the properties of the \hsd\ of NGC~4654 and NGC~4501 are compared in detail.
\paragraph{}
The \hsd\ of NGC~4654 also exceeds the usual $10$-$15$~\msunpc\ observed in spiral galaxies (\citealt{2008AJ....136.2782L}), with an even higher maximum \sga~=~40~\msunpc. The slight increase in the star-formation efficiency of NGC~4501 is comparable with the one obtained with a modified conversion factor for the \hsd\ of NGC~4654, namely 0.1~dex above the \sgm-\sfr\ correlation. Whereas the metallicity profile of NGC~4654 observed by \citet{1996ApJ...462..147S} and corrected following \citet{2020MNRAS.491..944C} suggests an increase in the conversion factor within the \hsd, this is not the case for NGC~4501. The overall metallicity profile of NGC~4501 is significantly higher and flatter than the profile of NGC~4654. At the outer radii the metallicity of NGC~4501 also measured by \citet{1996ApJ...462..147S} is about twice as high as that of NGC~4654. The metallicity within the \hsd\ of NGC~4501 remains therefore higher than solar. This suggests that the conversion factor within the \hsd\ of NGC~4501 is close to the Galactic value while for NGC~4654 a two times higher conversion factor seems appropriate. With such an increased conversion factor, the star-formation efficiencies with respect to the molecular gas of NGC~4654 and NGC~4501 are well comparable. This result supports the assumption of an increased \aco\ in the \hsd . As for NGC~4501, a significant drop of the $R_{\rm mol}/P_{\rm tot}$ of 0.2-0.3 dex is observed within the \hsd\ of NGC~4654. The Toomre $Q$ parameter in the northwestern region of NGC~4654 is half than that estimated by \citet{2016A&A...587A.108N} in the \hsd\ of NGC~4501, with a minimum of $Q \sim 0.5$. The total gas surface density of NGC~4501 in the \hsd\ is approximately $\Sigma_{\rm gas}=40-50$~\msunpc, while for NGC~4654, with a modified conversion factor, the total surface density reaches a local maximum around $\Sigma_{\rm gas}$~=~90~\msunpc. This difference may explain partially the lower Toomre $Q$ parameter found in NGC~4654. We produced a map of the observed ratio between the gas pressure and the pressure term due to the stellar gravitational potential $P_{\rm gas}/P_\star$ (\fig{fig:pp}). The map shows that the gas in the \hsd\ is self-gravitating with  $P_{\rm gas}/P_\star = 1.7$ at its maximum. This value is 2.5 times higher than that of the \hsd\ of NGC~4501. While in NGC~4501 the ISM
in the \hsd\ is approaching self-gravitation, the total mid-plane pressure of NGC~4654 in the corresponding region is dominated by $P_{\rm gas}$. The difference between the two \hsd\ of NGC~4501 and NGC~4654 might be also explained by the higher total gas surface density of NGC~4654 compared to that of NGC~4501. 

Since the two regions are comparable in all stated points, the \hsd\ of NGC~4654 being an enhanced version of that of NGC~4501, it can be assumed that the underlying physical processes are the same: a high gas surface density together
with a somewhat increased velocity dispersion and a low Toomre $Q$ parameter. Whereas we know that NGC~4654 undergoes both gravitational interaction and ram pressure stripping, NGC~4501 experiences only ram pressure stripping (\citealt{2008A&A...483...89V}). Therefore, we conclude that the combined effect of ram pressure stripping and gravitational interaction
gave rise to a higher gas surface density in the compressed region of NGC~4654. On the other hand, the reaction of the ISM
including star-formation is the same in the two galaxies. We speculate that the \hsd s\ in the gravitationally interacting
galaxy NGC~2207 (Elmegreen et al. 2016) are in the same physical state. 

\subsection{The radio spectral index}

\citet{2010A&A...512A..36V} studied the radio spectral index~$\upvarepsilon$ between $6$ and $20$~cm of 8 Virgo galaxies affected by ram pressure stripping. The spectral index along the northwestern stellar arm is significantly steeper than that of the rest of the disk ($\upvarepsilon \ga -0.8$). The spectral index reaches its maximum in the \hsd , with $\upvarepsilon \sim -0.5$, i.e, it is close to the typical value for a population of cosmic ray electrons at the time of injection. This suggests that the star-forming region is relatively young, only a few 10~Myr old (\citealt{2015A&ARv..24....4B}). 

\subsection{Joining between the analytical and dynamical models}

In both, the analytical or dynamical model, an increase in the velocity dispersion combined with a decrease in the Toomre $Q$ parameter are mandatory to reproduce the observations. Although the velocity dispersion increases consistently between these two models, the resulting SFR is significantly lower in the dynamical model than in the analytical model. 

The main difference between the two models lies in the cloud density $\rho_{\rm cl}$ within the \hsd. The dynamical model does not consider cooling and heating mechanisms of the ISM through self-gravitating collapse and stellar feedback. This means that the model does not allow clumping in regions that are presumed to be dense, leading to an underestimation of the local SFR. In the analytical and dynamical models, the decrease in the Toomre $Q$ parameter induces an increase in the global density. Whereas the volume filling factor is assumed to be constant in the dynamical model, it shows an increase by about a factor of two in the dynamical model leading to an increase in {the SFR (\eq{eq:5}).}
Despite the fact that the large-scale density distributions are consistent between the two models, the small-scale gas density and SFR are poorly reproduced by the dynamical model. In the latter, the star-formation is purely driven by collisions between gas particles, while in the analytical model supernova feedback determines the ISM properties at small-scales which have a significant impact on star-formation. 

\subsection{Reaching atomic gas surface densities in excess of 30~$M_\odot pc^{-2}$ in galaxies with and without stellar feedback} \label{sect:12.4}

Using the analytical model, we investigated how high \hi\ surface density regions with \sga~>~30~\msunpc\ can be created and maintained. The question is therefore to understand how far should the velocity dispersion be increased to reach such surface densities without any decrease in the Toomre $Q$ parameter or, conversely, how much must the Toomre $Q$ parameter be decreased below the critical value of one to reach such surface densities without increasing the velocity dispersion.

We first modeled an unperturbed disk with $\delta =  5$, a mean Toomre $Q$ parameter of $Q = 1.5$ and an accretion rate of $\dot{M}~=~0.1~\rm{M_\odot yr^{-1}}$. The additional free parameters $\xi$, $v_{\rm disp}^\star$ and $\gamma$ were not modified. We then created a \hsd\ with $\Sigma_{\rm HI} \sim 30$~\msunpc\ at a radius of 6~kpc by varying the gas density via the Toomre Q parameter or the velocity dispersion via the mass accretion rate. In both cases the stellar feedback is strongly increased. We found that: (i) To obtain a \hsd\ without an increase in the velocity dispersion, the Toomre $Q$ parameter must be $Q \le 0.3$. The resulting SFR in this region is about 60\% of the SFR in the galaxy center. The molecular gas surface density reaches high values, leading to a molecular fraction of \rmol~$\sim$~2 in this region ; (ii) {To obtain a \hsd\ without a decrease in the Toomre $Q$ parameter, the increase in the velocity dispersion in this region must be $\Delta v_{\rm disp}$~$\geq$~16~$\rm km\ s^{-1}$. The resulting SFR in this region is about 40\% of the SFR in the galaxy center. The molecular fraction is only \rmol~$\sim$~0.5 in this region. }

The analytical model therefore suggests that it is theoretically possible to reach \sga~>~30~\msunpc\ by modifying only the Toomre $Q$ parameter or the velocity dispersion. We found that
the star-formation rate does not allow us to discriminate between
these two extreme hypotheses. However, the resulting molecular
gas surface density of the two assumptions are strongly different. This suggests that CO observations are necessary to determine if stellar feedback can keep the density of the compressed ISM approximately constant or if the density increases during a compression phase, i.e, if the velocity dispersion is abnormally high or if the Toomre $Q$ parameter is particularly low.

For the determination of the molecular fraction a CO-to-H$_2$ conversion factor has to be applied.
In the case of a strong decrease in the Toomre $Q$ parameter a significant increase in the molecular fraction
can still be observed even when a Galactic conversion factor is assumed. 
We conclude that, in the absence of a reliable estimate of the intrinsic velocity dispersion, it is possible to roughly estimate the influence of stellar feedback in a \hsd\ located in the outer galactic disk, i.e, the balance between the decrease in the Toomre $Q$ parameter and the increase in the velocity dispersion by estimating the molecular fraction
assuming a Galactic CO-to-H$_2$ conversion factor.

\section{Conclusions}\label{sect:ccl}

New IRAM 30m HERA CO(2-1) data were combined with VIVA \hi\ data to investigate the distribution of the total gas within the disk of the Virgo spiral galaxy NGC~4654 and its ability to form stars. NGC~4654 undergoes both, a gravitational interaction with another massive galaxy and nearly edge-on ram pressure stripping. The combined effects of these interactions lead to the formation of an overdense stellar and molecular gas arm toward the northwest, ending with an abnormally high \hi\ surface density region. Previous studies (\citealt{2014PASJ...66...11C}) showed that within this region the \sfe\ is unusually high and the ratio of the molecular fraction to the total mid-plane pressure is significantly lower than that of the rest of the disk. With deeper CO observations of higher spatial resolution (12'') and a star-formation map based on GALEX FUV and SPITZER 24 $\rm{\mu m}$ data, we pursued this study to understand the physical properties of the ISM in the \hsd\ and their impact on the ability of the ISM to form stars. 

We applied two methods to determine the value of the CO-to-H$_2$ conversion factor within the disk of NGC~4654. Following \citet{2013ApJ...777....5S}, the first method consists in the simultaneous determination of the DGR and the conversion factor using Herschel 250$\rm{\mu m}$ FIR data as a tracer for total gas surface density. The second method is based on the determination of the DGR from direct metallicity measurements of \citet{1996ApJ...462..147S} corrected according to the calibration of \citet{2020MNRAS.491..944C}. Different results were found by the two models: a constant conversion factor and an increased conversion factor by a factor of two in the \hsd. The comparison of the metallicity and the SFE of the \hsd\ in NGC 4654 and NGC 4501, which has similar physical characteristics, supports the assumption of a modified conversion factor for NGC 4654 (see \sect{sect:12.1}). 

The radial profiles of the atomic gas, molecular gas and star-formation were compared to an analytical model of a star-forming turbulent clumpy disk (see \sect{sect:ana}). We simultaneously varied six free parameters to reproduce the available observations via two successive reduced-$\upchi^2$ minimizations. The radial profiles of the unperturbed southeastern and perturbed northwestern disk halves were fitted separately (see \sect{sect:10.1}).
Degeneracies between the free parameters were revealed, making it impossible to discriminate between the two assumptions on the CO-to-H$_2$ conversion factor (see \sect{sect:10.2}). However, regardless of the choice of the conversion factor used to compute the molecular gas surface density, the \hsd\ presents (i)
an increase in the intrinsic velocity dispersion by $\sim 5$~km\,s$^{-1}$ {(2-10~km\,s$^{-1}$) and (ii) a decrease in the Toomre $Q$ parameter (\fig{fig:qana}). For $\Delta v_{\rm disp}$~$\leq$~6~km\,s$^{-1}$, the Toomre Q parameters drops below unity, suggesting that the region is marginally unstable with respect to gas fragmentation.} The increase in velocity dispersion is compatible with the VIVA \hi\ observations.

The available observations were compared to a dynamical model that takes into account the gravitational interaction and ram pressure stripping (see \sect{sect:dyn}). The model atomic and molecular gas surface densities, velocity field, moment 2 map, slopes of the \rmol-$P_{\rm{tot}}$ and \sgm-\sfr correlations, and deviation of the \hsd\ from the \rmol-$P_{\rm{tot}}$ correlation are consistent with observations. However, the model is not able to reproduce the SFR in the \hsd\ because of the absence of physical processes such as gas cooling and stellar feedback preventing the formation of small-scale clumpy regions with high volume densities.

Using the analytical model, we examined the physical conditions required to maintain a \hsd\ with \sga\ > 30~\msunpc\ in the outer parts of galactic disks (see \sect{sect:12.4}). We found that it is possible to create such a peculiar H{\sc i} region by either strongly decreasing the Toomre $Q$ parameter or by strongly increasing the velocity dispersion. The CO surface brightness is a good criterion to discriminate between these two extreme solutions. The most realistic result, however, remains a combination of the two solutions, as observed in NGC 4654. 

Based on our results we suggest the following scenario for the \hsd :
during a period of gas compression through external interactions the gas surface density is enhanced leading to an increased SFR and stellar feedback. Our observations and subsequent modeling suggest that under the influence of stellar feedback the turbulent velocity dispersion 
significantly increases and hence the increase in the gas density is only moderate (less than a factor of two). 
{Thus, stellar feedback acts as a regulator of star-formation (see \citealt{2010ApJ...721..975O}, \citealt{2011ApJ...731...41O}).}

\bibliographystyle{aa}
\bibliography{aa}

\appendix

\section{CO(2-1) rms map}

\begin{figure}[ht!]
   \centering
   \includegraphics[width=\hsize]{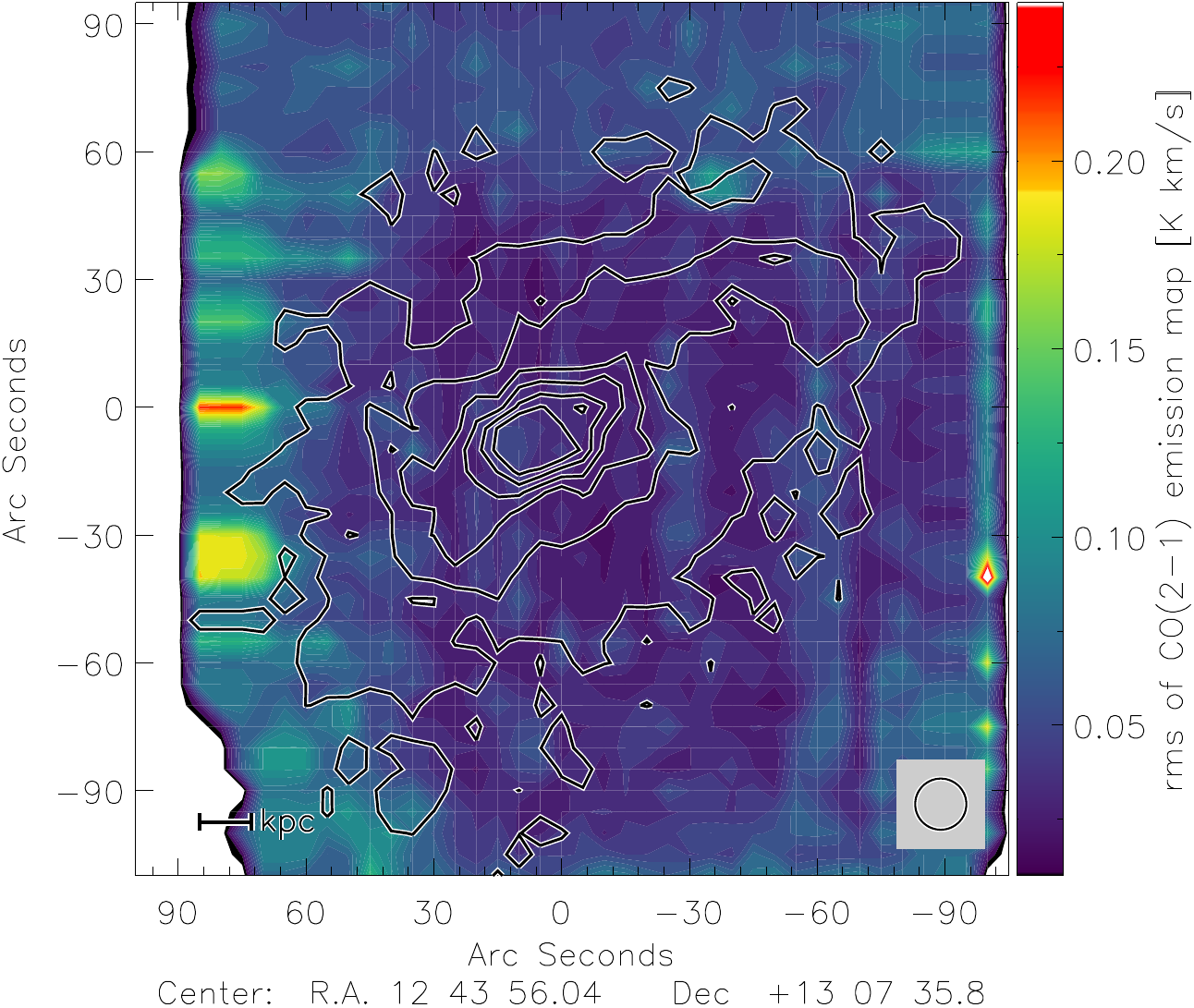}
   \caption{CO(2-1) rms noise with \sgm\ contours.}
   \label{fig:errco}
\end{figure}

\section{DGR minimization method - Other regions}

\begin{figure}[ht!]
   \centering
   \includegraphics[width=\hsize]{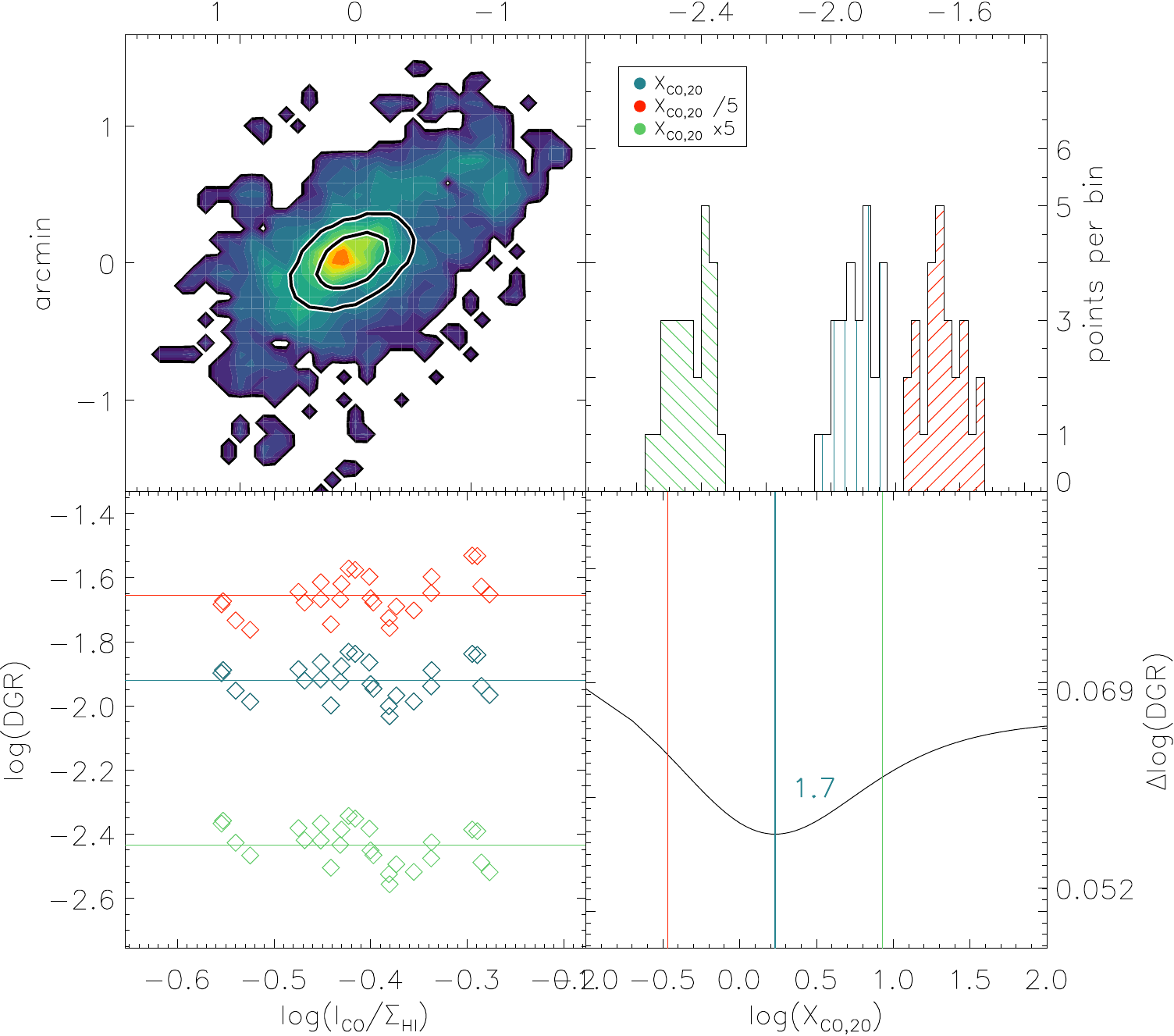}
   \caption{Same description as \fig{fig:sd4}.}
   \label{fig:sd1}
\end{figure}

\begin{figure}[ht!]
   \centering
   \includegraphics[width=\hsize]{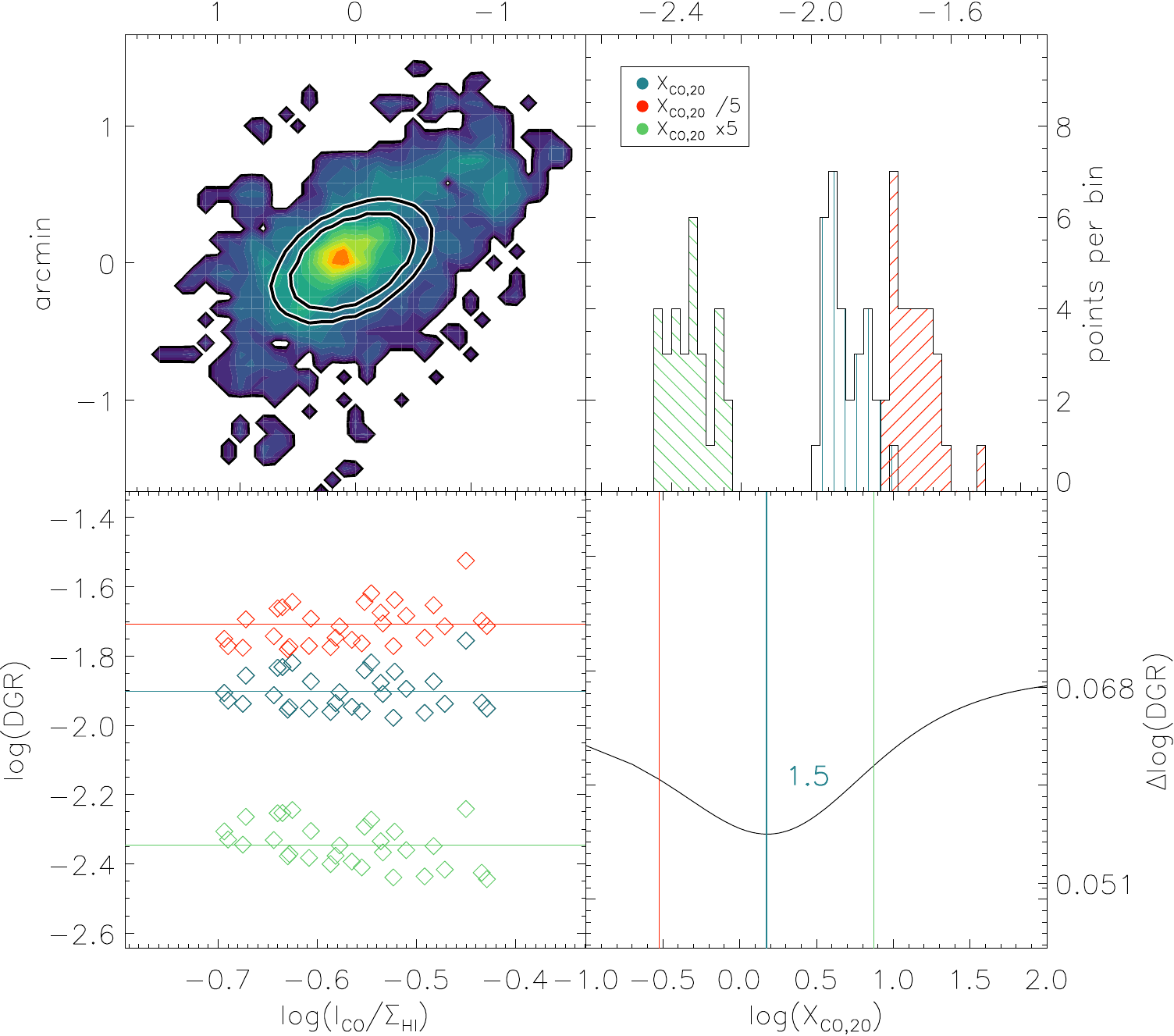}
   \caption{Same description as \fig{fig:sd4}.}
   \label{fig:sd2}
\end{figure}

\begin{figure}[ht!]
   \centering
   \includegraphics[width=\hsize]{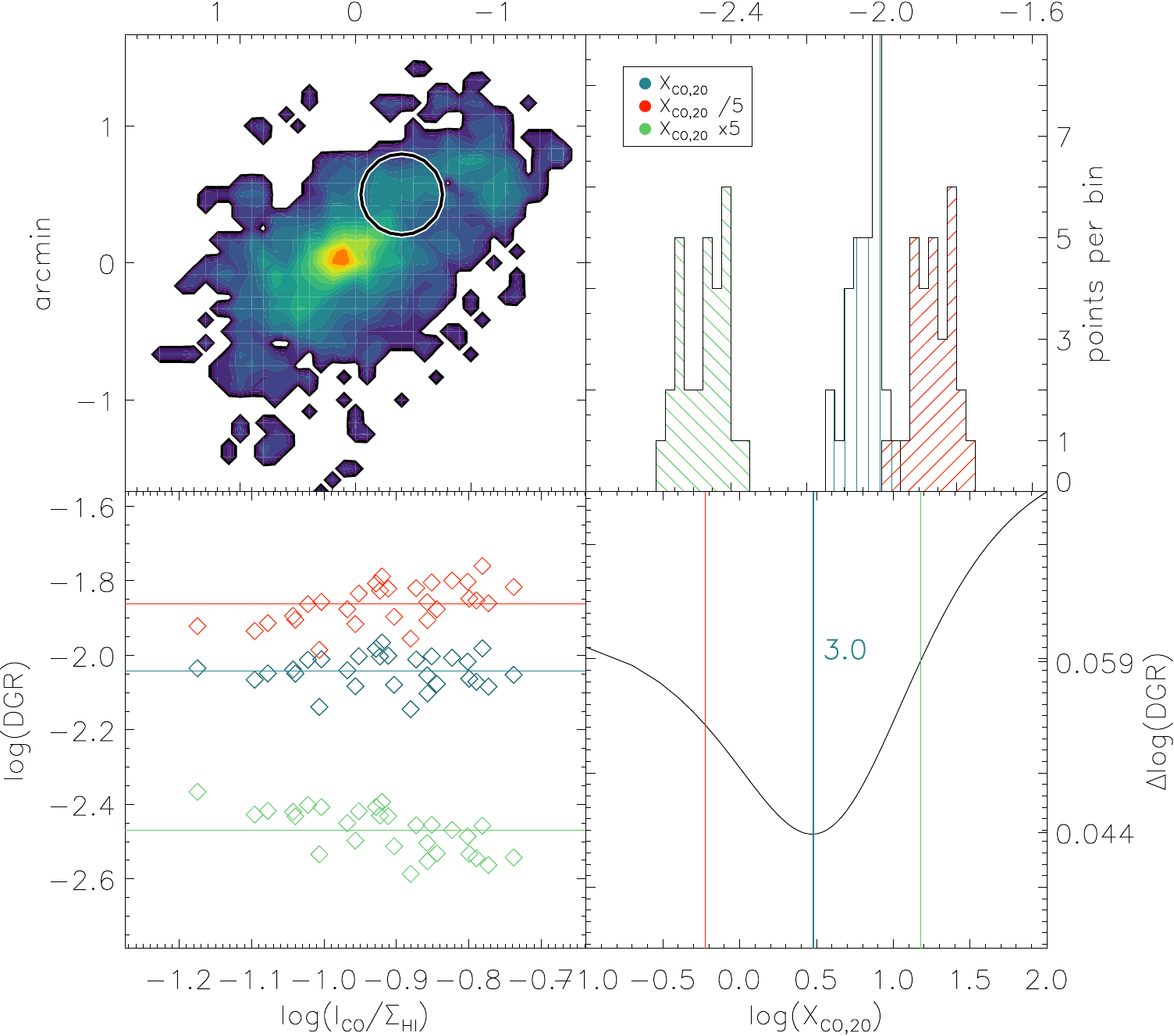}
   \caption{Same description as \fig{fig:sd4}.}
   \label{fig:sd3}
\end{figure}

\newpage

\section{Comparison between different metallicity estimation methods} \label{sect:dgrbis}

\begin{figure}[ht!]
   \centering
   \includegraphics[width=\hsize]{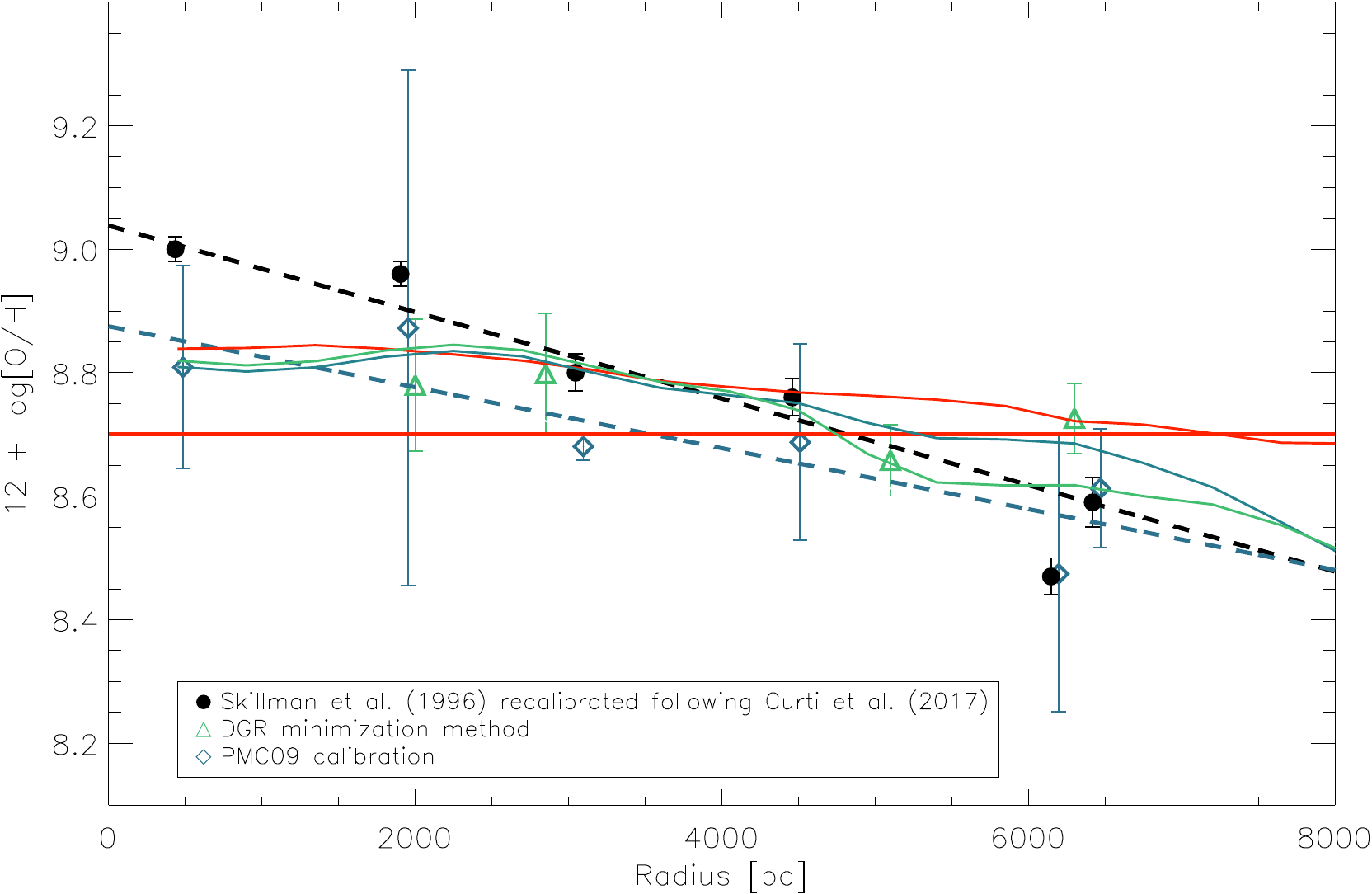}
   \caption{{Same description as \fig{fig:sk1}. The green, red, and blue lines correspond, respectively, to the metallicity obtained from the analytical models (1), (1a), and (1b) using the molecule formation timescale with the closed box model.}}
   \label{fig:sk1bis}
\end{figure}

\newpage

\section{Results with a constant CO-to-H$_2$ conversion factor} \label{cte}

In this section, we reproduce the same maps as in the Results sections. in order
to compare the results using a constant CO-to-H$_2$ conversion factor \aco\ =
4.36 \acou. 

\begin{figure}[ht!]
   \centering
   \includegraphics[width=\hsize]{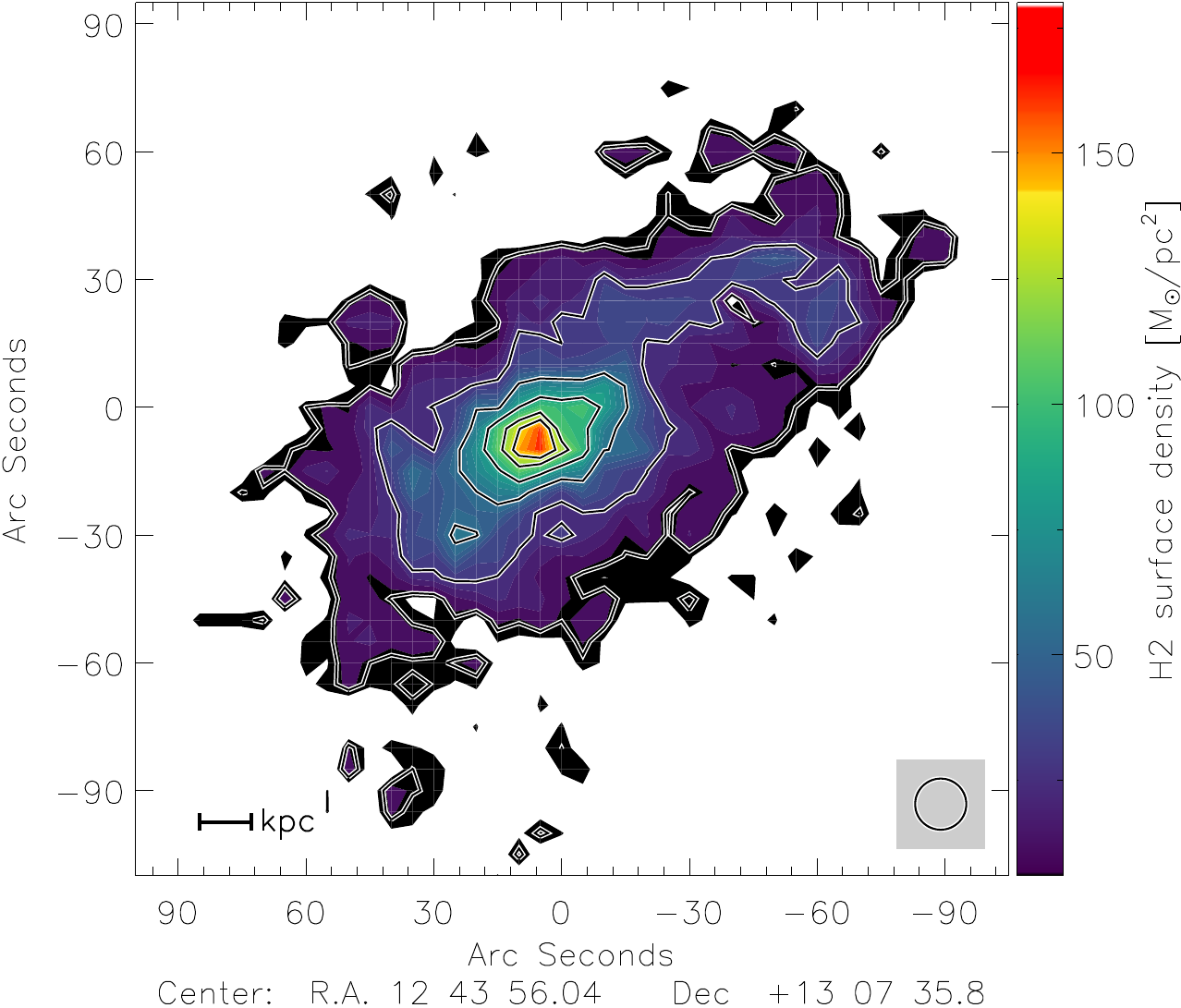}
   \caption{Molecular gas surface density. Contours levels are 10, 30, 60, 90, 120, and 140 $M_\odot pc^{-2}$. The resolution is 12''.}
   \label{fig:h2b}
\end{figure}

\begin{figure}[ht!]
   \centering
   \includegraphics[width=\hsize]{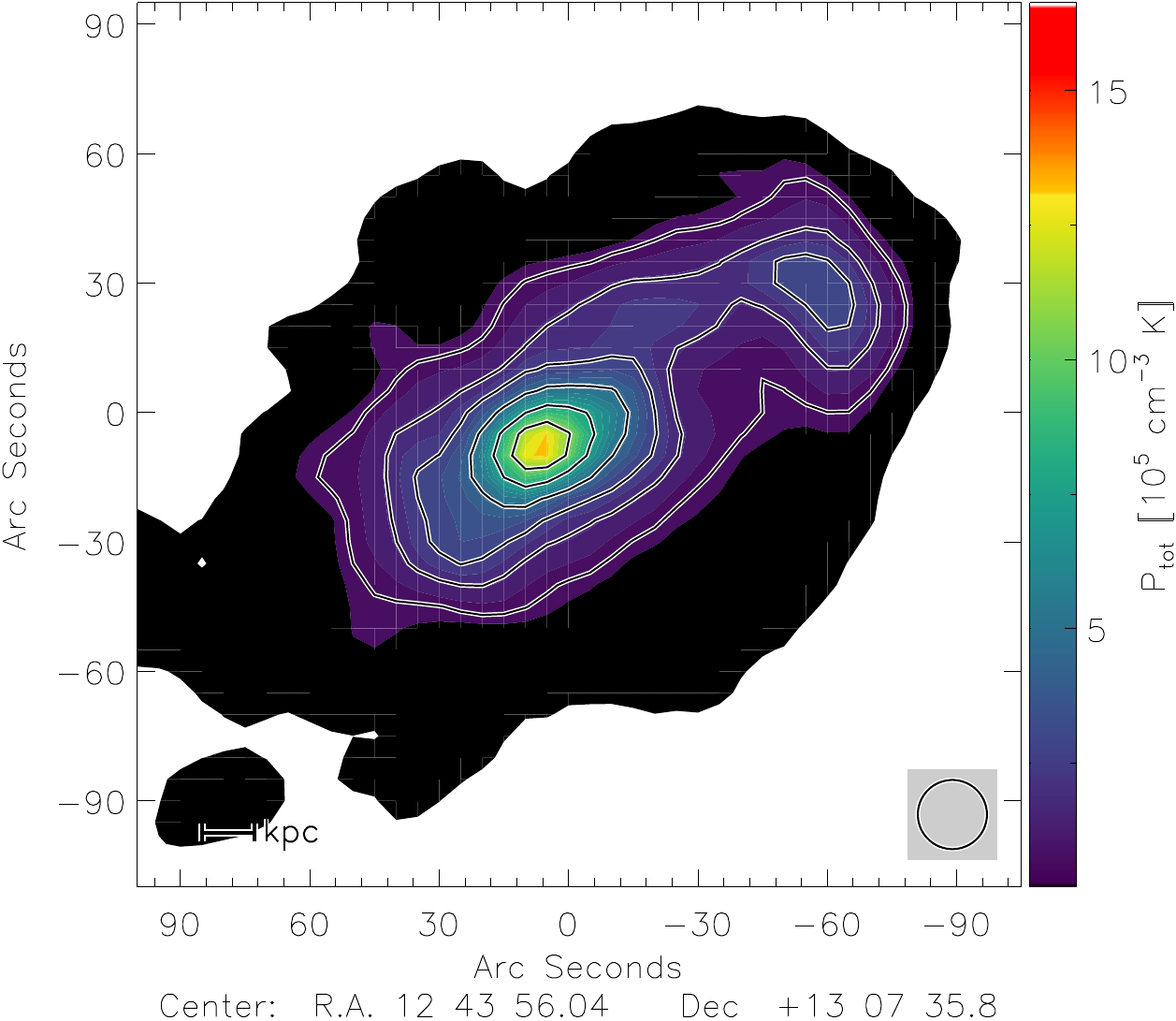}
   \caption{Total ISM mid-plane pressure. Contours levels are 1, 2, 3, 5, 8, and 11 $\times$ 10$^{5}$ cm$^{-3}$ K. The resolution is 16''.}
   \label{fig:pb}
\end{figure}

\begin{figure}[ht!]
   \centering
   \includegraphics[width=\hsize]{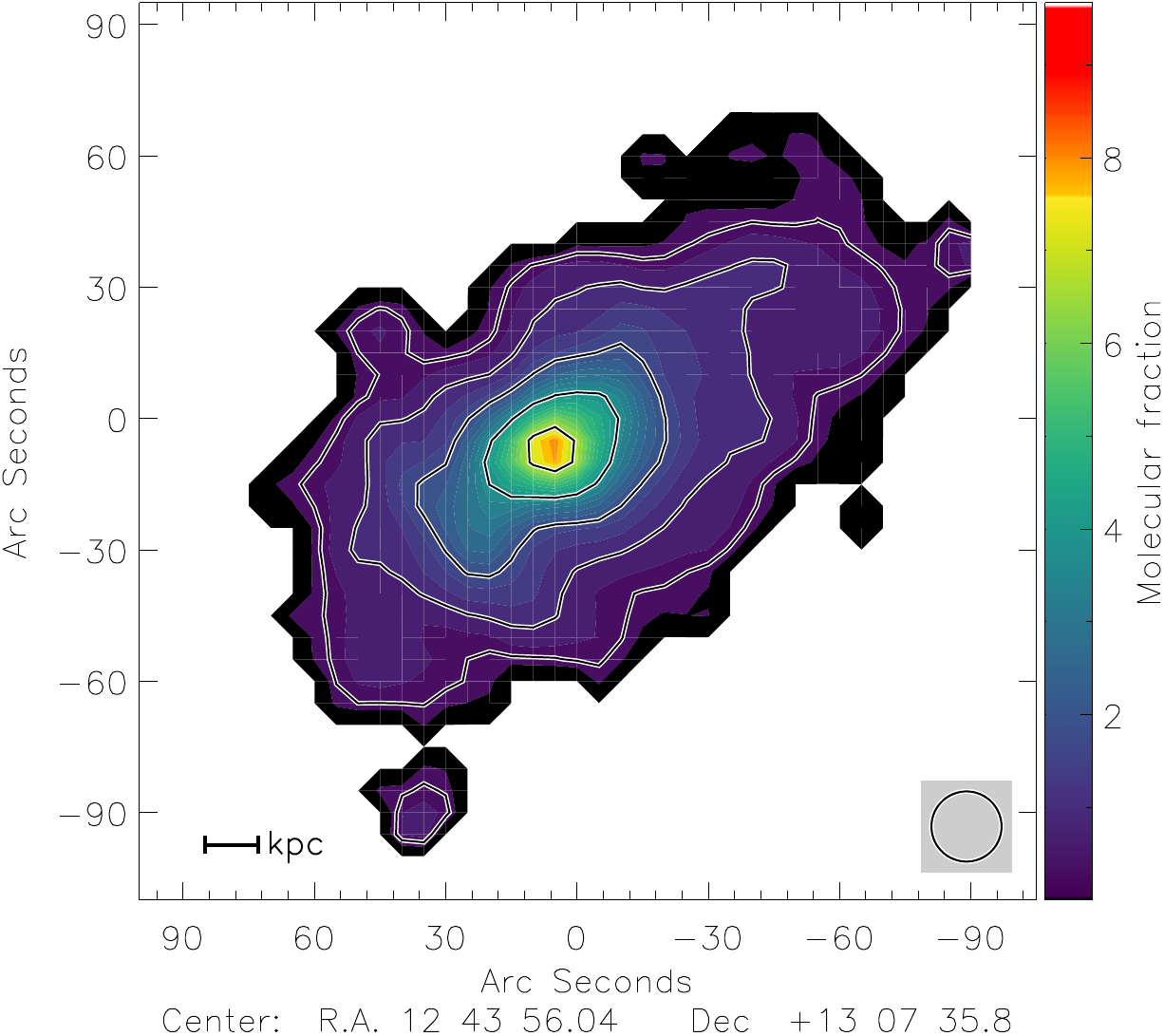}
   \caption{Molecular fraction R$_{\rm mol}$ = $\Sigma _{\rm{H_2}}$ /  $\Sigma _{\rm{HI}}$. Contours levels are 5, 10, 30, 60, 90, 120, and 140 $M_\odot pc^{-2}$. The resolution is 16''.}
   \label{fig:rmolb}
\end{figure}

\newpage

\section{Toomre $Q$ criterion alternative map}

\begin{figure}[ht!]
   \centering
   \includegraphics[width=\hsize]{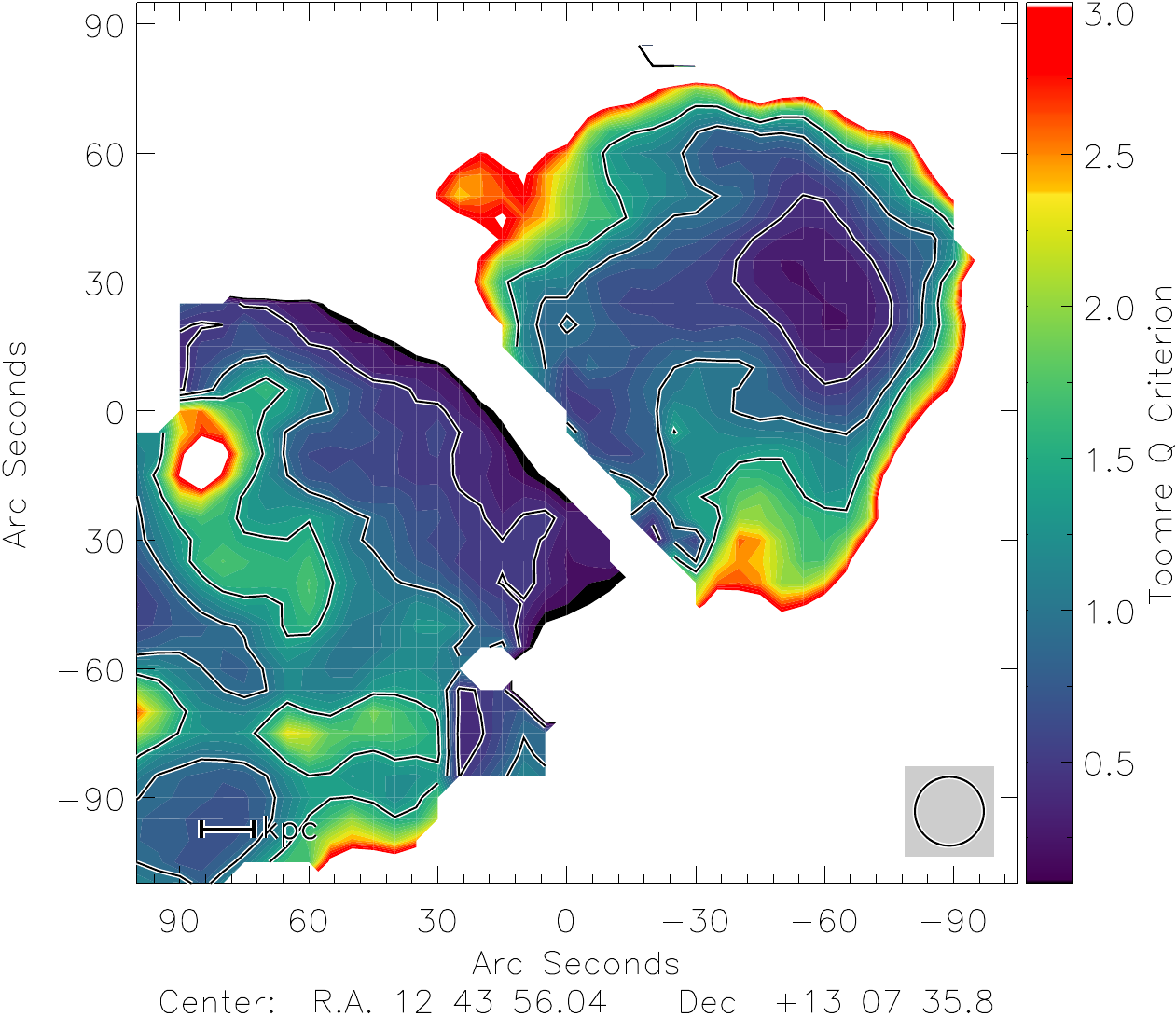}
   \caption{Toomre $Q$ parameter map obtained from the observational rotation curve. Contour levels correspond to 0.5, 1, and 1.5.}
   \label{fig:qobs}
\end{figure}

\newpage

\section{Star-formation rate comparison}

\begin{figure}[ht!]
   \centering
   \includegraphics[width=\hsize]{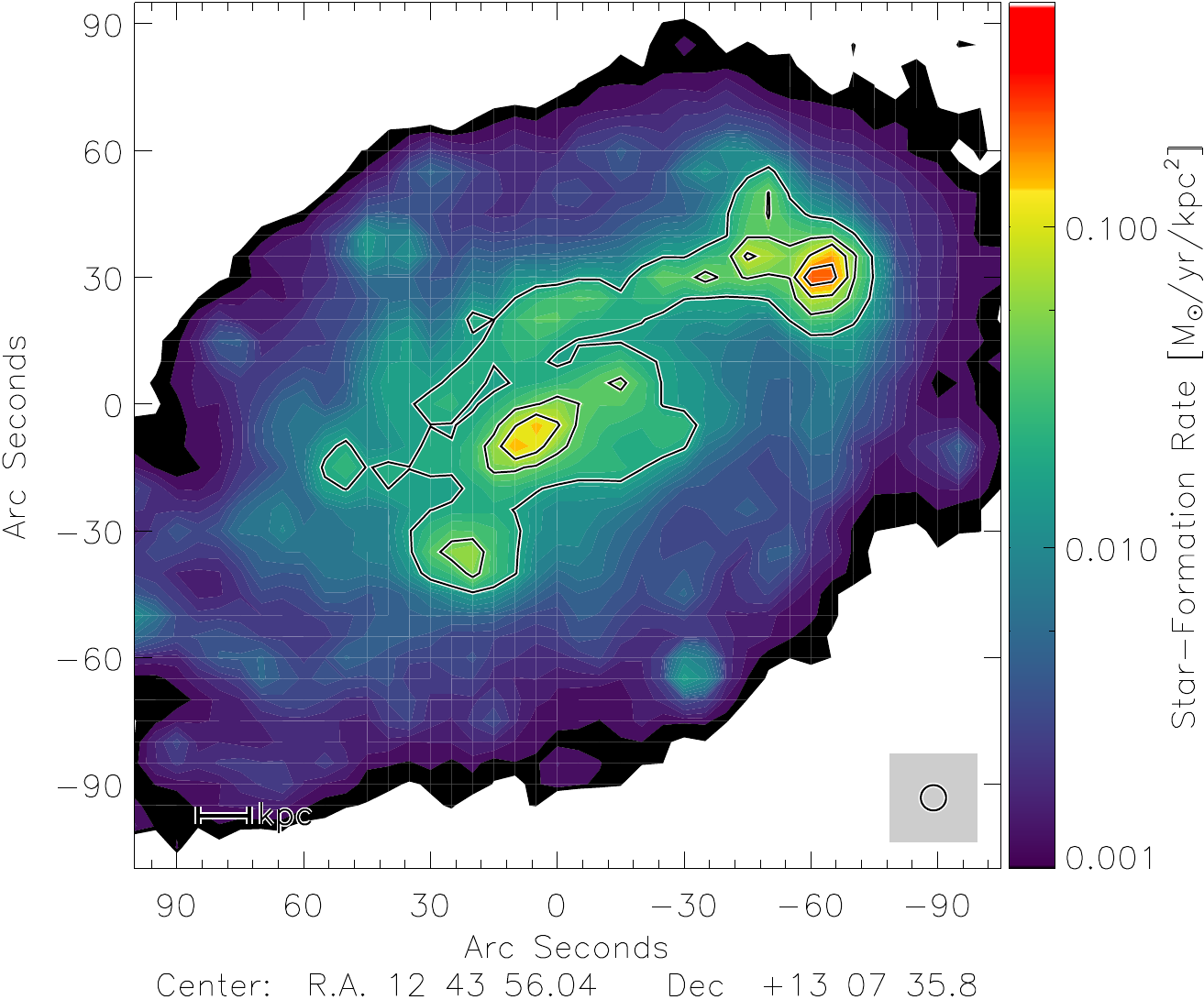}
   \caption{Star-formation map computed using $24$~$\mu$m and H$\alpha$ data.}
   \label{fig:sfrha}
\end{figure}

\begin{figure}[ht!]
   \centering
   \includegraphics[width=\hsize]{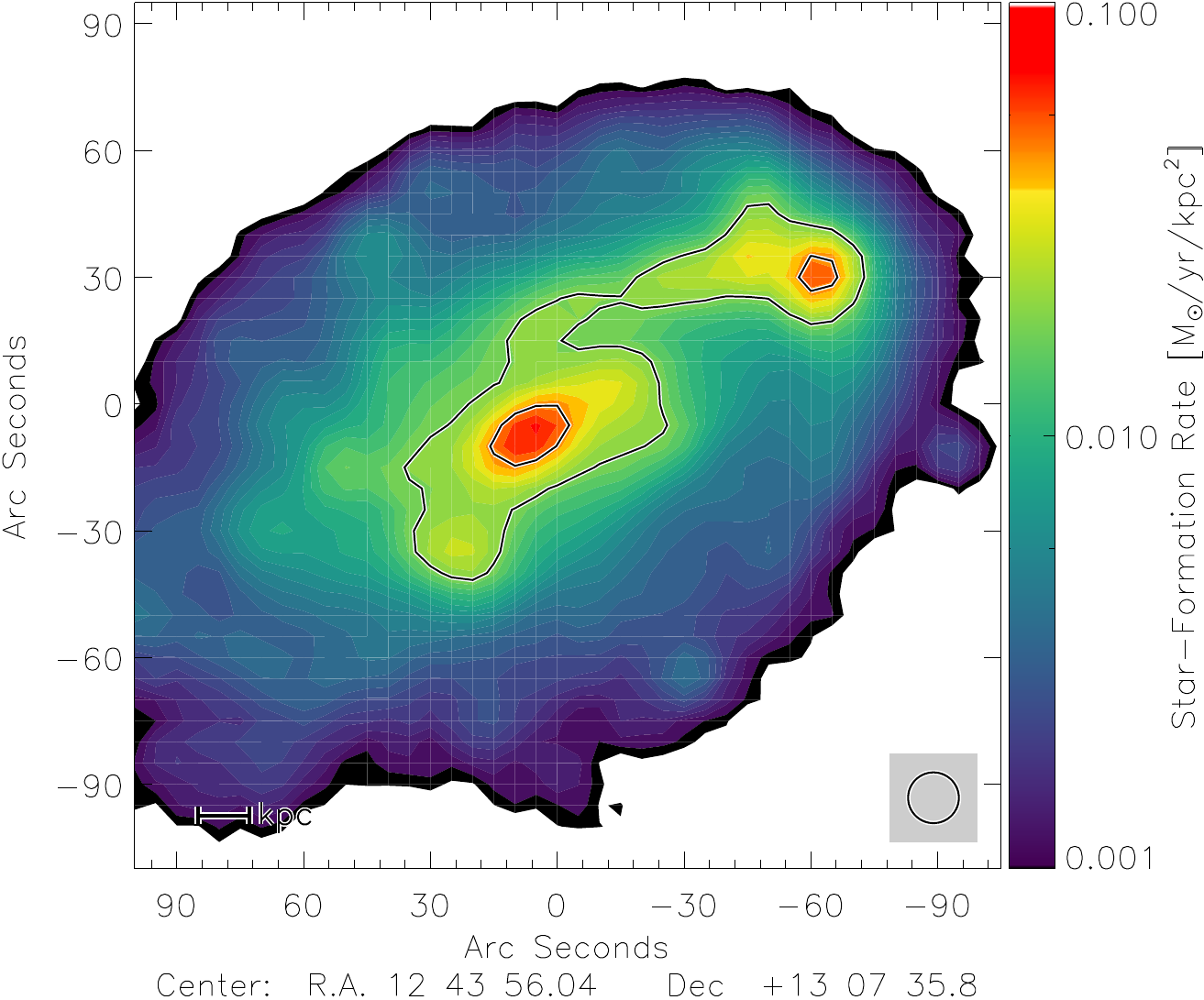}
   \caption{Star-formation map computed using FUV and TIR data.}
   \label{fig:sfrgal}
\end{figure}

\newpage

\newpage
\section{Alternative ram pressure stripping models} \label{others}

\begin{figure}[ht!] 
   \centering
   \includegraphics[width=\hsize]{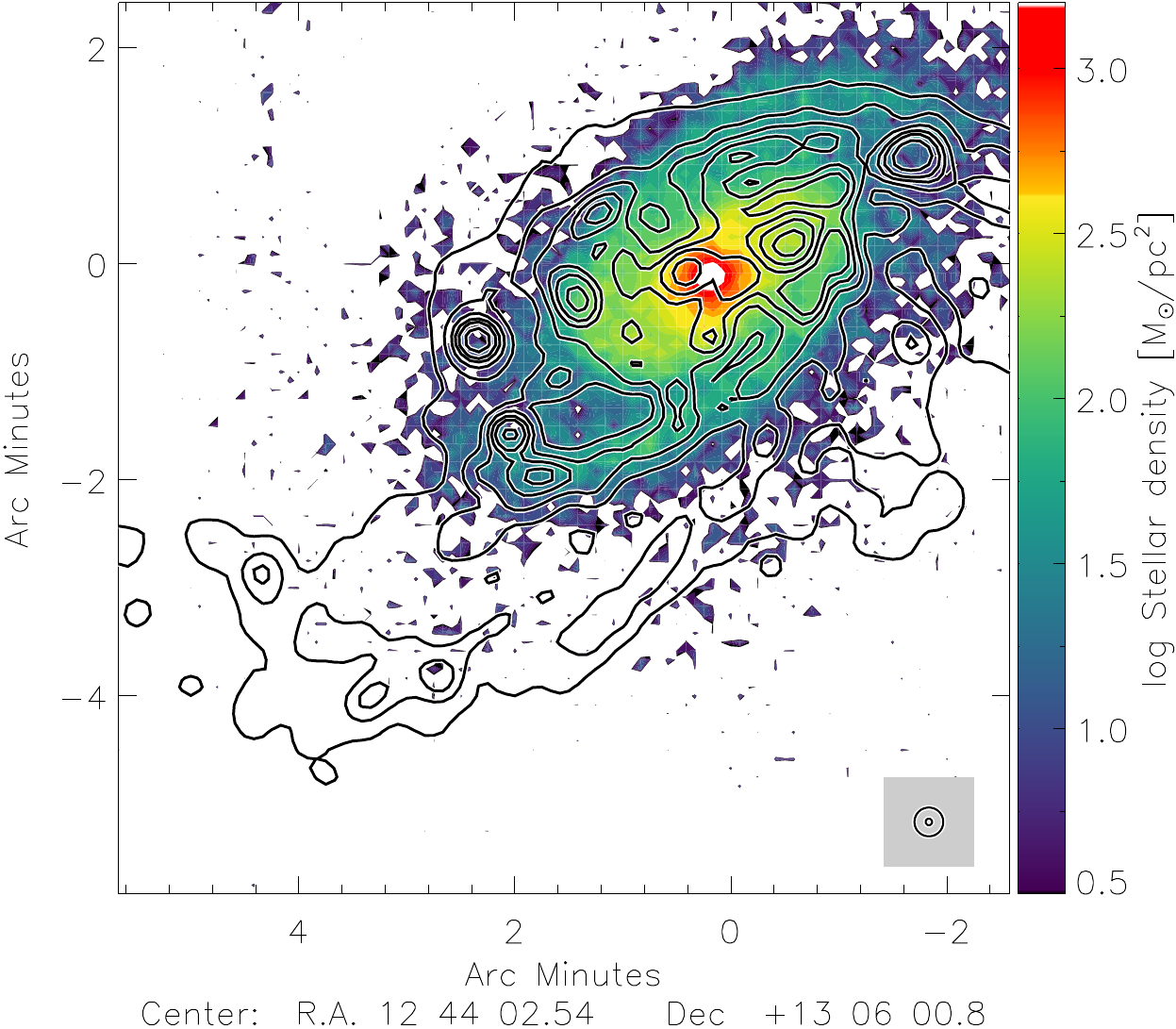}
   \caption{Half wind ram pressure model: stellar and atomic gas surface densities.}
   \label{fig:hisimub}
\end{figure}

\begin{figure}[ht!] 
   \centering
   \includegraphics[width=\hsize]{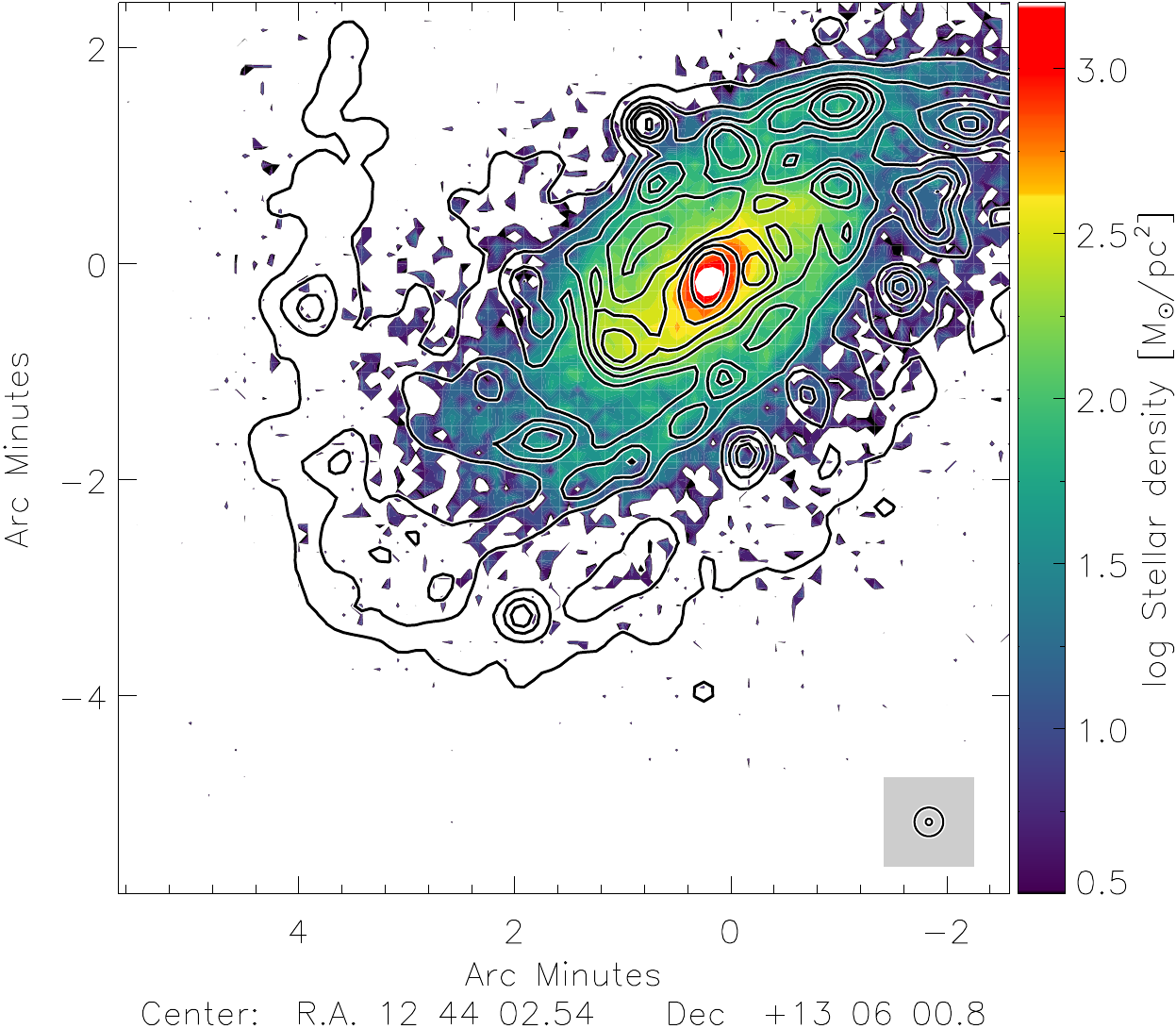}
   \caption{No wind ram pressure model: stellar and atomic gas surface density.}
   \label{fig:hisimuc}
\end{figure}

\begin{figure}[ht!] 
   \centering
   \includegraphics[width=\hsize]{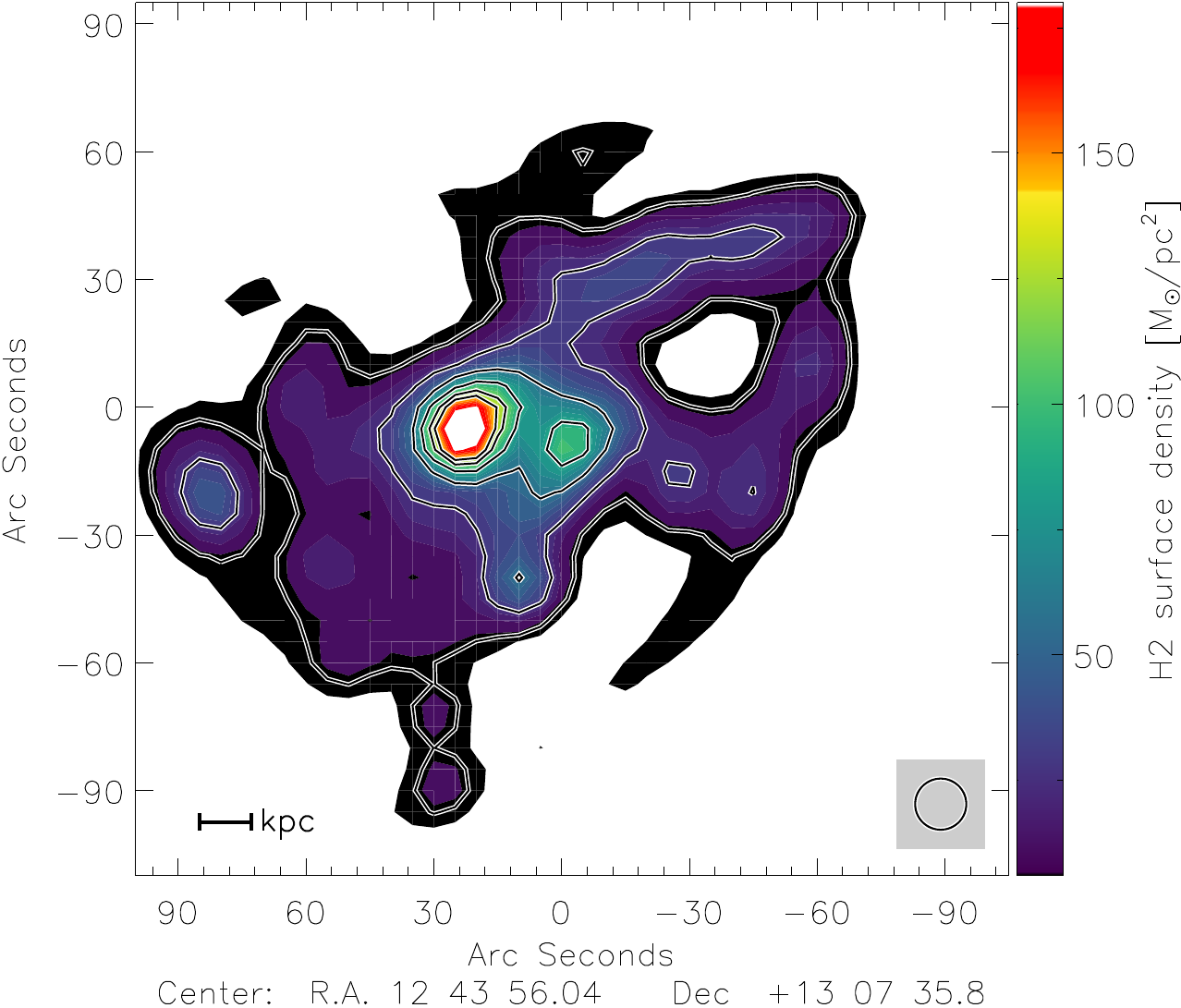}
   \caption{Half wind ram pressure model: molecular gas surface density.}
   \label{fig:h2simub}
\end{figure}

\begin{figure}[ht!] 
   \centering
   \includegraphics[width=\hsize]{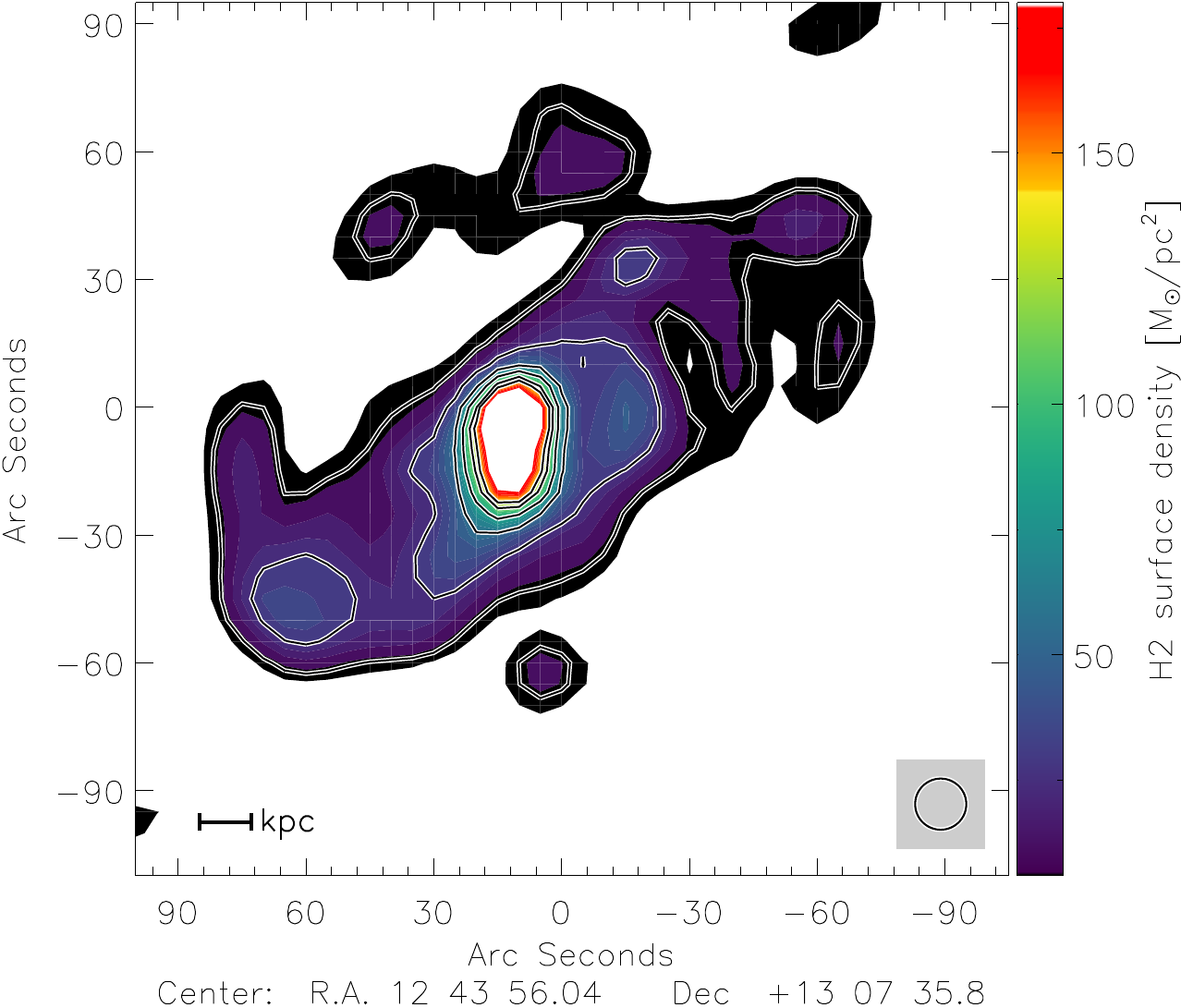}
   \caption{No wind ram pressure model: molecular gas surface density.}
   \label{fig:h2simuc}
\end{figure}

\begin{figure}[ht!] 
   \centering
   \includegraphics[width=\hsize]{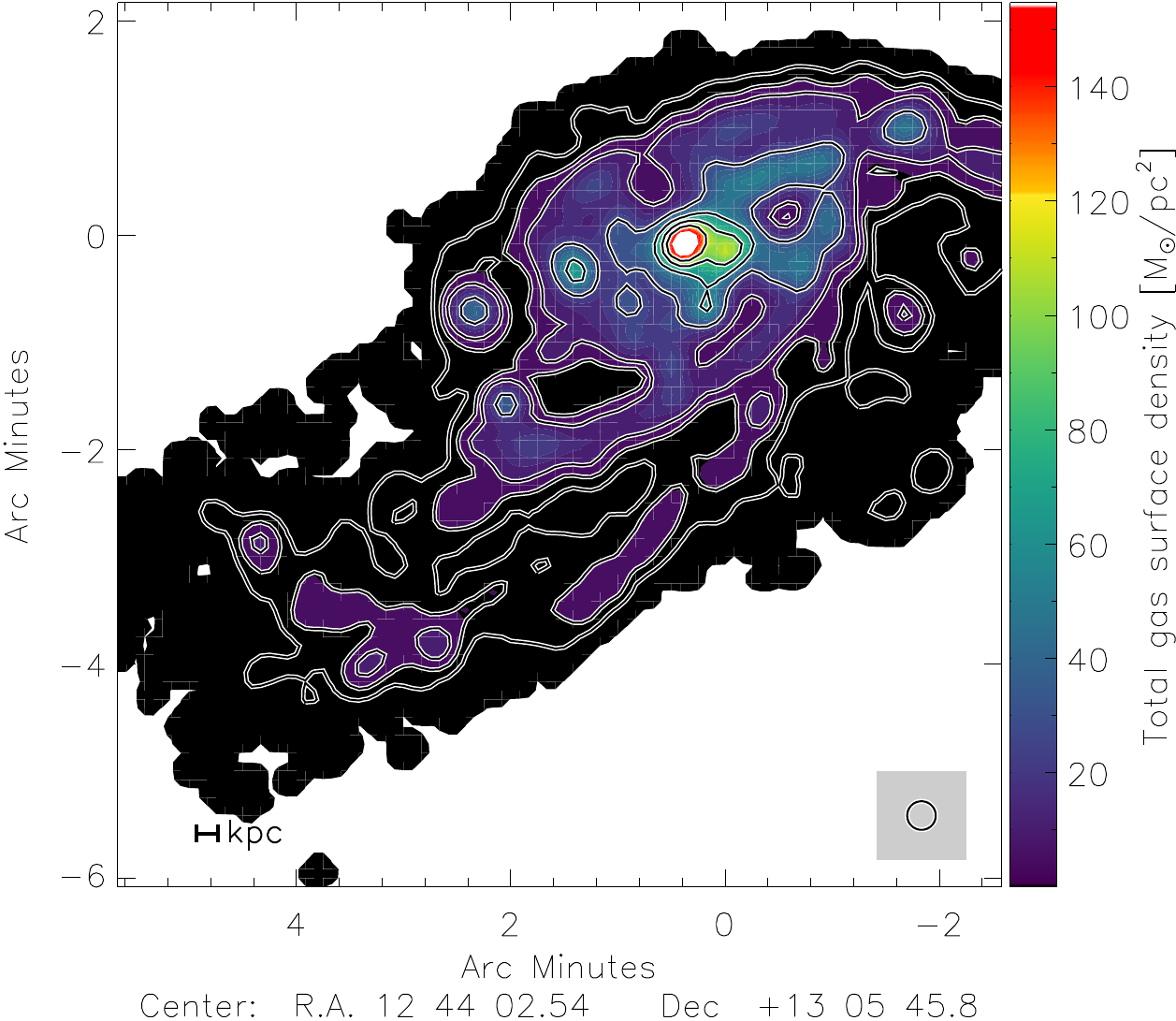}
   \caption{Half wind ram pressure model: total gas surface density.}
   \label{fig:gassimub}
\end{figure}

\begin{figure}[ht!] 
   \centering
   \includegraphics[width=\hsize]{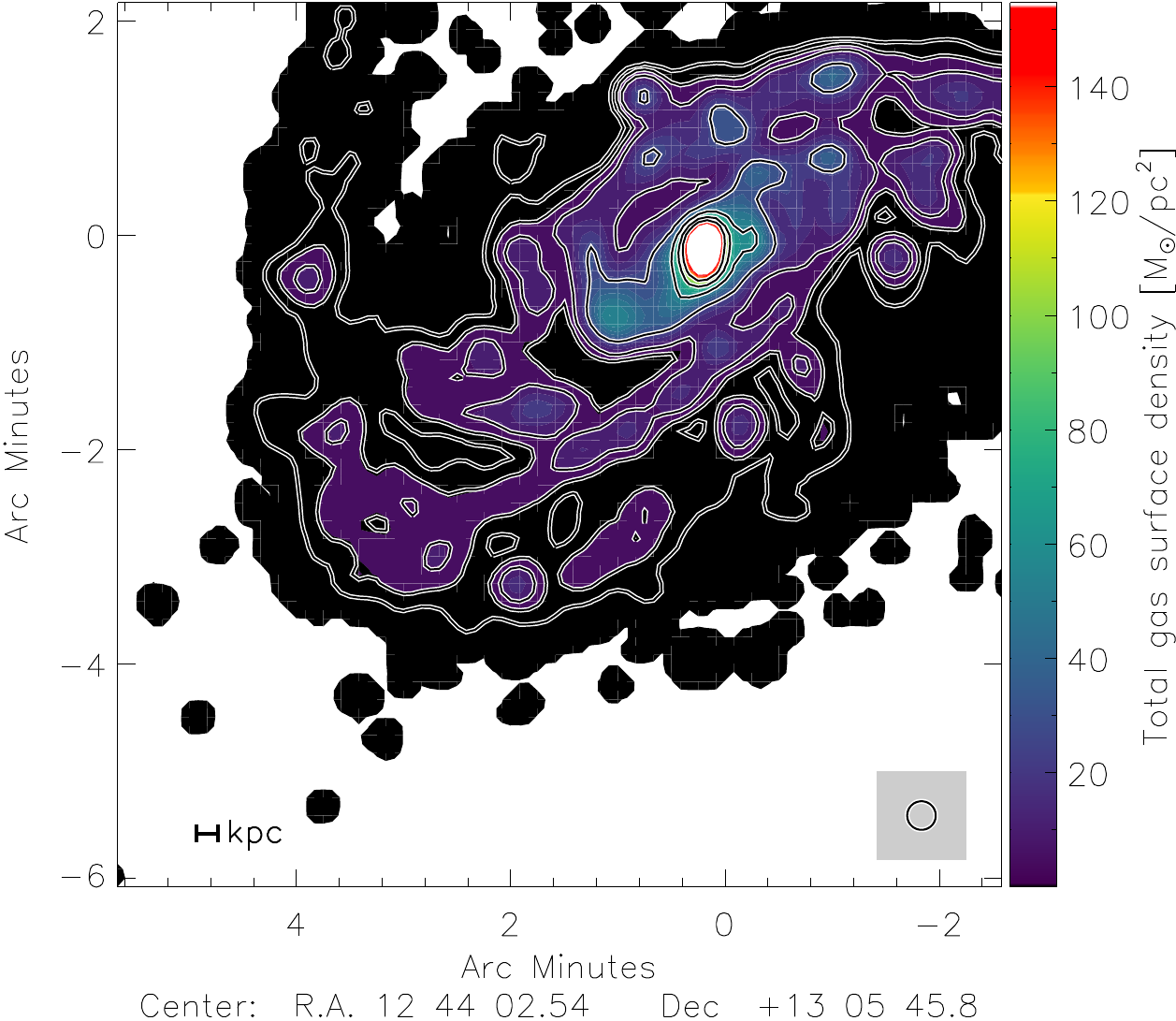}
   \caption{No wind ram pressure model: total gas surface density.}
   \label{fig:gassimuc}
\end{figure}

\begin{figure}[ht!] 
   \centering
   \includegraphics[width=\hsize]{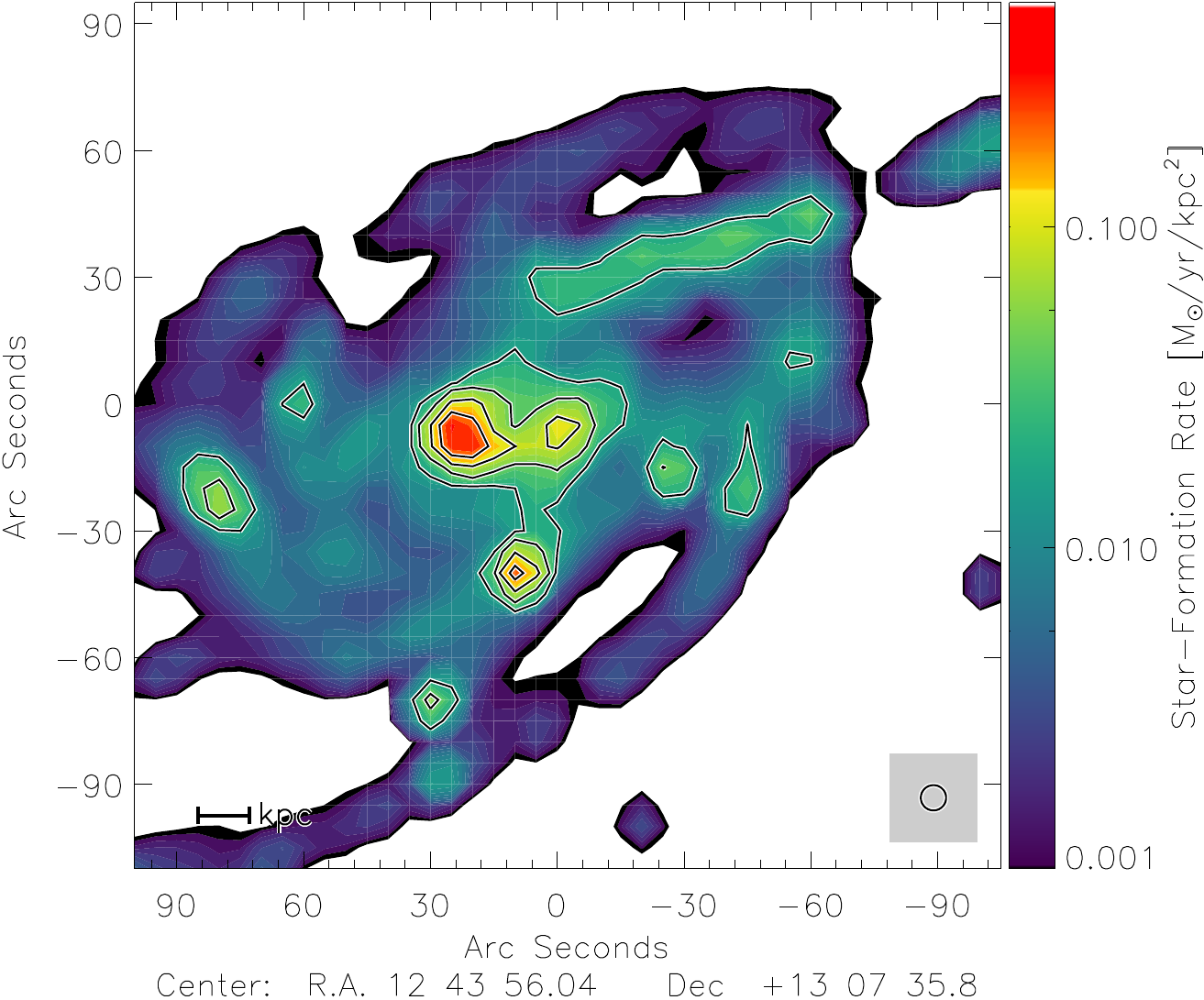}
   \caption{Half wind ram pressure model: SFR.}
   \label{fig:sfrsimub}
\end{figure}

\begin{figure}[ht!] 
   \centering
   \includegraphics[width=\hsize]{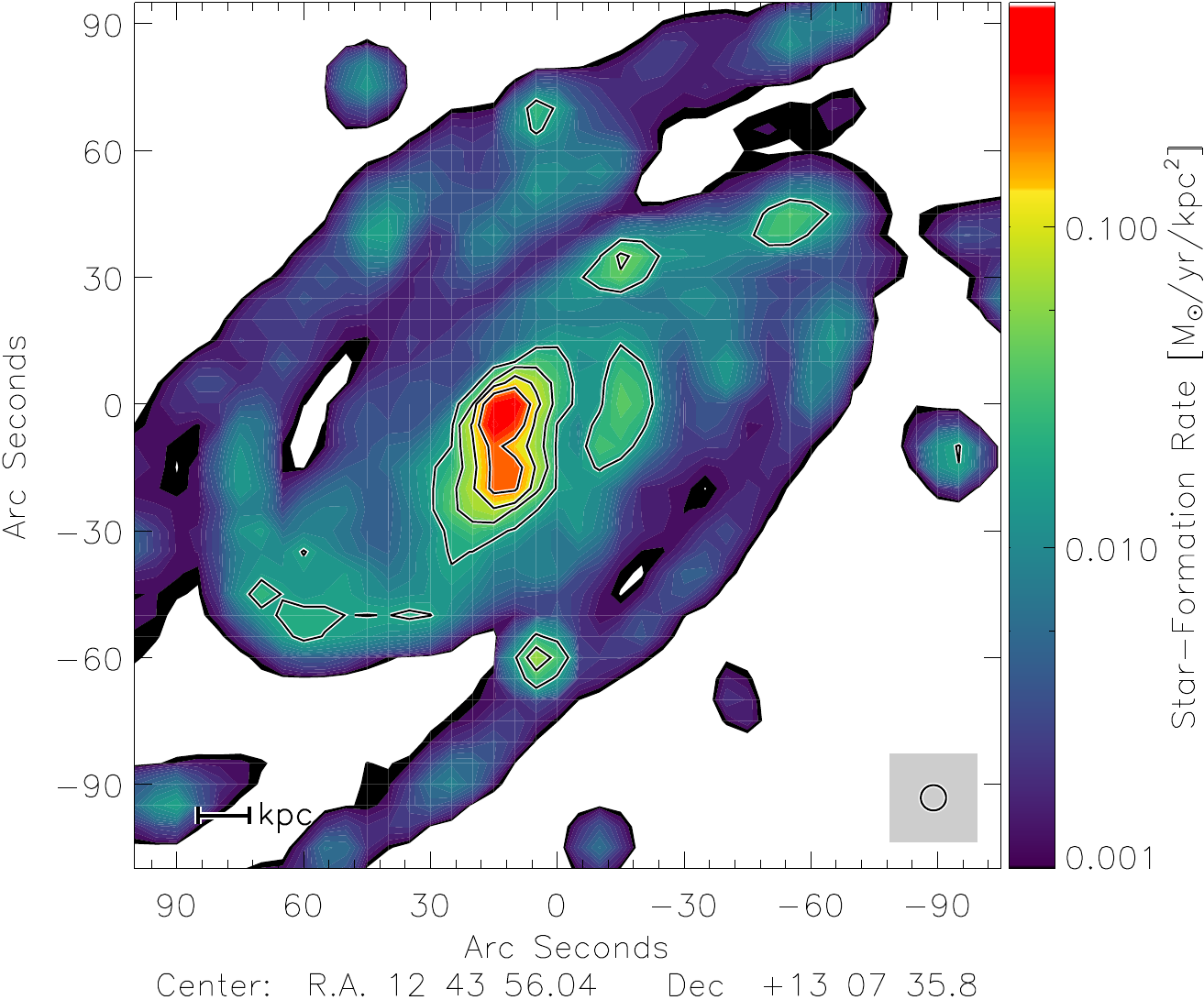}
   \caption{No wind ram pressure model: SFR.}
   \label{fig:sfrsimuc}
\end{figure}

\end{document}